\documentclass[11pt,epsf]{article}
\usepackage{subfigure}
\usepackage{graphics}
\usepackage{amssymb}
\usepackage{epsf}
\usepackage{rotating}
\usepackage{axodraw}
\usepackage{pstricks}

 \let\a=\alpha   \let\b=\beta   \let\g=\gamma   
          
       \let\l=\lambda  
 \let\n=\nu           \let\p=\pi

\def\a{\alpha}
\def\b{\beta}

\def\l{\lambda}

\def\g{\gamma}

\def\tr{\mathop{\rm Tr}}
\def\ds#1{#1\kern-1ex\hbox{/}}
\def\dsh{h\kern-1.2ex /}

\let\G=\Gamma

\newcommand{\bea}{\begin{eqnarray}}
\newcommand{\eea}{\end{eqnarray}}

\def\nn{\nonumber}
\def\beq{\begin{equation}}
\def\eeq{\end{equation}}

\def\beqn{\begin{eqnarray}}
\def\eeqn{\end{eqnarray}}
\def\ba{\begin{eqnarray}}
\def\ea{\end{eqnarray}}

\def\p{{\tt +}}

\def\slash#1{#1\hskip-6pt/\hskip6pt}

\newcommand{\beqa}{\begin{eqnarray}}
\newcommand{\eeqa}{\end{eqnarray}}

\begin{document}

\begin{center}
\vspace{1.5cm}
{\bf\large Axions from Intersecting Branes and Decoupled Chiral Fermions \\ at the Large Hadron Collider}

\vspace{1.5cm}
{\bf\large Claudio Corian\`{o} and Marco Guzzi} 

\vspace{1cm}

{\it  Dipartimento di Fisica, Universit\`{a} del Salento \\
and  INFN Sezione di Lecce,  Via Arnesano 73100 Lecce, Italy}\\
\vspace{.5cm}

~\\
~\\
\begin{abstract}
We present a study of a class of effective actions which show typical axion-like interactions, and of their possible effects at the Large Hadron Collider. One important feature of these models is the presence  of one pseudoscalar which is a generalization of the Peccei-Quinn axion. This can be very light and very weakly coupled, with a mass which is unrelated to its couplings to the gauge fields, described by  Wess Zumino interactions. 
 We discuss two independent realizations of these models, one derived from the theory of intersecting branes and the second one obtained by decoupling one chiral fermion per generation (one right-handed neutrino) from an anomaly-free mother theory. The key features of this second realization are illustrated using a simple example. 
 Charge assignments of intersecting branes can be easily reproduced by the chiral decoupling approach, which remains more general at the level of the solution of its anomaly equations. Using considerations based on its lifetime, we show that in brane models the axion can be dark matter only if its mass is ultralight ($\sim 10^{-4}$ eV), while in the case of fermion decoupling it can reach the GeV region, due to the absence of fermion couplings between the heavy Higgs and the light fermion spectrum. For a GeV axion derived from brane models we present a detailed discussion of its production rates at the LHC. 

\end{abstract}
\end{center}
\newpage
\section{Introduction}
The study of possible signatures of string/brane theory at lower energy has achieved a significant strength with the development, in the last few years, of several extensions of the Standard Model (SM) formulated in scenarios with intersecting branes and large extra dimensions \cite{Antoniadis:2001np,Antoniadis:2002qm,Antoniadis:2000ena,Ibanez:2001nd,Blumenhagen:2006ci,Kiritsis:2003mc}, which are characterized by quite distinct features compared to other constructions, such as those based on more traditional anomaly-free supersymmetric formulations. The latter include specific theories like the MSSM but also its 
further variants such as its next-to-minimal (nMSSM,NMSSM) extensions, eventually with the inclusion of a gauge structure 
enlarged by an extra anomaly-free $U(1)$ gauge symmetry (USSM) \cite{Cvetic:1997ky} (see \cite{Langacker:2008yv} for an overview). 

On the other hand, since anomalous $U(1)$'s are naturally produced in geometrical compactifications and are an important aspect of brane models, the search for possible signatures of string theory has necessarily to take into consideration the peculiarities of these anomalous extensions, which are characterized by anomalous extra neutral currents, contact interactions of Chern-Simons form at trilinear gauge level \cite{Coriano:2005js}
\cite{Anastasopoulos:2006cz}\cite{DeRydt:2007vg} and several axions of St\"uckelberg type. Supersymmetric extensions of these classes of models have also been investigated recently \cite{Anastasopoulos:2008jt,
Coriano:2008aw}.

One of the most demanding feature of these formulations, in regard to possible experimental searches, is to clarify the 
role of gauge anomalies on a substantial sector of collider phenomenology, from precision measurements of leptoproduction to double prompt-photon production \cite{Armillis:2008vp}, just to mention a few processes. These and many more are all affected 
by the new anomalous trilinear gauge vertices \cite{Armillis:2007tb} which appear in these models, although their studies are expected to be quite difficult experimentally.

In fact, the limited accuracy of hadron colliders might reduce the expectations in regard to the possible experimental identification of subtle effects due to the mechanism(s) which underline the cancellation of the gauge anomalies. Nevertheless, the presence of an axion-like particle in the spectra of these theories is an important feature of intersecting brane models, which represents a serious departure from the typical anomaly-free formulations - both for supersymmetric and non-supersymmetric models -  and provides a natural justification for a light pseudoscalar state.

The higher perturbative order  at which these effects start to appear in the perturbative expansion and the limitations of the parton model description seem to indicate that the analysis of anomalous effects are more likely to be the goal of a linear collider rather than that of the LHC, nevertheless the signatures of new physics are manyfold and are not limited to collider physics, but have remarkable implications also in astroparticle physics and cosmology.  

Among the aspects that can be addressed within these new formulations are those related to the flavour sector and the connection between these constructions and the traditional solutions of the strong CP-problem, previously addressed with the help of global $U(1)$ symmetries, such as in the invisible axion model 
 \cite{Peccei:1977ur,Shifman:1979if, Kim:1979if, Zhitnitsky:1980tq, Dine:1981rt}. We recall that studies of the flavour sector of the SM in the presence of gauged anomalous $U(1)$'s are not new, having been used in the past in a variety of cases, for example in the construction of realistic scenarios for neutrino mixing \cite{Irges:1998ax}. At the same time, the study of axion-like particles is at the center of new important proposals for their detection which are now under an intense investigation at DESY \cite{Ahlers:2007qf,Ahlers:2007rd}. Other interesting proposals consider the possible implications of axion-like particles in the propagation of gamma rays 
 \cite{DeAngelis:2007dy}.  We believe that these motivations are sufficient to justify generalized searches of pseudoscalars as a possible solution of the dark matter problem. At the same time anomalous gauge interactions, in combination with quantum gravitational effects, show puzzling features, due to the presence of phantom fields \cite{Coriano:2008pg,Armillis:2009sm} in the local formulation of the trace anomaly \cite{Thomas:2009uh} which deserve a closer look.

\subsection{An axion with independent gauge couplings and mass} 
An axion-like particle is characterized by the usual pseudoscalar couplings to the gauge fields (the 
$b F\tilde{F}$ term, where $b$ is an axion) but has a mass which is 
unrelated to its coupling. Different mass ranges for the axion have quite different implications at a phenomenological level. For instance, for a very light axion ($\approx 10^{-4}-10^{-6}$ eV), as for the PQ case, the pseudoscalar can 
mix with the photon and can generate, in the presence of background galactic magnetic fields, the usual phenomena of birifringence and dichroism for light propagation \cite{Maiani:1986md}, with important effects at astrophysical level \cite{DeAngelis:2007dy,DeAngelis:2007yu} and other experimental signatures \cite{Ahlers:2007qf,Ahlers:2007rd}. The optical activity of the intergalactic medium due to the presence of background axions, also in this generalized case, is essentially caused by the $b F\tilde{F}$ coupling in the equations of motion of the lagrangean \cite{Harari:1992ea}\cite{Coriano:1992bh}.

 In the invisible axion model astrophysical arguments bound the mass of the axion (and its interaction to the gauge fields) requiring its suppression by a large scale $f$. All the axion couplings and the axion mass
\beq
 m\approx  6 \cdot 10^{-6} \textrm{eV} \frac{10^{12}}{f} \textrm{GeV}
\eeq
are inversely proportional to $f$, where $f$ is arbitrary ($f\approx 10^{9}$ GeV experimentally) and makes the axion, indeed, very light. In general, a very light axion, being a quasi-goldstone mode of a global symmetry, is produced copiously  at the center of the sun and escapes after its production, with a mean free path which is larger than the radius of the sun. 
The failure by existing ground-based helioscopes to detect this particle in a detector of Sikivie type 
\cite{Sikivie:1983ip}\cite{Hagmann:1998cb,Sikivie:2007qm} has been used to bound its mass and its interaction with the gauge fields. The bound can be evaded if the axion has 
a mass larger than the temperature at the center of the sun, since in this case would not be produced at its center, mass which is not allowed for the invisible axion according to current constraints. For a very light axion interesting effects are allowed, such as its non-relativistic decoupling, 
since its average momentum at the
QCD phase transition is not of the order of the associated temperature, which is in the GeV range,  
but of the Hubble expansion rate ($3 \cdot 10^{-9}$ eV), and the formation of Bose-Einstein condensates \cite{Sikivie:2009qn}.

 As we have mentioned, the gauging of the axionic symmetries can lift the typical constraints of the invisible axion model, allowing a wider parameter space, which is the main motivation for our study.
  In principle, in the extensions that we consider, this pseudoscalar can be very light, while its gauge interactions can be suppressed by a scale which is given  by the mass of the lightest extra $Z^\prime$ present in the neutral sector of these models. For this reason, these types of pseudoscalars are naturally associated to the neutral current sector, with new implications at the level of the trilinear gauge interactions.

So far, two models have been developed in which the structure of the effective action allows a physical axion: the MLSOM (the Minimal Low-Scale Orientifold Model) \cite{Coriano':2005js} and the USSM-A \cite{Coriano:2008xa,Coriano:2008aw}, the first being a non-supersymmetric model, the second a supersymmetric one. In the first model, motivated by a construction based on intersecting branes, the scalar sector involves beside the St\"uckelberg axions, 2 Higgs doublets. At the same time, the gauge structure of the Standard Model is corrected by the presence of extra neutral currents due to the extra $U(1)$. 

To date, a detailed analysis of these models is contained in  
\cite{Coriano:2007xg}, worked out for a single extra $U(1)$. In the supersymmetric case the presence of a physical axion is guaranteed if the superpotential allows extra superfields which are singlet respect to the Standard Model but are charged under the anomalous $U(1)$'s. The field content of the superpotential of the nMSSM is sufficient to have a physical axion in the spectrum \cite{Coriano:2008xa,Coriano:2008aw}.

\subsection{Gauged axions} 
The gauging of axionic symmetries is realized in the low energy effective lagrangeans by introducing (shifting) axions, one for each anomalous $U(1)$ present in the gauge structure of a given model. These are accompanied by Wess-Zumino terms in order to restore the gauge invariance of the theory due to the chiral anomalies present in these constructions. These axions (St\"uckelberg axions) are not all physical fields. In fact, the only physical axion, called the "axi-Higgs" in \cite{Coriano':2005js}, is identified in the CP-odd sector of the scalars by a joint analysis of the potential and of the bilinear mixing terms $(B_i\partial b_i )$ generated by the 
St\"uckelberg mass terms which are present for each anomalous $U(1)$. We are going to summarize below the scalar of the scalar potentials which allow a physical axion in the spectrum, either massless or massive. Since the mass of this particle is expected to obtain small non-perturbative corrections due to the instanton vacuum, as for the invisible axion,  these small corrections are described by extra terms in the scalar potential which are allowed by the symmetry. These terms make the physical axion part of the scalar potential, but their size remains, in the class of theories that we analyze, essentially unspecified. In supersymmetric models they are expected to correspond to non-holomorphic corrections to the superpotential \cite{Coriano:2008xa} which involve directly the axion/axino superfield.
The size of these corrections depends on the way the fundamental symmetry is broken, and the appearance of the axion in the scalar potential just parameterizes our ignorance of the fundamental mechanism which is responsible for these corrections. For this reason, we focus our analysis on several mass windows for this particle, although the most relevant mass range for collider studies is the GeV region.

\subsection{Organization of this work}
The analysis presented in this work concerns the phenomenology of the axi-Higgs in anomalous abelian models with a single anomalous extra $U(1)$ and in the non-supersymmetric case. The construction, therefore, is the one typical of the MLSOM, formulated in the context of intersecting branes. A similar analysis can be performed in the supersymmetric case, although it is more complex and will be presented elsewhere. Our analysis, however, is not limited to models of intersecting branes, but to the entire class of effective actions which are characterized by axion-like interactions at low energy, independently from their high energy completion. Typical charge embeddings of brane constructions, as we are going to show, can nevertheless be obtained in our approach starting from an anomaly-free spectrum and decoupling some chiral fermions. Some differences between the two realization remain, at phenomenological level, since the corresponding axion, in the case of decoupled fermions, does not couple to the light fermions which are part 
of the low-energy spectrum.

Our motivations for working within this more general framework has been motivated by scenarios where a heavy fermion, for instance a right-handed neutrino, decouples from the low energy spectrum leaving one St\"uckelberg axion (the phase of a Higgs field) in the effective lagrangean. We will come to discuss these points in more detail in one of the sections below. The different completions of these lagrangeans start differing at the level of operators whose mass dimensions is larger than 5, the five dimensional ones being the Wess-Zumino terms.  

After reviewing briefly these models in order to make our analysis self-contained, we illustrate how their anomalous content can be obtained by requiring that only some of the anomaly equations are satisfied, taking as a starting point an anomaly-free chiral spectrum  and decoupling some chiral fermions. Typical brane models such as the Madrid model \cite{Ibanez:2001nd} are obtained for a particular choice of the free charges allowed by the decoupling of the heavy chiral fermions and are just particular solutions of the anomaly equations. We then move towards a phenomenological analysis of the axi-Higgs in the MLSOM, selecting the GeV mass range for the axion. This region is the most promising one for collider studies of this particle, although in this range, as we are going to show, it is not long-lived. A GeV axion can be long lived, but must have suppressed couplings to the fermions of the low-energy spectrum, and one way of getting this lagrangean is via the mechanism of decoupling of heavy fermions (and of the radial excitations of the associated Higgs field) from the low energy theory. 
We show, using a simple toy model, how this can occur.

Production and decay rates for this particle are studied for all the mass windows in the MLSOM for typical LHC searches. We give in an appendix a summary of the scalar sector of the lagrangean and the determination of coefficients of the Wess Zumino terms. We have also included a section where we present a discussion and a comparison of the effective action of intersecting brane models versus the analogous one obtained by decoupling a chiral fermion, illustrating briefly the origin of the various operators left in the low energy formulation, with the axion interpreted as the phase of a second Higgs sector, partially decoupled from the 2 Higgs doublets included in the electroweak sector.

\section{The model: overview of its general structure}
We analyze a class of models characterized by a gauge structure of the form $SU(3)\times SU(2)\times U(1)_Y\times U(1)_B$, defined in \cite{Coriano:2007xg}, where the $U(1)_B$ gauge symmetry is anomalous and the corresponding gauge boson $(B)$ undergoes mixing with the rest of the gauge bosons of the Standard Model. Details can be found in \cite{Coriano:2007xg,Coriano:2007xg,Coriano:2007fw}; here we just 
summarise the main features of this construction for which we will define rather general charge assignments.  As we have already stressed, the reason for keeping our analysis quite general is motivated by the observation that effective actions of intersecting brane models are not uniquely identified. Various completions can generate the same low energy signatures, at least up to operators of dimension 5, which, for anomalous gauge theories, are the Wess-Zumino terms. These points will be illustrated in a section below, where we will solve the basic equations that characterize the charge assignments of the anomalous model, under some assumptions on the fermion spectrum which are essential in order to make our analysis concrete.

\subsection{The structure of the effective action}
The effective action has the structure given by
\beqn
{\mathcal S} &=&   {\mathcal S}_0 + {\mathcal S}_{Yuk} +{\mathcal S}_{an} + {\mathcal S}_{WZ} + {\mathcal S}_{CS}
\label{defining}
\eeqn

where ${\mathcal S}_0$ is the classical action which is given in an appendix. It contains the usual gauge degrees of freedom of the Standard Model plus the extra anomalous gauge boson $B$ which is already massive, before electroweak symmetry breaking, via a St\"uckelberg mass term. The scalar potential is the maximal one permitted  by the symmetry and allows electroweak symmetry breaking. The structure of the Yukawa sector ${\mathcal S}_{Yuk}$ is very close to that of the Standard Model. In one of the sections below we identify the fundamental physical degrees of freedom of this sector after electroweak symmetry breaking, which, in our analysis, is based on the choice of the largest potential allowed by the symmetry. The model is a canonical gauge theory with dimension-4 operators plus dimension 5 counterterms of Wess-Zumino type. 

In Eq. (\ref{defining}) the anomalous contributions coming from the 1-loop triangle diagrams involving abelian and non-abelian gauge interactions are summarized by the expression
\beqn
{\mathcal S}_{an}&=& \frac{1}{2!} \langle T_{BWW} BWW \rangle +  \frac{1}{2!} \langle T_{BGG} BGG \rangle
 + \frac{1}{3!} \langle T_{BBB} BBB \rangle     \nonumber\\
&&+ \frac{1}{2!} \langle T_{BYY} BYY \rangle + \frac{1}{2!} \langle T_{YBB} YBB \rangle,
\eeqn
where the symbols $\langle \rangle$ denote integration. For instance, 
the anomalous contributions in configuration space are given explicitly by 
\ba
\langle T_{B W W} B W W \rangle&\equiv& \int dx\, dy \, dz
  T^{\lambda \mu \nu, ij}_{BWW}(z,x,y) B^{\lambda}(z) W^{\mu}_{i}(x)
W^{\nu}_{j}(y)
\ea
and so on, where $T_{BWW}$ denotes the anomalous triangle diagram with 
one $B$ field and two $W$'s external gauge lines. The gluons  are denoted by $G$.

In the same notations
the Wess Zumino (WZ) counterterms are given by
\beqn
{\mathcal S}_{WZ}&=& \frac{C_{BB}}{M} \langle b  F_{B} \wedge F_{B}  \rangle
+ \frac{C_{YY}}{M} \langle b F_{Y} \wedge F_{Y}  \rangle + \frac{C_{YB}}{M} \langle b F_{Y} \wedge F_{B}  \rangle \nonumber\\
&&+ \frac{F}{M} \langle b Tr[F^W \wedge F^W]  \rangle  +  \frac{D}{M} \langle b Tr[F^G \wedge F^G] \rangle,
\eeqn
while the gauge dependent CS abelian and non abelian counterterms \cite{Anastasopoulos:2006cz} needed to cancel
the mixed anomalies involving a B line with any other gauge interaction of the
SM take the form
\beqn
{\mathcal S}_{CS}&=&+ d_{1} \langle BY \wedge F_{Y} \rangle + d_{2} \langle YB \wedge F_{B} \rangle  \nonumber\\
&&+ c_{1} \langle \epsilon^{\mu\nu\rho\sigma} B_{\mu} C^{SU(2)}_{\nu\rho\sigma} \rangle
+ c_{2} \langle \epsilon^{\mu\nu\rho\sigma} B_{\mu} C^{SU(3)}_{\nu\rho\sigma} \rangle,
\eeqn
with the non-abelian CS forms given by
\beqn
C^{SU(2)}_{\mu \nu \rho} &=&  \frac{1}{6} \left[ W^{i}_{\mu} \left( F^W_{i,\,\nu \rho} + \frac{1  }{3} \, g^{}_{2}  
\, \varepsilon^{ijk} W^{j}_{\nu} W^{k}_{\rho}  \right) + cyclic   \right]              ,    \\
C^{SU(3)}_{\mu \nu \rho} &=&  \frac{1}{6} \left[ G^{a}_{\mu} \left( F^G_{a,\,\nu \rho} + \frac{1 }{3} \, g^{}_{3}
\, f^{abc} G^{b}_{\nu} G^{c}_{\rho}  \right) + cyclic  \right].
\eeqn
The only constraint which fixes the coefficients in front of the WZ counterterms is gauge invariance. Specifically, the anomalous variation of $\mathcal{S}_{an}$ is compensated by the variation of $\mathcal{S}_{WZ}$. Imposing this condition one discovers that the scale of the WZ counterterms (M) becomes the St\"uckelberg mass term $M_{St}\equiv M_1$. This is found in the defining phase of the model, in which the realization of the gauge symmetry is 
in the St\"uckelberg form. Obviously, in this phase only the $B$ gauge boson is massive (in a St\"uckelberg phase). The breaking of the electroweak symmetry, triggered by the Higgs potential and the transition to the mass eigenstates determines a rotation of the St\"uckelberg axion $b$ into a physical axion $\chi$ plus some Nambu-Goldstone modes. This rotation brings in a redefinition of the suppression scale $M$, which now coincides with the mass of the extra $Z^\prime$ gauge boson, as shown in an appendix. 

\subsection{The scalar potentials and their axion-dependent phases}

In previous studies it has been shown that anomalous abelian models, realized in the case of potentials with 
2 Higgs doublets, both in the non-supersymmetric and in the supersymmetric cases, are characterized by the presence of an axion-like particle in the spectrum. In the context of the 2 Higgs doublets model shown in detail in \cite{Coriano':2005js, Coriano:2007xg} the presence of PQ-breaking terms in the scalar potential allows the axion to become massive. 
The PQ symmetric contribution is given by
\bea
V_{PQ}(H_u, H_d) = \sum_{a=u,d} \Bigl(  \mu_a^2  H_a^{\dagger} H_a + \l_{aa} (H_a^{\dagger} H_a)^2\Bigr)
-2\l_{ud}(H_u^{\dagger} H_u)(H_d^{\dagger} H_d)+2{\l^\prime_{ud}} |H_u^T\tau_2H_d|^2,
\eea
which is a pure Higgs scalar potential, while in the PQ-breaking terms we introduce a dependence on the axion field $b$ by means of explicit phases 
\begin{eqnarray}
V_{\slash{P}\slash{Q}}(H_u, H_d, b) &=&  b_{1} \, \left( H_u^{\dagger} H_d \, e^{-i  \Delta q^B \frac{b}{M_1}}  \right)
+ \lambda^{}_1 \left( H_u^{\dagger}H_d \,e^{-i \Delta q^B \frac{b}{M_1}} \right)^2  \nonumber\\
&&+ \lambda^{}_2 \left( H_u^{\dagger}H_u \right) \left( H_u^{\dagger}H_d \,e^{-i \Delta q^B \frac{b}{M_1}} \right)
+ \lambda^{}_3 \left( H_d^{\dagger}H_d \right) \left( H_u^{\dagger}H_d \,e^{-i \Delta q^B  \frac{b}{M_1}} \right) + h.c. \nonumber \\
\label{PQbreak}
\end{eqnarray}
where $\Delta q^B=q^B_u-q^B_d$, $b^{}_{1}$ has mass squared dimension, while $\lambda^{}_{1}$, $\lambda^{}_{2}$, $\lambda^{}_{3}$ are dimensionless couplings. 
In the scalar potential we can isolate three sectors, namely, two neutral and one charged sector, which are described by the quadratic expansion of the potential around its minimum 
\beqa
&&V_{CP-even}(H_u,H_d) + V_{CP-odd}(H_u,H_d,b)+ V_{\pm}(H_u,H_d)=
\nonumber\\\\
&&\left({H_u}^-, {H_d}^-\right){\cal N}_1\left(\begin{array}{c}
{H_u}^+\\
{H_d}^+ \\
\end{array}\right) + \left(Re{H_u}^0, Re{H_d}^0\right){\cal N}_2\left(\begin{array}{c}
Re{H_u}^0\\
Re{H_d}^0 \\
\end{array}\right) 
\nonumber \\
&& + \left(Im {H_u}^0, Im{H_d}^0, a_I^\prime \right){\cal N}_3\left(\begin{array}{c}
Im{H_u}^0\\
Im{H_d}^0 \\
b\\
\end{array}\right).
\eeqa

\begin{itemize} 
\item{\bf The Charged Sector}
\end{itemize}
In the charged sector we find a zero eigenvalue of the mass matrix, corresponding to the Goldstone mode $G^+$
and the nonzero eigenvalue
\beqa
m^2_{H^+} &=&
4{\lambda'}_{ud}v^2 -2 \left(
\frac{2b }{v^2\sin{2\b}} + 2\l_1 + \tan{\b} \l_2 + \cot{\b} \l_3 \right)v^2,
\eeqa
corresponding to the charged Higgs mass. 
The two vevs of the Higgs sector are defined by $v_{d}=v \cos\beta;  v_{u}=v \sin\beta$, with $v^2=v_u^2 + v_d^2$. The rotation matrix into the physical eigenstates is
\beq
\left(\begin{array}{c}
{H_u}^+\\
{H_d}^+ \\
\end{array}\right)=
\pmatrix{ \sin \beta  & -\cos {\beta} \cr
 \cos {\beta} & \sin {\beta}}
\left(\begin{array}{c}
{G}^+\\
{H}^+ \\
\end{array}\right).
\label{chargedrot}
\eeq

\begin{itemize}
\item{\bf The CP-even Sector}
\end{itemize}
In the neutral sector both a CP-even and a CP-odd subsectors are present.
The CP-even sector is described by ${\cal N}_2$ which can be diagonalized by an appropriate rotation matrix
in terms of CP-even mass eigenstates $(h^0,H^0)$ as
\beq
\left(\begin{array}{c}
{Re H_u}^0\\
{Re H_d}^0 \\
\end{array}\right)=
\pmatrix{ \sin \a  & -\cos {\a} \cr
 \cos {\a} & \sin {\a}}
\left(\begin{array}{c}
{h}^0\\
{H}^0 \\
\end{array}\right),\label{chargedrot1}
\eeq
with
\beq
\tan\a= \frac{{\cal N}_2(1,1) - {\cal N}_2(2,2) - \sqrt{\Delta}}{2 {\cal N}_2(1,2)}\label{alpha}
\eeq
and
\beq
\Delta=\left({\cal N}_2(1,1)\right)^2 - 2 {\cal N}_2(2,2){\cal N}_2(1,1)
+ 4 \left({\cal N}_2(1,2)\right)^2 +
\left({\cal N}_2(2,2)\right)^2.
\eeq
The definition of these matrix elements is left to an appendix.
The eigenvalues corresponding to the physical neutral Higgs fields are given by
\beqa
m_{h^0}^2 &=& \frac{1}{2}\left( {\cal N}_2(1,1)+{\cal N}_2(2,2) - \sqrt{\Delta}\right)\label{SMHiggs}
\nonumber \\
m_{H^0}^2 &=& \frac{1}{2}\left( {\cal N}_2(1,1)+{\cal N}_2(2,2) + \sqrt{\Delta}\right).
\eeqa
We refer to \cite{Coriano':2005js} for a more detailed discussion of the
scalar sector of the model with more than
one extra $U(1)$.
\begin{itemize}
\item{\bf The CP-odd sector}
\end{itemize}
The symmetric matrix describing the mixing of the CP-odd Higgs sector with 
the axion field $b$ is given by ${\cal N}_3$. After the diagonalization we can construct the orthogonal matrix $O^\chi$ that rotates the 
St\"uckelberg field and the CP-odd phases of the two Higgs doublets into the mass eigenstates 
$(\chi, G^{\,0}_1, G^{\,0}_2)$
\bea
\left(
\begin{array}{c} ImH^{0}_{u} \\
                 ImH^{0}_{d} \\
					  b
\end{array} \right)=O^{\chi}
\left(
\begin{array}{c} \chi\\
                 G^{0}_1\\
					  G^{0}_2
\end{array} \right).
\label{CPodd}
\eea
The mass matrix of this sector exhibits two zero eigenvalues 
corresponding to the Goldstone modes $G^{0}_1, G^{0}_2$ and a mass eigenvalue, 
that corresponds to the physical axion field $ \chi$, with a value
\bea
m_{\chi}^2 =
-\frac{1}{2} \, c^{}_{ \chi} \, v^2  \left[ 1 + \left(  \frac{q_u^B-q_d^B}{M_1}\,
 \frac{v \, \sin{2\b} }{2} \right)^2 \right] = 
-\frac{1}{2} \, c^{}_{ \chi} \, v^2  \left[ 1 +  \frac{ ( q_u^B - q_d^B )^{2}}{M^{2}_{1}} \,
\frac{v^{2}_{u} v^{2}_{d} }{v^2}  \right],
\label{axionmass}
\eea
with the coefficient
\bea
c^{}_{\chi} = 4 \left( 4 \lambda_1 +  \lambda_3 \cot \beta + \frac{ b_{1} }{ v^2 } \frac{ 2 }{ \sin 2\beta } 
+  \lambda_2  \tan\beta  \right).
\eea
The mass of this state is positive if $c_{\chi} < 0$. The Goldstone bosons $(G_Z,G_{Z^\prime})$ are obtained by orthonormalizing $(G^0_1,G_2^0)$ that 
span a two dimensional space.  Notice that, in general, the mass of the axi-Higgs is the result of two effects: 
the presence of the Higgs vevs and the presence of the St\"uckelberg mass via the PQ-breaking 
potential. In the particular case of a charge assignment such that $q^B_u =q^B_d$, in the PQ-breaking potential the dependence on the axion field disappears ($V_{\slash{P}\slash{Q}}(H_u, H_d, b) \rightarrow V_{\slash{P}\slash{Q}}(H_u, H_d)$) and the rotation matrix simplifies to
\bea
\left(
\begin{array}{c} ImH^{0}_{u} \\
                ImH^{0}_{u} \\
					  b
\end{array} \right)=
\pmatrix{
- \cos \beta          &     \sin \beta  &      0   \cr
       \sin \beta      &    \cos \beta  &       0   \cr
           0               &    0               &   1      
             } 
\left(
\begin{array}{c} A^0\\
                 G^{0}_1\\
                 G^{0}_2
\end{array} \right)\,.
\eea
For this particular assignment of the Higgs charges the $Z$ and $Z^\prime$ bosons are still massive, as can be seen from eqs.~(\ref{massZ}, \ref{massZp}).  A brief counting of the physical degrees of freedom shows, also in this case, that we expect only one physical particle in the CP-odd sector. 
Then, in this particular case, it is easily found that the model doesn't exhibit Higgs-axion mixing because the physical degree of freedom $A^0$, as identified by the scalar potential, is a combination of the imaginary parts of the two Higgs ${Im}H^{0}_{u}, {Im}H^{0}_{u}$, while the axion is only part of the Goldstones modes $G_Z$ and $G_{Z^\prime}$, identified by an inspection of the derivative couplings. 

\section{ Axions from the decoupling of a chiral fermion} 
Other realizations of these effective 
models are obtained by studying the decoupling of a chiral fermion from an 
original anomaly-free theory, due to large Yukawa couplings \cite{Coriano:2006xh}. The remnant axion, 
in this particular realization, is the surviving massless phase of a heavy Higgs. We will illustrate briefly this approach sketching the derivation, though in the case of a simple model, in a section below. Obviously, in these types of completions of the anomalous theory, the challenge of the construction would consist in the identification of a pattern of 
sequential breaking of the underlying anomaly-free theory in order to generate suitable axion-like Wess-Zumino 
interactions, which are not part of our simple example. 

For instance, considerable motivations for this reasoning comes from unified models based on an anomaly-free fermion spectrum assigned to special representations of the gauge symmetry. Specifically, one could consider the \underline{{\bf 16}} of $SO(10)$ in which find accommodation the fermions of an entire generation of the Standard Model plus a right handed neutrino. The decoupling of a right handed neutrino could leave a remnant pseudoscalar in the spectrum with axion-like couplings. While the explicit realization of this  construction and the (sequential) breaking of the original GUT towards the spectrum of the Standard Model is rather complex, the implications of these assumptions can be grasped by a simple model.

To illustrate these points, we introduce a simple toy model and show step by step that a specific form of the decoupling can generate a certain dynamics at low energy which is completely described by an effective action 
with St\"uckelberg and a Higgs-Stuckelberg phases, Wess Zumino interactions and higher dimensional operators suppressed by the St\"uckelberg mass. 
 It should be mentioned that in our example, the low energy gauge boson $B$, which has anomalous effective interactions, would be massive in the St\"uckelberg form. We recall that 
the study of the St\"uckelberg construction  has been discussed recently in several 
works \cite{Feldman:2006wb,Feldman:2007wj} (see also \cite{Ruegg:2003ps}) for non-anomalous theories, with its possible experimental signatures.

The model requires two Higgs fields, here assumed to be two complex
 scalars, and a potential characterized by a first breaking of the anomaly-free gauge symmetry at a certain scale ($v_\phi$), followed by a second breaking at a lower scale $v_H $ 
($v_H << v_{\phi}$). The heavy Higgs is assumed to decouple (partially) after the first breaking. Specifically, the decoupling involves the radial fluctuations ($\rho$) of the field $\phi$, and all the interactions which are characterized by operators which are suppressed by a certain power of $\rho/v_{\phi}$. We expand the heavy Higgs $\phi$ as 

\beq
\phi\sim \left( \frac{v_{\phi}+\rho }{\sqrt{2}}\right) e^{i\theta}
\eeq
with $\theta$ denoting a massless phase that may be rendered massive during the process of decoupling of the radial excitation by some small tilting, as it occurs for the ordinary Peccei-Quinn axion (PQ). 
The (almost massless) phase remains in the low energy theory.  The St\"uckelberg axion is identified from $\theta$ in a certain way, that will be specified below. Also we assume, for simplicity, that only one chiral fermion becomes heavy in the course of decoupling of the heavy Higgs,  and is integrated out of the low energy spectrum.  As we have already stressed, our approach can be made more realistic, but we expect that the crucial steps that bring to its specific effective action at low energy can be part of a more complete theory.
 
 The Yukawa couplings, expanded around the vacuum of the heavy Higgs, show the presence of a complex phase ($\theta$) that we try to remove by a chiral redefinition of the integration measure before we integrate out the heavy fermion. It is this chiral redefinition of the fermionic measure which induces, by Fujikawa's approach, typical Wess-Zumino terms in the low energy effective theory. This theory, obviously, admits a derivative expansion in terms of the large scale $v_{\phi}$, which can be systematically captured by a derivative expansion in $1/v_{\phi}$, or equivalently, the St\"uckelberg mass, since the two scales are related ($M_1\sim g_B v_\phi$).

\subsection{Partial integration } 
To be specific, we consider a model with 2 fermions 
and a gauge symmetry of the form $U(1)_A\times U(1)_B$, where $A$ is vector like and $B$ is the anomalous gauge boson. We define the lagrangean 

\beqa
\mathcal{L}& =& -\frac{1}{4} F^A_{\mu\nu}F^{A\, \mu\nu}  -\frac{1}{4} F^B_{\mu\nu}F^{B\, \mu\nu} 
+ \sum_{i=1}^{2} \left( \bar{\psi}^{(i)}_L \slash{D} \psi^{(i)}_L + \bar{\psi}^{(i)}_R \slash{D} \psi^{(i)}_R \right)
\nonumber \\
&&+ \lambda  \bar{\psi}^{(1)}_L \phi \psi^{(1)}_R + \lambda  \bar{\psi}^{(1)}_R \phi^* \psi^{(1)}_L + 
|D_\mu H|^2 + |D_\mu \phi|^2 - V(\phi,H)
\eeqa 
where we have neglected the Yukawa coupling of the light fermion(s) $\psi^{(2)}_L, \psi^{(2)}_R$, which are proportional to the vev of the light Higgs $v_H$.  
For simplicity we may consider a simple scalar potential function of the two Higgs $\phi$ and $H$, such as $V(\phi,H)$, that as we have mentioned, admits vacua which are widely separated. While this would induce a hierarchy between the two vevs, and could be the real difficulty in the realization of this scenario, one possible way out would be to 
consider $V(\phi,H)$ to be the sum of two separate potentials. Since the phase of the heavy Higgs survives 
in the low energy theory as a pseudo-goldstone mode,  it may acquire a mass if the potential in which it appears is tilted. 

\begin{table}[t]
\begin{center}
\begin{tabular}{|c|c|c|c}
\hline
 $field$ & $U(1)_A$ & $U(1)_B$  \\
\hline
$\psi^{(1)}_L $ & $q^{(1)}_{A L} $ & $q^{(1)}_{B L} $ \\ 
\hline 
$\psi^{(1)}_R $ & $q^{(1)}_{A R} $ & $q^{(1)}_{B R} $ \\ 
\hline
$\psi^{(2)}_L $ & $q^{(2)}_{A L} $ & $q^{(2)}_{B L} $ \\ 
\hline 
$\psi^{(2)}_R $ & $q^{(2)}_{A R} $ & $q^{(2)}_{B R} $ \\ 
\hline
$H$ & $q_A^H $ & $ q_B^H $ \\
\hline
$\phi$ & $q_A^{\phi} $ & $ q_B^{\phi} $ \\
\hline
\end{tabular}
\end{center}
 \caption{\small Charge assignments for the A-B toy model.}
\label{chargesPsi}
\end{table}
 We show in Tab. \ref{chargesPsi} the charge assignments of the model. We define 
\beqa
D_{\mu} H &=&\left( \partial_{\mu} + i q_B^H g_B B_\mu \right) H \nonumber \\
D_{\mu} \phi &=&\left( \partial_{\mu} + i q_B^{\phi} g_B B_\mu \right) \phi \nonumber \\
D_{\mu} \psi^{(i)}_L &=& \left(\partial_\mu + i q^{(i)}_{A L} g_A A_\mu + i q_{B L} g_B B_\mu\right) \psi^{(i)}_L .\\
\eeqa
Under a gauge transformation we have $\psi\to \psi^\prime$
\beqa
\psi^{\prime   (i)}_L &=& e^{-i q^{(i)}_L g_B \theta }\psi^{(i)}_L \nonumber \\
\psi^{\prime (i)}_R &=& e^{-i q^{(i)}_R g_B \theta} \psi^{(i)}_R 
 \eeqa
with $\delta B_\mu= B^\prime _\mu - B_\mu = - \partial_\mu \theta$.

We assume that the charge assignments are such that the model is anomaly-free. Notice also that $B$, in this realization, becomes massive via a first breaking at the large scale $v_\phi$ and then its mass gets corrected by the second breaking, characterized  by the scale $v_H$. 

We parameterize the fluctuations of the field $\phi$ around the first vacuum in the form

\beq
\phi=\frac{v_\phi + \rho}{\sqrt{2}} e^{-i q_B^\phi g_B\theta}
\eeq
from which we obtain the first contribution to the mass of the $B$ gauge boson in the form 
$M_1=q_B^\phi g_B v_\phi$. As we are going to show next, this mass can be taken to be the St\"uckelberg mass of a reduced Higgs system if we neglect the radial excitations. In fact we have 

\beq
|D_\mu \phi|^2 = \frac{1}{2} \left(\partial_\mu \phi\right)^2 + \left(\frac{v_\phi + \rho}{\sqrt{2}}\right)^2 
(q_B^\phi g_B)^2 \left( - \partial_\mu \theta + B_\mu\right)^2,
\label{covder}
\eeq
and we isolate from the phase $\theta$ of this exact relation a dimensionful field $b$ which will be taking the role of a 
St\"uckelberg mass term as
\beq
\theta=\frac{b}{q_B^\phi g_B v_\phi}.
\eeq
We can expand (\ref{covder}) in the form 
\beq
|D_\mu \phi|^2= \frac{1}{2}\left( \partial_\mu - M_1 B_\mu\right)^2 + O(\rho/v),
\label{stmass}
\eeq
with $M_1\equiv q_B^\phi g_B v_\phi$, defined to be the St\"uckelberg mass.  
The decoupling of the radial excitations of the very heavy Higgs from the low energy lagrangean generates a St\"uckelberg mass term on the rhs of (\ref{stmass}), whose phase $\theta$ is at this stage massless. Notice that after the second symmetry breaking, the mass of the $B$ gauge boson will acquire an additional contribution proportional to 
$g_B q_B^H v_H$, in analogy to the first breaking, that is 
\beq
M_B=\sqrt{ M_1^2 + (g_B q_B^H v_H)^2}.
\eeq
Notice also that after the first radial decoupling of the heavy Higgs $\phi$, the Yukawa mass terms are affected by a phase dependence that can be eliminated from the effective lagrangean via an anomalous transformation. 
To illustrate this point consider the expansion of the Yukawa term around the vacuum of the heavy Higgs 
 
\beq
\lambda \bar{\psi}^{(1)}_L \phi \psi^{(1)}_R= \lambda\frac{1}{\sqrt{2}}(v_\phi +\rho) \bar{\psi}^{(1)}_L \psi^{(1)}_R
e^{-i q_B^\phi g_B \theta}\eeq
which is affected by a phase that we will try to remove in the course of the elimination of the heavy degrees of freedom of the mother theory.
Notice that in this case we do not take a large Yukawa coupling ($\lambda$), as in previous analysis 
\cite{D'Hoker:1984ph,D'Hoker:1984pc}, since the large fermion mass of $\psi^{(1)}$ is instead obtained via 
the large vev of the heavy Higgs, $v_\phi$. For this reason, having defined the St\"uckelberg mass $M_1$ in terms of the same vev, after neglecting the radial contributions we obtain 
\beq
\lambda \bar{\psi}^{(1)}_L \phi \psi^{(1)}_R=\kappa M_1 \bar{\psi}^{(1)}_L \psi^{(1)}_R
e^{-i q_B^\phi g_B \theta}, \,\,\,\,\,\,\, \kappa= \frac{\lambda}{\sqrt{2}}q_B^\phi g_B.
 \eeq
Before performing the partial integration on the heavy fermion $\psi^{(1)}$, it is convenient to define a change of variables in the functional integral, in order to remove the phase-dependence on $\theta$ present in the Yukawa couplings. For this reason, let's consider the part of the partition function directly related to the heavy fermion $\psi^{(1)}$, which is involved in the procedure of partial integration. This is given by 

\beq
\mathcal{Z}^{(1)}(A,B)=\int \mathcal{D} \psi^{(1)}_L \mathcal{D} \bar{\psi}^{(1)}_L
\mathcal{D} \psi^{(1)}_R \mathcal{D} \bar{\psi}^{(1)}_R
e^{i \int d^4 x \mathcal{L}^{(1)}} 
\eeq
where
\beq
\mathcal{L}^{(1)} =\psi^{(1)}_L\slash{D} \psi^{(1)}_L + \bar{\psi}^{(1)}_R\slash{D} \psi^{(1)}_R + 
\kappa M_1 \bar{\psi}^{(1)}_L \psi^{(1)}_R
e^{-i q_B^\phi g_B \theta} + h.c. 
\eeq
and we have neglected the contributions proportional to the radial excitation of the heavy Higgs. 
At this point we try to remove the phase $\theta$ from the Yukawa couplings by performing a field redefinition in the functional integral of the heavy fermion. We set 

\beqa
\psi^{(1)}_{B L}= e^{-i q^{(1)}_{B L} g_B \theta} \psi^{\prime(1)}_{B L}\nonumber \\
\psi^{(1)}_{B R}= e^{-i q^{(1)}_{B R} g_B \theta} \psi^{\prime(1)}_{B R},
\eeqa
where from gauge invariance we have 
\beq
q^{(1)}_{B R} + q^\phi_B - q^{(1)}_{B L}=0.
\eeq
The field redefinition induces in the integration measures two jacobeans 
\beqa
\mathcal{D}\psi^{(1)}_L \mathcal{D}\bar{\psi}^{(1)}_L &=& \mathcal{J}_L \mathcal{D}\psi^{\prime (1)}_L 
\mathcal{D}\bar{\psi}^{\prime (1)}_L \nonumber \\
\mathcal{D}\psi^{(1)}_R \mathcal{D}\bar{\psi}^{(1)}_R &=& \mathcal{J}_R \mathcal{D}\psi^{\prime(1)}_L \mathcal{D}\bar{\psi}^{\prime(1)}_R
\eeqa
which are computed using Fujikawa's approach (see for instance \cite{Panagiotakopoulos:1985fc}). We obtain 
\beqa
\mathcal{J}_L &=& e^{-i q_{B L}^{(1)}\frac{1}{32 \pi^2}\langle\theta F\wedge F\rangle_L} \nonumber \\
\mathcal{J}_R &=& e^{-i q_{B R}^{(1)}\frac{1}{32 \pi^2}\langle\theta F\wedge F\rangle_R}.  
\eeqa
In this case $F_{\mu\nu L,R}=\left[D_\mu,D_\nu\right]_{L,R}$ contains both gauge fields $(A, B)$ and the corresponding gauge charges of the heavy $(L,R)$ fermions such as, for instance, 
\beq
F_{\mu\nu L,R}=  i q^{(1)}_{A L,R} F^A_{\mu\nu} + i q^{(1)}_{B L,R} F^B_{\mu\nu}. 
\eeq
The structure of the effective action after the field redefinition takes the form 
\beq
\mathcal{Z}^{(1)}(A,B)=\int \mathcal{D} \psi^{\prime(1)}_L \mathcal{D} \bar{\psi}^{\prime(1)}_L
\mathcal{D} \psi^{\prime(1)}_R \mathcal{D} \bar{\psi}^{\prime(1)}_R
e^{i \int d^4 x \mathcal{L}^{\prime(1)} + \mathcal{L}_{WZ}} 
\eeq
where 

\beqa
\mathcal{L}^{\prime(1)} =\psi^{\prime(1)}_L\left(\slash{D} - i q_{BL}^{(1)}\slash{\partial}\theta \right)\psi^{\prime(1)}_L + \bar{\psi}^{\prime (1)}_R\left(\slash{D}- i q_{BL}^{(1)}\slash{\partial}\theta \right) \psi^{\prime(1)}_R + 
\kappa M_1 \bar{\psi}^{\prime(1)}_L \psi^{\prime (1)}_R + h.c.
\eeqa

with the Wess-Zumino (WZ) lagrangean obtained from the expansion of the $\theta F\wedge F$ terms. These are suppressed by the St\"uckelberg mass term $M_1$  $(\theta=b/M_1)$.

At this point we can perform the Grassmann integration over the heavy fermion, which trivially gives the functional determinant of an operator, $\mathcal{\bf \large P}$, explicitly given by 
\beq
\mathcal{\bf P}=v^{\prime}_\phi
\left( \begin{array}{cc}
\frac{\slash{D} - i q_{B L}^{(1)}g_B \slash{\partial}\theta}{v^{\prime}_ \phi} & 1 \\
1 & \frac{\slash{D} - i q_{B R}^{(1)} g_B\slash{\partial}\theta}{v^{\prime}_ \phi }\\
\end{array} \right),
\eeq
where $v^{\prime}_\phi\equiv v^\phi/\sqrt{2}$. The remaining terms in the total partition function of the model can 
be obtained from the functional integral
\beq
\mathcal{Z}_{eff}\sim\int \mathcal{D} \psi^{(2)}_L \mathcal{D} \bar{\psi}^{(2)}_L
\mathcal{D} \psi^{(2)}_R \mathcal{D} \bar{\psi}^{(2)}_R  \mathcal{D} H \mathcal{D} b \mathcal{D}\theta
\,\,e^{i \int d^4 x \mathcal{L}_{eff}}
\eeq
 where 
 \beq
\mathcal{L}_{eff}= \mathcal{L}^{\prime(2)} + \mathcal{L}_{WZ} + \tr\log \mathcal{\,\bf P} +   \frac{1}{2}\left( \partial_\mu - M_1 B_\mu\right)^2 + | D_\mu H|^2 - V(H,\theta)
 \eeq
 with
 \beq
 \mathcal{L}^{\prime(2)}= -\frac{1}{4}F_A^2 -\frac{1}{4}F_B^2 + 
\psi^{(2)}_L\slash{D} \psi^{(2)}_L + \bar{\psi}^{(2)}_R\slash{D} \psi^{(2)}_R.
\eeq

The derivative expansion of the effective action can be organized in terms of corrections in the St\"uckelberg mass. Obviously, a similar approach can be followed for the integration of a Majorana fermion, which is slightly more involved. The basic physical principle, however, remains the same also in this second variant. In this case the functional determinant can be organized as in \cite{Goity:1988ey}. 

There are some implications concerning the two realizations of this class of effective actions, especially in regard to the possible mass of the axion as a dark matter candidate in the various models that share the effective actions that we have presented.
The first observation concerns the absence of a direct Yukawa coupling between the heavy Higgs and the light fermion spectrum, which is part of the effective action after partial integration on the heavy fermion modes. 
This feature is absent in the MLSOM, and turns out to be rather important since it affects drastically the lifetime of the axion, as we are going to elaborate in the following sections. We will find that a GeV axion is favoured by the mechanism of partial decoupling but is not allowed in the MLSOM. In this second case a very light axion is necessary in order to have a state which is long lived and that can be a good dark matter candidate.
 
\subsection{Parametric solutions of the anomaly equations}
It is clear that the typical effective action isolated by the decoupling of (one or more) chiral fermions can be organized in terms of the defining lagrangean plus the WZ counterterms, which restore the gauge invariance of the model. Therefore, up to operators of mass dimension 5, the two lagrangeans are quite overlapping at operatorial level. For this reason, we will construct a complete charge assignments for these models, starting from an anomaly-free theory, with a spectrum that we deliberately choose to include one right-handed neutrino per generation,  and which we will decouple from 
the low energy dynamics according to the procedure described above. Of course, other choices are also possible. As we have already stressed, the motivations for selecting this approach are not just of practical nature,  although it allows to generate effective anomalous models with ease. For instance, one could envision a scenario, inspired by leptogenesys, which could offer a realization of this decoupling mechanism, although its details remain, at the moment,  rather general. We will not pursue the analysis of this point any further, and leave it as an interesting possibility for future studies.  
However, we will discover, by using the decoupling approach, that a significant class of charge assignments of intersecting brane models can be easily reproduced by the free gauge charges which parametrize the violation of the conditions of cancellation of the anomaly equations. We should also mention that the dependence of our results on the various charge assignments is truly small, showing that the relevant parameters of the models are the St\"uckelberg mass, the anomalous coupling and the parameters of the potential, which control the axion mass in each realization.

To proceed, we impose first the conditions of cancellation of the gauge and of the 
mixed gravitational-$U(1)_B$  anomalies, thereby fixing the $U(1)_B$ charges, followed by the conditions of invariance of the Yukawa couplings, in order to determine the charges of the two Higgs \cite{Appelquist:2002mw}.
We take the $U(1)^{}_B$ fermion charges to be family-independent in order to avoid possible
constraints from flavor-changing neutral current processes.
We label the generic fermion charges under the additional group $U(1)_B$ as shown
in Table~\ref{chargesB}.

\begin{table}[h]
\begin{center}
\begin{tabular}{|c|c|c|c|c|c|c|}
\hline
$Q_L$ & $u_R $ & $ d_R $ &    $L$  &  $e^{}_R$ & $\nu^{}_R$ \\
\hline
$q_{Q_L}^B$   & $q_{u_R}^B$  & $q_{d_R}^B$   & $q_{L}^B$   & $q_{e^{}_R}^B$  & $q_{\nu^{}_R}^B$ \\
\hline \end{tabular}
\end{center}
 \caption{\small Labels for the gauge charges of the fermion spectrum.}
\label{chargesB}
\end{table}
For every anomalous triangle we allow, in general, a
WZ counterterm whose coefficient has to be tuned in order to satisfy the conditions for anomaly cancellation.   
For the fermion charges $q_{L}^B, q_{d_R}^B,  q_{e^{}_R}^B $ we find the following constraints
\bea
BSU(2)SU(2):&&  q_{L}^B + 3 q_{Q_L}^B - C_{BWW}=0,   \nn\\
BSU(3)SU(3):&&   q_{d_R}^B + q_{u_R}^B - 2 q_{Q_L}^B  - C_{Bgg}=0, \nn\\
BYY:&&  3 q_{e_R}^B + 6 q_{Q_L}^B +3 q^B_{u_R} - \frac{3}{2}C_{BWW} - C_{BYY} 
-C_{Bgg},
\label{solve_charge}
\eea
where the coefficients appearing in front of the WZ counterterms are proportional to
the charge asymmetries
\ba
&&C_{BWW}\propto \sum_f \theta_{f L},\nn\\
&&C_{Bgg}\propto \sum_Q \theta_Q^B,\nn\\
&&C_{BYY}\propto \sum_f \theta_f^{BYY},\nn\\
&&C_{BBB}\propto \sum_f \theta_f^{BBB}\,,
\ea
which are detailed in an appendix, and with the hypercharges of $U(1)_Y$ given in Tab.~(\ref{solve_q}).

If we consider the charges $q_{Q_L}^B, q_{L}^B$ as free parameters of the model, 
$C_{BWW}\,, C_{Bgg}\,, C_{BYY}$ can be in principle expressed in terms 
of these parameters. The other three conditions coming from the 
gauge invariance give the following further constraints
\bea
YBB:&& -3 (q^B_{d_R})^2 - 3(q^B_{e_R})^2 +3 (q^B_L)^2 -3 (q^B_{Q_L})^2 
+6(q^B_{u_R})^2 -C_{YBB}=0 
\nn\\
BBB:&& 9 (q^B_{d_R})^3 +3(q^B_{e_R})^3 -6 (q^B_L)^3 -18 (q^B_{Q_L})^3
+9(q^B_{u_R})^3 -C_{BBB}=0
\nn\\
BRR:&& 9 q^B_{d_R} +9 q^B_{u_R}+ 3 q^B_{e_R} -6q^B_L -18q^B_{Q_L} -3 C_{BGG} =0, 
\label{solve_counter}
\eea
where the condition on the BRR triangle comes from the mixed gravitational-$U(1)_B$ 
anomaly cancellation. From the gauge invariance of the Yukawa 
couplings (see Lagrangian~(\ref{yukawa_utile})), we obtain
\bea
\label{yuk_constraint}
q^B_{Q_L} - \frac{q^B_d}{2} - q^B_{d_R}&=&0, \nn\\
q^B_{Q_L} + \frac{q^B_u}{2} - q^B_{u_R} &=& 0, \nn\\
q^B_L- \frac{q^B_d}{2} - q^B_{e_R}&=&0, \nn\\
q^B_L+ \frac{q^B_u}{2}&=&0,
\eea
which can be used to constrain the charges of the two Higgs
doublets $q^B_u, q^B_d$ and the counterterms $C_{BWW}, C_{Bgg}$. 
Collecting the constraints in eqs.~(\ref{yuk_constraint}), (\ref{solve_charge}) 
and~(\ref{solve_counter}) we obtain a set of ten equations whose solution allows us to
identify a class of charge assignments that we call $f$  
\bea
f(q_{Q_L}^B, q_{L}^B, \Delta q^B)=
(q_{Q_L}^B, q_{u_R}^B; q^{B}_{d_R}, q^{B}_{L}, q^{B}_{e_R}, q^{B}_{u},q^{B}_{d}).
\eea
These depend only upon the three free parameters $q^B_{Q_L}$, $q^B_{L},\Delta q^B$, 
where $\Delta q^B=q^B_u-q^B_d$.
The explicit dependences are shown in Table~\ref{solve_q},
while the related WZ counterterms  take the form 
\bea
&&C_{BYY} =  -\frac{3}{2} (q^B_L-5 q^B_{Q_L}) +2\Delta q^B,
\\
&&C_{YBB} = 3(q^B_L)^2 -\frac{3}{2} \left[ 18 (q^B_{Q_L})^2 
+8 q^B_{Q_L}\Delta q^B +(\Delta q^B)^2 \right],
\\
&&C_{BBB} = -6(q^B_L)^3 +78(q^B_{Q_L})^3 + 72 (q^B_{Q_L})^2 \Delta q^B
+18 q^B_{Q_L} (\Delta q^B)^2 + \frac{3}{2}(\Delta q^B)^3,
\\
&&C_{Bgg}=\frac{1}{2}\Delta q^B,
\\
&&C_{BWW} = q^B_{L} + 3q^B_{Q_L},
\label{charge_asy}
\eea
where in particular, from the charge assignment shown in Table~\ref{solve_q}, 
we identify the counterterm for the mixed gravitational-$U(1)_B$
anomaly with
\bea
C_{BGG} = 2(-q^B_L + q^B_{Q_L} +\Delta q^B).
\label{difference}
\eea
Then the WZ counterterms, as defined in general in eqs.~(\ref{WZcoeff}),
can now be specialized in terms of the different charge assignments
$f(q^B_{Q_L}, q^B_{u_R}, \Delta q^B)$, just by substituting the corresponding chiral asymmetries. This function will appear in several of our plots. 

\begin{table}[t]
\centering
\renewcommand{\arraystretch}{1.2}
\begin{tabular}{|c|c|c|c|c|}\hline
$f$ & $SU(3)^{}_C$ & $SU(2)^{}_L$ & $U(1)^{}_Y$ & $U(1)^{}_B$ \\ \hline \hline
$Q_L$ &  3 & 2 & $ 1/6$ & $q_{Q_L}^B$\\
$u_R$ &  3 & 1 & $ 2/3$ & $0$\\
$d_R$ &  3 & 1 & $ -1/3$ & $2 q_{Q_L}^B + \frac{1}{2} \Delta q^B  $\\
$L$ &  1 & 2 & $ -1/2$ & $q_{L}^B $\\ 
$e_R$ &  1 & 1 & $-1$ & $ 2 q_{Q_L}^B + \frac{1}{2} \Delta q^B $\\ 
$\nu_R$&  1 & 1 & 0 & $0$\\ \hline
$H^{}_u$ &  1 & 2 & $1/2$ & $ - 2q_{Q_L}^B $\\ 
$H^{}_d$ &  1 & 2 & $1/2$ & $ - 2q_{Q_L}^B - \Delta q^B  $\\ \hline
\end{tabular}
\caption{\small The three-parameter family $f(q_{Q_L}^B, q_{L}^B, \Delta q^B)$ 
of solutions for fermion and scalar charges. 
\label{solve_q}}
\end{table}
Finally, since in the case  $q^B_u-q^B_d=0$ the $O^\chi$ matrix 
would become trivial, we require the following relation between the Higgs charges
\bea
q^B_u-q^B_d \neq 0 
\eea
where, in particular, $q^B_u-q^B_d = 4$ is exactly the value implied by 
the charge assignment derived from the Madrid Model (see Table~\ref{charge_higgs}) for the two Higgs. We will be using this value to constrain the chiral asymmetry 
$\theta^B_f$ by means of eq.~(\ref{difference}), and will be taken as the starting value for all our 
comparisons. Notice that the family $f(q^B_{Q_L}, q^B_{L}, \Delta q^B)$ 
for the particular choice $q_{Q_L}^B=-1, q_{L}^B=-1$ reproduces the entire charge assignment of the Madrid Model
\bea
f(-1, -1, 4)=(-1, 0, 0, -1, 0, +2, -2).
\eea

\subsection{The Madrid model}
We just recall, as already mentioned, that the charge assignment for our anomalous (brane) model that we consider is obtained from the 
intersection of 4 branes $(a,b,c,d)$ with generators $(q_a, q_b, q_c, q_d)$ which are rotated on the hypercharge basis $U(1)_{X_i}$ with $i=A,B,C$ and $U(1)_Y$, with an anomaly free hypercharge.  The $U(1)_{a}$ and $U(1)_{d}$ symmetries are proportional to the baryon number and the lepton number respectively. The $U(1)_{c}$ symmetry can be identified as the third component of the right-handed weak isospin, while the $U(1)_{b}$ is a PQ-like symmetry. A detailed discussion of this construction can be found in \cite{Ibanez:2001nd} and 
\cite{Ghilencea:2002da}. The identification of the generators involve the solution of some constraint 
equations. In general, for a simple $T^6$ compactification the solutions of these equations are parametrized by a phase $\epsilon =\pm1$, the Neveu-Schwarz background
on the first two tori $\beta_i=1-b_i=1, 1/2$, the four integers
$n_{a2}, n_{b1}, n_{c1}, n_{d2}$ which are the wrapping numbers of the branes around the extra (toroidal) manifolds of the compactification, and finally a parameter $\rho=1, 1/3$.  One of the possible choices for these parameters is reported in Table \ref{parameters} which identifies a particular class of models, the so called Class A models.
\begin{table}[h]
\begin{center}
\begin{tabular}{|c|c|c|c|c|c|c|c|c|}
\hline
    $\nu$  & $\beta_1$ & $\beta_2 $ & $n_{a2}$  &  $n_{b1}$ & $n_{c1}$ & $n_{d2}$ \\
\hline  1/3 & 1/2  & $  1 $ &  $n_{a2}$ &  -1 & 1 & 1 - $n_{a2}$\\
\hline
\end{tabular}
\end{center}
\caption{\small Parameters for a Class A model with a D6-brane .}
\label{parameters}
\end{table}
\begin{table}[h]
\begin{center}
\begin{tabular}{|c|c|c|c|c|c|c|}
\hline
                $f$         & $Q_L$ & $u_R $ & $d_R $ &    $L$  &  $e^{}_R$ & $\nu^{}_R$ \\
\hline  $q^{}_{Y}$  &  1/6     &   2/3     &     - 1/3    &  -1/2   &    - 1       &  0  \\
\hline   $q^{}_{B}$  & -1    & 0  & 0   & -1   & 0  & 0 \\
\hline \end{tabular}
\end{center}
\caption{\small Fermion spectrum charges in the $Y$-basis for the Madrid model \cite{Ghilencea:2002da}.
\label{charges}}
\end{table}
The result of this D-brane construction is the charge assignment specified in Table~\ref{charges} whose corresponding fermion spectrum is anomalous under the extra $U(1)_B$ abelian symmetry. Imposing the gauge invariance of the Yukawa couplings, see eq.~(\ref{yukawa_utile}), we constraint the charges of the Higgs doublets to the values specified in Table~\ref{charge_higgs}.
\begin{table}[h]
\begin{center}
\begin{tabular}{|c|c|c|c|c|}
\hline
   &  Y &$ X^{}_A$  & $X^{}_{B} $    \\
\hline $H^{}_{u}$ & 1/2 & 0  & 2     \\
\hline  $H^{}_{d}$   & 1/2  &0  & -2     \\
\hline \end{tabular}
\end{center}
\caption{\small Higgs charges in the Madrid model. 
\label{charge_higgs}}
\end{table}

\

\section{Trilinear and quadrilinear interactions of the axi-Higgs
from the MLSOM scalar potential}

One of the objectives of this work is to quantify the decay rates in the 
various channels of the axi-Higgs $\chi$ and of the two Higgs bosons $H_0$ and $h_0$ of the CP-even sector,
and to explore some possible channels in which the production of an axi-Higgs can be realized at the 
LHC. For this goal we proceed with a careful inspection of the interaction lagrangian, in order to extrapolate 
all the relevant couplings and interactions of the axi-Higgs and of the CP-even sector with the other particles.
We start this analysis by collecting first all the trilinear and quadrilinear interactions of the axi-Higgs
that emerge from the scalar potential and then move to the mixed vertices which involve both the CP-even and CP-odd sectors. 

Collecting the quadrilinear vertices we obtain
\ba
{\cal L}_{\chi^4}= \left[R_1^{\chi^4} + R_2^{\chi^4} + R_3^{\chi^4} + R_4^{\chi^4}\right]\chi^4 ,
\ea
where we have defined 
\ba
&&R_1^{\chi^4}=\frac{1}{4} \l_{uu} (O^{\chi}_{11})^4 + \frac{1}{4} \l_{dd} (O^{\chi}_{21})^4
\nonumber\\
&&R_2^{\chi^4}=-\frac{1}{2} \l_{ud} (O^{\chi}_{11})^2 (O^{\chi}_{21})^2
\nonumber\\
&&R_3^{\chi^4}=\frac{1}{2} \l_1 (O^{\chi}_{11})^2 (O^{\chi}_{21})^2 
-2\frac{v_d}{M_1}\Delta q^B\l_1 (O^{\chi}_{11})^2 O^{\chi}_{21}  O^{\chi}_{31}
+2\frac{v_u}{M_1}\Delta q^B\l_1 (O^{\chi}_{11}) (O^{\chi}_{21})^2  O^{\chi}_{31} + O(1/M^2)
\nonumber\\
&&R_4^{\chi^4}=\frac{1}{2} \l_2 (O^{\chi}_{11})^3 O^{\chi}_{21} 
+ \frac{1}{2} \l_3 (O^{\chi}_{21})^3 O^{\chi}_{21}
+\frac{v_u}{2 M_1}\Delta q^B \left[ \l_2 O^{\chi}_{21} O^{\chi}_{31} (O^{\chi}_{11})^2
+\l_3 O^{\chi}_{31} (O^{\chi}_{21})^3\right]
\nonumber\\
&&\hspace{1cm} -\frac{v_d}{2 M_1}\Delta q^B \left[ \l_3 O^{\chi}_{11} O^{\chi}_{31} (O^{\chi}_{21})^2
+\l_2 O^{\chi}_{31} (O^{\chi}_{11})^3\right].
\ea
The first contribution $(R_1)$ is extracted from the  diagonal part of the Higgs potential (i.e $\sim \l_{aa} (H_a^{\dagger} H_a)^2$), 
the second originates from the non-diagonal u-d terms  ($\sim \l_{ud}(H_u^{\dagger} H_u)(H_d^{\dagger} H_d)$), 
the third comes from the 
contribution of the $PQ$-breaking potential proportional to $\l_1$,
while $R_4^{\chi^4}$ is the contribution of the last two pieces of the same potential which are 
proportional to $\l_2$ and $\l_3$. 

The quadrilinear couplings of the axi-Higgs with the neutral Higgs sector involve 
interactions between two axions and the two neutral states $(H^0,h^0)$. We can write
the interaction lagrangian as follows
\ba
&&{\cal L}_{\chi^2 H^0 h^0}= \left[R_1^{\chi^2 H^0 h^0} + R_2^{\chi^2 H^0 h^0} 
+ R_3^{\chi^2 H^0 h^0} + R_4^{\chi^2 H^0 h^0}\right]\chi^2 H^0 h^0
\nonumber\\ 
&&\hspace{1.5cm}
+ \left[R_1^{\chi^2 H^0 H^0}+R_2^{\chi^2 H^0 H^0}+ R_3^{\chi^2 H^0 H^0}
+ R_4^{\chi^2 H^0 H^0}\right] \chi^2 H^0 H^0 
\nonumber\\
&&\hspace{1.5cm} + \left[R_1^{\chi^2 h^0 h^0} + R_2^{\chi^2 h^0 h^0}+ R_3^{\chi^2 h^0 h^0}
+ R_4^{\chi^2 h^0 h^0}\right]\chi^2 h^0 h^0
\ea
where the coefficients $R_i^{\chi^2 H H }$ are defined in an appendix.

The trilinear interactions of the axi-Higgs with the neutral Higgs sector exhibit
couplings with two axions and one Higgs state $H^0,h^0$. The interaction lagrangian can be written as
\ba
{\cal L}_{\chi^2 higgs}={\cal L}_{\chi^2 H^0}+{\cal L}_{\chi^2 h^0}
\ea
where we have defined 
\ba
{\cal L}_{\chi^2 H^0}=\left[\sum_{i=1}^5 R_i^{\chi^2 H^0}\right]\chi^2 H^0 \,,
&&{\cal L}_{\chi^2 h^0}=\left[\sum_{i=1}^5 R_i^{\chi^2 h^0}\right]\chi^2 h^0\,.
\ea
Again, the $R_i^{\chi^2 h^0/H^0}$ coefficients are listed in an appendix.
It is important to note that these couplings are also present in a general 2HDM,
while they are absent in the MSSM due to the strong constraints obtained by imposing supersymmetry.

\subsection{ Self interactions in the CP-even sector }

The self interactions of $H_0$ and $h_0$ can be described as above, by analyzing the quadrilinear 
and trilinear vertices generated by the rotation of the fields in the physical basis after electroweak symmetry breaking (EWSB).
Starting from the quadrilinear interactions we can write 
\ba
{\cal L}_{H^4}={\cal L}_{{H_0}^4}+{\cal L}_{{h_0}^4}+{\cal L}_{{h_0}^2 {H_0}^2}
+{\cal L}_{{h_0}{H_0}^3}+{\cal L}_{{H_0}{h_0}^3},
\ea
where
\ba
&&{\cal L}_{{H_0}^4}=\left[R_1^{H_0^4} + R_2^{H_0^4} + R_3^{H_0^4} + R_4^{H_0^4}\right]H_0^4
\nonumber\\
&&{\cal L}_{{h_0}^4}=\left[R_1^{h_0^4} + R_2^{h_0^4} + R_3^{h_0^4} + R_4^{h_0^4}\right]h_0^4
\nonumber\\
&&{\cal L}_{{h_0}^2 {H_0}^2}=\left[R_1^{h_0^2 H_0^2} + R_2^{h_0^2 H_0^2} + R_3^{h_0^2 H_0^2}\right]H_0^2 h_0^2
\nonumber\\
&&{\cal L}_{{h_0}{H_0}^3}=\left[R_1^{h_0 H_0^3} + R_2^{h_0 H_0^3}\right]H_0^3 h_0
\nonumber\\
&&{\cal L}_{{H_0}{h_0}^3}=\left[R_1^{H_0 h_0^3} + R_2^{H_0 h_0^3}\right]h_0^3 H_0.
\ea
The coefficients $R_i^{H^4}$ can be found in an appendix. 
Also here it is interesting to observe that $R_1$ and $R_2$ are in general related to the $PQ$ symmetric part
of the scalar potential, while $R_3$ and $R_4$ come from the $PQ$-breaking terms.

The trilinear interaction lagrangian can be written as
\ba
{\cal L}_{H^3}={\cal L}_{{H_0}^3}+{\cal L}_{{h_0}^3}+{\cal L}_{{h_0}^2 {H_0}}+{\cal L}_{{h_0}{H_0}^2}
\ea
where we have defined 
\ba
&&{\cal L}_{{H_0}^3}=\left[R_1^{H_0^3} + R_2^{H_0^3} + R_3^{H_0^3} + R_4^{H_0^3}\right]H_0^3
\nonumber\\
&&{\cal L}_{{h_0}^3}=\left[R_1^{h_0^3} + R_2^{h_0^3} + R_3^{h_0^3} + R_4^{h_0^3}\right]h_0^3
\nonumber\\
&&{\cal L}_{{h_0}^2 {H_0}}=\left[R_1^{h_0^2 H_0} + R_2^{h_0^2 H_0} + R_3^{h_0^2 H_0}\right]H_0 h_0^2
\nonumber\\
&&{\cal L}_{{h_0}{H_0}^2}=\left[R_1^{h_0 H_0^2} + R_2^{h_0 H_0^2}\right]H_0^2 h_0.
\ea
All the coefficients $R_i^{H^3}$ are given in an appendix.

\subsection{Trilinear interactions of the CP-even sector with the $W^{\pm}$ and $Z$ gauge bosons}

Since, in general, the branching ratios for the decay of the Higgs into a pair of vector bosons $W^{\pm}$ or 
$ZZ$ are relevant in a certain kinematical regime, it is important to quantify the tree level decay rate for this channel, and to give an estimate of the coefficients of 
the trilinear interactions of $H_0$ and $h_0$ with two gauge bosons $W^{+}W^{-}$ and $ZZ$.
For the charged $W^{\pm}$ it is straighforward to obtain the corresponding coefficients 
\ba
&&C^{H0}_{WW}=\frac{g_2^2}{2}\left(\sin{\alpha}~v_d  - \cos{\alpha}~v_u \right),
\nonumber\\
&&C^{h0}_{WW}=\frac{g_2^2}{2}\left(\sin{\alpha}~v_d  + \cos{\alpha}~v_u \right).
\ea
The calculation of the coefficients for the analogous interactions with the $Z$'s is more complicated
because of the structure of the model. For this purpose it is useful to introduce the following coefficients
\ba
&&f_1 = 2 M_1^2 - g^2 v^2 + N_{BB},
\nonumber\\
&& \xi_1 = \frac{f_1^2 +f_1\left(\sqrt{f_1^2 +4 g^2 x_B^2} -2 g_B q_B^d x_B \right)
+2 x_B\left[x_B g^2 +g_B q_B^d\left(g_B q_B^d x_B - \sqrt{f_1^2 +4 g^2 x_B^2}\right) \right]}
{2\sqrt{2} \left(4 g^2 x_B^2 +f_1\sqrt{f_1^2 +4 g^2 x_B^2} \right)}
\nonumber\\
&& \xi_2 = \frac{f_1^2 +f_1\left(\sqrt{f_1^2 +4 g^2 x_B^2} -2 g_B q_B^u x_B \right)
+2 x_B\left[x_B g^2 +g_B q_B^u\left(g_B q_B^u x_B - \sqrt{f_1^2 +4 g^2 x_B^2}\right) \right]}
{ 2\sqrt{2} \left(4 g^2 x_B^2 +f_1\sqrt{f_1^2 +4 g^2 x_B^2} \right)}
\nonumber\\
\ea
and the interactions $H$-$Z$-$Z$ at tree level - summarized by 
the coefficients $C^{H}_{ZZ}$ - are given by
\ba
&&C^{H0}_{ZZ}=\frac{1}{\sqrt{2}}\left(v_d~g^2~\xi_1^2 \sin{\alpha}   - v_u~g^2~\xi_2^2 \cos{\alpha}   \right),
\nonumber\\
&&C^{h0}_{ZZ}=\frac{1}{\sqrt{2}}\left(v_u~g^2~\xi_2^2\sin{\alpha}  + v_d~g^2~\xi_1^2 \cos{\alpha}  \right),
\ea
where $g^2=g_Y^2+g_2^2$.

\section{The Yukawa couplings and the axi-Higgs }

The couplings of the two Higgs and of the axi-Higgs to the fermion sector 
are entirely described by the Yukawa lagrangian.
The Yukawa couplings of the model are given by
\bea
{\cal L}_{\rm Yuk}^{unit.} 
&=& - \Gamma^{d} \, \overline{Q}_{L} H_{d} d_{R} - \Gamma^{d} \, \overline{d}_R H^{\dagger}_{d} Q_{L} - 
\Gamma^{u} \, \overline{Q}_{L} (i \sigma_2 H^{*}_{u}) u_{R} 
- \Gamma^{u} \, \overline{u}_R (i \sigma_2 H^{*}_{u})^{\dagger} Q_{L} \nonumber\\
&&-   \Gamma^{e} \, \overline{L} H_{d} {e}_{R} - \Gamma^{e} \, \overline{e}_R H^{\dagger}_{d} L 
- \Gamma^{\nu} \, \overline{L} (i \sigma_2 H^{*}_{u}) \nu_{R} 
- \Gamma^{\nu}  \, \overline{\nu}_R (i \sigma_2 H^{*}_{u})^{\dagger} L  \nonumber\\
&=& - \Gamma^{d} \, \overline{d} H^{0}_{d} P_{R} d - \Gamma^{d} \, \overline{d}  H^{0*}_{d} P_{L} d  
- \Gamma^{u} \, \overline{u}  H^{0*}_{u} P_{R} u - \Gamma^{u} \, \overline{u}  H^{0}_{u} P_{L} u \nonumber\\
&&- \Gamma^{e}  \, \overline{e} H^{0}_{d} P_{R} e - \Gamma^{e} \, \overline{e}  H^{0*}_{d} P_{L} e 
- \Gamma^{\nu} \, \overline{\nu} H^{0*}_{u} P_{R} \nu - \Gamma^{\nu} \, \overline{\nu}  H^{0}_{u} P_{L} \nu,
\label{yukawa_utile}
\eea
where the Yukawa coupling constants $\Gamma^{d}, \Gamma^{u}, \Gamma^{e}$ and 
$\Gamma^{\nu}$ run over the three generations, i.e. $u = \{u, c, t\}$, $d = \{d, s, b\}$, 
$\nu$ = \{$\nu_{e}$, $\nu_{\mu}$, $\nu_{\tau}$\} 
and $e$ = \{$e$, $\mu$, $\tau$\}.
Rotating the CP-odd and CP-even neutral sectors into the mass eigenstates and 
expanding around the vacuum  we obtain 
\bea
H_u^0 &=& v_u + \frac{  Re{H^0_{u}} + i \, Im{H^0_u}}{\sqrt{2}}  \nonumber\\
&=&  v_u + \frac{  (h^0 \sin\a  - H^0 \cos\a ) 
+ i \, \left(O^{\chi}_{11}\chi + O^{\chi}_{12}G^{\,0}_1 + O^\chi_{13} G^{\,0}_2  \right) }{\sqrt{2}}  
\label{Higgs_up}   
\eea
\bea     
H_d^0 &=&  v_d + \frac{ Re{H^0_d} + i \, Im{H^0_d}}{\sqrt{2}}   \nonumber\\
&=&  v_d  +  \frac{  (h^0 \cos\alpha  + H^0 \sin\alpha ) + i  \, \left( O^{\chi}_{21}\chi 
+ O^{\chi}_{22}G^{\,0}_1 + O^\chi_{23} G^{\,0}_2 \right) }{\sqrt{2}}  
\label{Higgs_down}
\eea  
so that in the unitary gauge we obtain
\bea
H_u^0 &=& v_u + \frac{1}{\sqrt{2}} \left[ (h^0\; \sin{\a}  - H^0\; \cos{\a})  +
i \, O^{\chi}_{11} \;\chi \right]                   \nonumber\\
      &=&   v_u + \frac{1}{\sqrt{2}} \left[  (h^0\; \sin{\a}  - H^0\; \cos{\a})  -
i \,  N \cos\beta  \;\chi  \right] \\
H_d^0 &=& v_d +  \frac{1}{\sqrt{2}}  \left[ (h^0\; \cos{\a}  + H^0\; \sin{\a})  +
i \, O^{\chi}_{21}\;\chi \right] \,\,\,\;\;                \nonumber\\
&=&     v_u + \frac{1}{\sqrt{2}}  \left[ (h^0\; \cos{\a}  + H^0\; \sin{\a})  +
i \,  N \sin\beta  \;\chi \right],  
\label{Higgsdec}
\eea
where the vevs of the two neutral Higgs bosons $v_u=v \sin \beta $ and $v_d= v \cos \beta $ satisfy 
\ba
\tan \beta = \frac{v_u}{v_d}, \qquad  v=\sqrt{v_u^2+v_d^2}.
\ea
We have also relied on the definitions of 
$O^{\chi}$ introduced in a previous work \cite{Coriano:2007xg}
\ba
O^{\chi}_{11}= -N \cos \beta,
\ea
\ba 
O^{\chi}_{21}=N \sin \beta,
\ea
that we have reported in an appendix. For convenience we have introduced the following normalization coefficient
\ba
N = \frac{1}{ \sqrt{ 1+ \frac{  ( q_u^B - q_d^B )^2 }{ M^{\,2}_1 }  \frac{ v_d^2 v_u^2 }{ v^2 } } }.
\label{norm}
\ea
The fermion masses are given by
\bea
&& m_{u} =  {v_u \G^{u}},\hskip 1cm  m_{\n } =  {v_u \G^\n},  \nonumber\\
&& m_{d} =  {v_d \G^d},\hskip 1cm  m_{e} =  {v_d \G^e},
\label{f_masses}
\eea
where the generation index has been suppressed for brevity. The fermion masses, defined in terms of the two expectation values $v_u,v_d$ of the model, show an enhancement of the down-type Yukawa couplings for large values of $\tan \beta$ while at the same time the up-type Yukawa couplings get a suppression. The couplings of the $h^0$ boson to fermions are given by
\ba
{\cal L}_{\rm Yuk}(h^0) &=&  -  \Gamma^d \, \overline{d}_{L} d_R
\left( \frac{ \cos\a}{\sqrt{2}} h^0 \right) - \Gamma^u \, 
\overline{u}_{L} u_{R}   \left( \frac{ \sin\a }{\sqrt{2}} h^0  \right)  
-  \Gamma^e \, \overline{e}_{L}e_R  \left( \frac{ \cos\a}{\sqrt{2}} h^0 \right)    \nn\\
&&-  \Gamma^{\nu} \, \overline{\nu}_{L} \nu_{R}   \left( \frac{ \sin\a }{\sqrt{2}} h^0  \right)
+ c.c. 
\ea
The couplings of the $H^0$ boson to the fermions are 
\ba
{\cal L}_{\rm Yuk}(H^0) &=&  - \Gamma^d \, \overline{d}_{L} d_R 
\left( \frac{ \sin\a}{\sqrt{2}} H^0 \right) -   \Gamma^u \, 
\overline{u}_{L} u_{R}   \left( - \frac{ \cos\a }{\sqrt{2}} H^0  \right)
 - \Gamma^e \, \overline{e}_{L}e_R  \left( \frac{ \sin\a}{\sqrt{2}} H^0 \right)   \nn\\
&&-  \Gamma^{\nu} \, \overline{\nu}_{L} \nu_{R}   \left( - \frac{ \cos\a }{\sqrt{2}} H^0  \right)
+ c.c. 
\ea
For later reference we group together the couplings of the axi-Higgs $\chi$ with the fermion sector  
\bea
{\cal L}_{\rm Yuk}(\chi) &=& - \Gamma^d \, \overline{d}_{L} d_R \left(i\frac{ N \sin\b}{\sqrt{2}}\chi  \right) 
-   \Gamma^u \, \overline{u}_{L} u_{R}  \left(-i \frac{N \cos\beta}{\sqrt{2}}\chi  \right) 
-  \Gamma^e \, \overline{e}_{L}e_R \left( i \frac{N \sin\beta}{\sqrt 2}\chi \right)   \nonumber\\
&& \,\,\,-  \Gamma^{\nu} \, \overline{\nu}_{L} \nu_{R} \left(- i \frac{N \cos\beta}{\sqrt 2}\chi  \right)  +c.c.  
\label{chif}
\eea
We have listed these couplings in Tab.~(\ref{higgs_fermion}) where the normalization 
coefficient $N$ is defined in (\ref{norm}).
\begin{table}[h]
\begin{center}
\begin{tabular}{|c|c|c|}
\hline
                                                  &          up-fermion    &     down-fermion       \\
\hline        Higgs SM                  &  $- \frac{m_f}{ v}$      &   $ -  \frac{ m_f}{ v}$          \\
\hline
\hline      Lighter Higgs $h^0$    &  $-  \frac{m_f}{ v}  \, \sin \alpha / \sin \beta  $ &  $ - \frac{m_f}{ v}  \, \cos \alpha / \cos \beta  $     \\
\hline     Heavier Higgs $H^0$   &   $  \frac{m_f}{v}  \, \cos \alpha / \sin \beta  $  &  $ -  \frac{m_f}{v}  \,  \sin \alpha / \cos \beta$     \\
\hline     axi-Higgs   $\chi$         &   $   i  \frac{m_f}{v}  \, N / \tan \beta  $  & $ -  i  \frac{m_f}{ v}  \,  N  \tan \beta $  \\
\hline 
\end{tabular}
\end{center}
\caption{\small Couplings of the neutral MLSOM Higgs bosons to up- and down-type fermions, and comparison with the fermion couplings of the SM Higgs boson.}
\label{higgs_fermion}
\end{table}
From the Yukawa couplings of eq.~(\ref{yukawa_utile}) and relations~(\ref{Higgs_up}), (\ref{Higgs_down}) we can extract 
the coupling of the Goldstone boson $G^{0}_{2}$ to the fermions
\beqn
{\cal L}_{\rm Yuk}(G^{\,0}_{2})  &=& - \Gamma^{d} \, \overline{d} 
\left( i \frac{O^{\chi}_{23}}{\sqrt{2}} G^{0}_{2} \right) P_{R} d 
- \Gamma^{d} \, \overline{d} \left(- i \frac{O^{\chi}_{23}}{\sqrt{2}} G^{0}_{2} \right)  P_{L} d 
- \Gamma^{u} \, \overline{u}  \left( - i \frac{O^{\chi}_{13}}{\sqrt{2}} G^{0}_{2} \right) P_{R} u    \nonumber\\
&&- \Gamma^{u} \, \overline{u} \left( i \frac{O^{\chi}_{13}}{\sqrt{2}} G^{0}_{2} \right)  P_{L} u 
- \Gamma^{e} \, \overline{e} \left( i \frac{O^{\chi}_{23}}{\sqrt{2}} G^{0}_{2} \right)  P_{R} e 
- \Gamma^{e} \, \overline{e}  \left(- i \frac{O^{\chi}_{23}}{\sqrt{2}} G^{0}_{2} \right) P_{L} e \nonumber\\
&&-\Gamma^{\nu} \, \overline{\nu} \left( - i \frac{O^{\chi}_{13}}{\sqrt{2}} G^{0}_{2} \right) P_{R} \nu
 - \Gamma^{\nu} \, \overline{\nu}  \left( i \frac{O^{\chi}_{13}}{\sqrt{2}} G^{0}_{2} \right) P_{L} \nu. 
\eeqn
Using the expression of $O^{\chi}$ we can compute the coupling between the Goldstone boson $G^{0}_{2}$ and the down-like quarks that 
takes the form 
\beqa
- \, \Gamma^{d} \, \overline{d} H^{0}_{d} P_{R} \, d - \Gamma^{d} \,\overline{d} H^{0*}_{d} P_{L} \, d  
&=&    \frac{m_d}{ \sqrt 2} \left\{ - N \left[  - \frac{(q^{B}_{u} - q^{B}_{d})}{M_{1}} 
\frac{v_{u}^{2}}{v^2}  \right] \right\}  i \overline{d} \gamma^{5}  d  G^{\,0}_{2}
\eeqa
and similarly for the other generations.
These expressions  have been used in order to fix the explicit form of the  Wess-Zumino (WZ) counterterms using the condition of gauge invariance.

\section{Decay rates of the axi-Higgs}
We proceed to compute  the partial decay widths and the branching ratios of 
the axi-Higgs for different decay modes in the CP-odd 
sector of the MLSOM, taking the mass of the axion as a free parameter. As we have already mentioned, in the case of the MLSOM, there is an interesting window in which the axion acquires a lifetime typical of a good dark matter candidate. This mass value, which is the same as that of a traditional Peccei-Quinn axion ($\sim 10^{-4}$ eV, or in the ultralight mass window), 
is  not the most interesting one for studies of this particle at the LHC. The reason of this result has to be found in the fact that the most relevant channels for the production of a particle of this mass are 1) the pseudoscalar vertex with a top or bottom quark loop (the dominance of one or the other fermion contribution depends closely on the value of $\tan\beta$); 2) the direct WZ vertex in which the axion is radiated off by a gauge field. The WZ term is quite small compared to the contribution from the fermion loop, which is instead dependent on the mass of the axion. For an ultralight axion the loop contribution is rather small and the chances of producing a particle of such a mass by gluon fusion or in $q\bar{q}$ annihilation of light quarks are quite small. For this reason, if we are interested in the study of a GeV axion, which is the goal of the numerical sections that follow,  we are automatically excluding a long-lived particle. On the other end, in this mass region, we are instead analyzing a particle whose behaviour  is Higgs-like but with a direct (although small) direct coupling to the gauge fields. At the same time,  the Higgs-like nature of the axion can be investigated by taking its mass in the several GeV region, say in the 100-120 GeV range.  Our results, however, are quite general, in this respect, and can be used for direct studies of this particle in any mass range. As we have already stressed, what makes a distinction between a "standard" CP-odd Higgs state and the axion of the MLSOM are the WZ interactions, which are, in any case, subdominant compared to the triangle diagram in any mass range. 

In the case of fermion decoupling one can proceed with similar considerations, although the conclusions are rather different and will be addressed below. We will describe in a final section the main properties of the axion if its origin is to be traced back to a decoupled Higgs  sector, which show, in this second realization, that the axion can be long lived and with a mass in the GeV range.

The relevant parameters which appear in the decay are the following coefficients 

\ba
c^{\, \chi, u} &=&  - i \frac{m^{}_{u}}{v_u} \,  O^{\chi}_{11} =  i \frac{m^{}_{u}}{v} \,  \frac{N}{\tan \beta},  \qquad 
c^{\, \chi, d} =  - i \frac{m^{}_{d}}{v^{}_{d}} \, O^{\chi}_{21}= - i \frac{m^{}_{d}}{v} \, N \tan \beta,    \nonumber\\
c^{\, \chi, \nu} &=& - i \frac{m^{}_{\nu}}{v^{}_u} \,  O^{\chi}_{11}  = i \frac{m^{}_{\nu}}{v} \,  \frac{N}{\tan \beta},  \qquad 
c^{\, \chi, e} = - i  \frac{m^{}_{e}}{v^{}_{d}} \, O^{\chi}_{21} =   - i \frac{m^{}_{e}}{v} \, N \tan \beta, 
\label{chi_fermions} 
\ea
which will be essential in order to establish the size of the various decay channels. 

Since we are interested in a relatively light axi-Higgs, we have focused our study
on a kinematical mass range going from $1$ to $100$ GeV. The fermionic decay channels that we consider 
are the $b\bar{b}$, $c\bar{c}$, $s\bar{s}$ for the tree level decays into quarks, $\tau\bar{\tau}$ and 
$\mu\bar{\mu}$ for the decays into leptons. At one-loop order we consider the decay into two photons, two gluons and 
in one photon and one $Z$ boson. We have added both the massless contribution 
coming from the WZ counterterm and the fermion loop 
contribution from a pseudoscalar triangle. 
The total decay rate of the axi-higgs in this approximation is given by
\ba
\Gamma^{\chi}_{tot}=\Gamma^{\chi}_{g g}+\Gamma^{\chi}_{\g\g}+\Gamma^{\chi}_{\g Z} 
+ \sum_{q=s,c,b}\Gamma^{\chi}_{q\bar{q}}+
\sum_{l=\mu,\tau}\Gamma^{\chi}_{l\bar{l}}.
\ea

\begin{itemize}
\item{\bf The tree level decays into fermions: $\chi\rightarrow f\bar{f}$} 
\end{itemize}
At leading order, for the tree-level process $\chi \rightarrow f \bar f$, 
we obtain the decay rate
\ba
\Gamma(\chi \rightarrow f \bar f) = \frac{m_\chi}{8 \pi} e^2 Q_f^2 (c^{\chi, f})^2 N_c(f) 
\sqrt{1 - \left( \frac{2 m_f}{m_\chi} \right)^2 }, 
\ea
for a value of the fermion mass below the pair production 
threshold ($4m_f^2 < m^2_\chi$). The pseudoscalar couplings to the 
fermions ($c^{\chi, f}$) have been defined in Eq. (\ref{chi_fermions}).

The leading decay is $\chi \to b\bar{b}$, due to the suppression of the fermion couplings of 
the up-type fermions (clearly shown in Table~\ref{charge_higgs}). 
We show the variations of the branching ratio (BR) of the pseudoscalar for different 
charge assignments $f(-1,-1,\Delta q^B)$, and as observed before, there are no substantial
differences induced by the selection of different assignments.

\begin{figure}[h]
\begin{center}
\includegraphics[width=6cm]{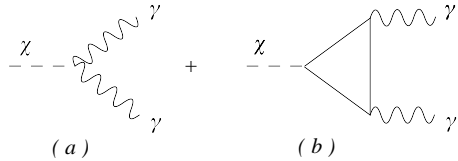}
\caption{\small Massless plus massive contributions to the $\chi \rightarrow \g \g$ process.
\label{chi_decay}}
\end{center}
\end{figure}

\begin{itemize}
\item{\bf One-loop decays into photons and gluons: $\chi \rightarrow \g\g$ and $\chi \rightarrow gg$}
\end{itemize}
We now compute the partial decay width of the axi-Higgs 
boson into two photons $\chi \rightarrow \g \g$. The invariant matrix element considered 
for the process is the sum of the two contributions shown in Fig.~\ref{chi_decay}. 
The first amplitude (Fig.~\ref{chi_decay}a) is a massless WZ vertex   
\bea
{\mathcal M}^{\mu \nu}_{WZ}(\chi \rightarrow \g \g) = 4 g^{\chi}_{\g\g} \varepsilon[\mu,\nu,k_1,k_2], 
\eea
where the coefficient $g^{\chi}_{\g\g}$ comes from the counterterm given in formula (\ref{phys_couplings}).
The second amplitude (Fig.~\ref{chi_decay}b) is a pure massive contribution
\bea
{\mathcal M}^{\mu \nu}_{f}(\chi \rightarrow \g \g) = \sum_f N_c(f) \,i  C_0(m^2_\chi,m_f) c^{\chi, f}_{\g\g} \varepsilon[\mu,\nu,k_1,k_2],   
\qquad f=\{u,d,\nu, e\}
\label{pseudo}
\eea
where $N_c(f)$ is the color factor, 1 for leptons and 3 for quarks.
In the domain $0< m_\chi < 2  m_f $ the pseudoscalar triangle when both 
photons are on mass-shell $k_1^2=k_2^2=0$ is given by the expression
\bea
C_0(m^2_\chi,m_f) = -\frac{m_f}{\pi^2 m_\chi^2}  \arctan^2 \frac{1}{ \sqrt{ \left( \frac{2 m_f}{m_\chi} \right)^2-1}}  = -\frac{m_f}{\pi^2 m_\chi^2}  \arctan^2 \frac{1}{ \sqrt{- \rho_{f \chi}^2}},
\label{region1}
\eea
with
\ba
\rho^{}_{f \chi} = \sqrt{1 -  \left( \frac{2 m_f}{m_\chi} \right)^2 },
\ea
while in the domain $2  m_f < m_\chi$ it becomes 
\bea
C_0(m^2_\chi,m_f)= \mbox{Re} C_0(m^2_\chi,m_f) + i \mbox{Im} C_0(m^2_\chi,m_f).
\label{region2}
\eea
Here we have set
\ba
\mbox{Re} C_0(m^2_\chi,m_f) &=&  \frac{m_f}{ \pi^2 m^2_\chi}  \left[ \frac{1}{4}\log^2 \left( \frac{1+\rho^{}_{f \chi}}{1-\rho^{}_{f \chi} } \right) - \frac{\pi^2}{4}  \right], \\
\mbox{Im} C_0(m^2_\chi,m_f) &=& \frac{m_f}{ \pi^2 m^2_\chi}\left[ \frac{\pi}{2} \log \left( \frac{1+\rho^{}_{f \chi}}{1-\rho^{}_{f \chi} } \right) \right].
\ea
In the numerical analysis presented below, we have introduced the function $f(\tau)$, defined in any kinematic domain, whose real part is given by 
\ba
\textrm{Re}[f(\tau)] = \left\{ \begin{array}{ll}
(\arcsin{1/\sqrt{\tau}})^2 & \textrm{if}\, \tau \geq 1\\
-\frac{1}{4}\left[\log^2\left(\frac{1+\sqrt{1-\tau}}{1-\sqrt{1-\tau}} \right) -\pi^2 \right] 
& \textrm{if}\, \tau < 1
\end{array} \right.
\ea
while its imaginary part is
\ba
\textrm{Im}[f(\tau)] = \left\{ \begin{array}{ll}
0 & \textrm{if}\, \tau \geq 1\\
\frac{\pi}{2}\left[\log\left(\frac{1+\sqrt{1-\tau}}{1-\sqrt{1-\tau}} \right)\right] 
& \textrm{if}\, \tau < 1
\end{array} \right.
\ea
where $\tau=4m_f^2/m_{\chi}^2$.

x
Finally, the 1-loop decay $\chi \rightarrow \g\g$ is given by the following amplitudes
\bea
{\mathcal M}^{\mu \nu}(\chi \rightarrow \g \g) = {\mathcal M}^{\mu \nu}_{WZ}+{\mathcal M}^{\mu \nu}_{f}
\eea
and the rate computed from the two contributions shown in Fig.~\ref{chi_decay} is
\bea
\Gamma(\chi \rightarrow \g\g) &=&  \frac{m^3_\chi}{32 \pi}  \left\{ 8 (g^\chi_{\g\g})^2 
+ \frac{1}{2} \left| \sum_f N_c(f) i \frac{\tau_f ~f(\tau_f)}{4\pi^2 m_f} e^2 Q_f^2 c^{\chi, f} \right|^2  \right.  
\nonumber\\
&& \left. \,\,\,\,\,\,\, \,\,\,\,\,\,\, \,\,\,\,+ \,\,4 g^\chi_{\g\g} 
\sum_f N_c(f) i \frac{\tau_f ~f(\tau_f)}{4\pi^2 m_f} e^2 Q_f^2 c^{\chi, f} \right\}. 
\eea
In Fig.~\ref{chi_decay}a we have isolated the massless contribution to 
the decay rate coming from the WZ counterterm $\chi F^{}_\g F^{}_\g$ whose expression is 
\bea
\Gamma^{}_{WZ}(\chi \rightarrow \g\g)= \frac{m^3_\chi}{4 \pi}(g^\chi_{\g\g})^2.
\eea

We should notice that the massive contribution from amplitude (\ref{pseudo}) 
is completely independent of the anomalous coupling $g_B$, which does not appear 
in the coefficients $c^{\chi, f}$, as can be seen from Eq.~(\ref{chi_fermions}). 
For the decay into two gluons we proceed in a similar manner (see Fig.~\ref{chi_production}) 
and the amplitude is given by
\bea
{\mathcal M}^{\mu \nu}_{\small WZ}(gg \rightarrow \chi) = 4 g^{\chi}_{gg} \varepsilon[\mu,\nu,k_1,k_2],
\eea
where the coefficient $g^{\chi}_{\,gg}$ is given in Eq. (\ref{phys_couplings}).
The second amplitude (Fig.~\ref{chi_production}b) is a pure massive contribution
\bea
{\mathcal M}^{\mu \nu}_{q}(gg \rightarrow \chi)=   \sum_q i C_0(m^2_\chi,m_q) \mbox{Tr}[T^a T^b] c^{\chi, q}_{gg} 
\varepsilon[\mu,\nu,k_1,k_2],   \qquad q=\{u,d\}
\label{pseudogg}
\eea
with u = \{$u$, $c$, $t$\} and d = \{$d$, $s$, $b$\}, and the coefficients 
$c^{\chi,q}$ are defined in relations (\ref{chi_fermions}).
The decay rate is then given by
\bea
\Gamma(\chi \rightarrow gg) &=&  \frac{m^3_\chi}{16 \pi}   \left[ 8 (g^\chi_{gg})^2 
+\frac{1}{2} \left| \sum_q i \frac{N_c\tau_f ~f(\tau_f)}{4\pi^2 m_f} 4\pi\alpha_s c^{\chi, q} \right|^2    \right. \nonumber\\
&&\left. \,\,\,\,\,\,\, \,\,\,\,\,\,\, \,\,\,\,+ \,\,4 g^\chi_{gg} \sum_q i \frac{N_c\tau_f ~f(\tau_f)}{4\pi^2 m_f} 4\pi\alpha_s c^{\chi, q}   \right],
\eea
while the expression of the isolated contribution from the corresponding WZ counterterm is instead given by
\bea
\Gamma_{WZ}(\chi \rightarrow gg)=\frac{m^3_\chi}{2 \pi}(g^\chi_{gg})^2.
\eea

\begin{figure}[t]
{\includegraphics[
width=6cm,
angle=-90]{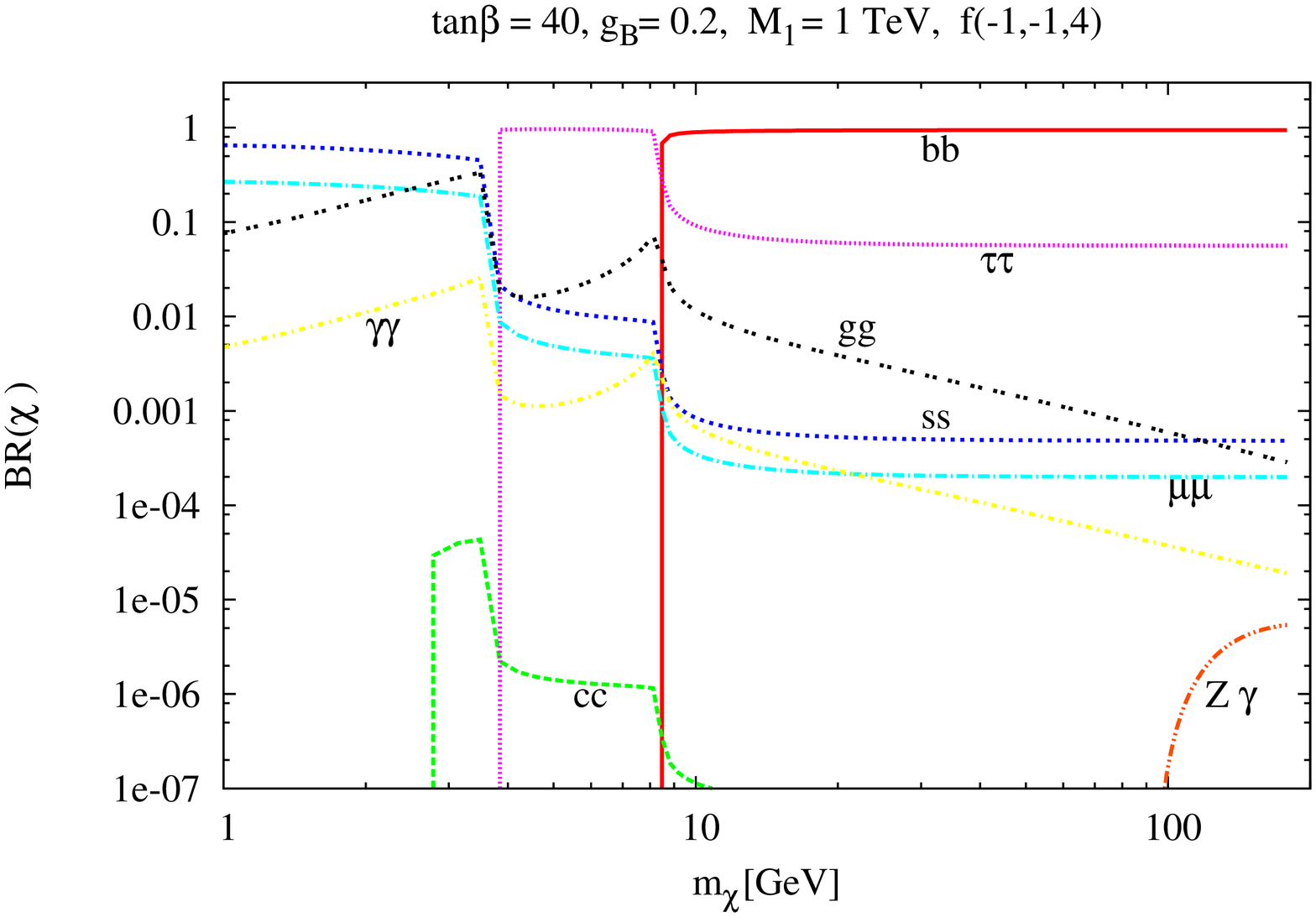}}
{\includegraphics[
width=6cm,
angle=-90]{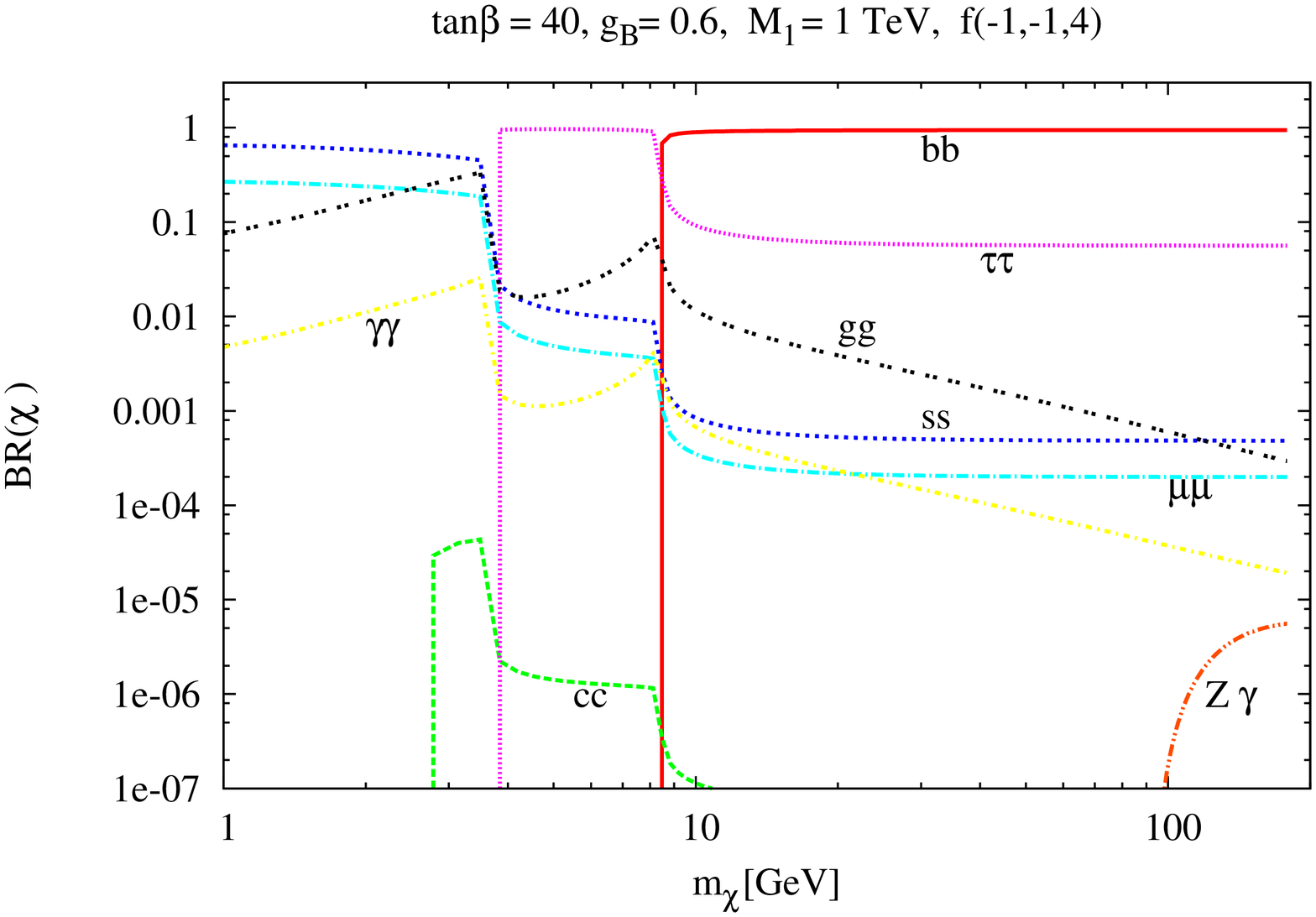}}
{\includegraphics[
width=6cm,
angle=-90]{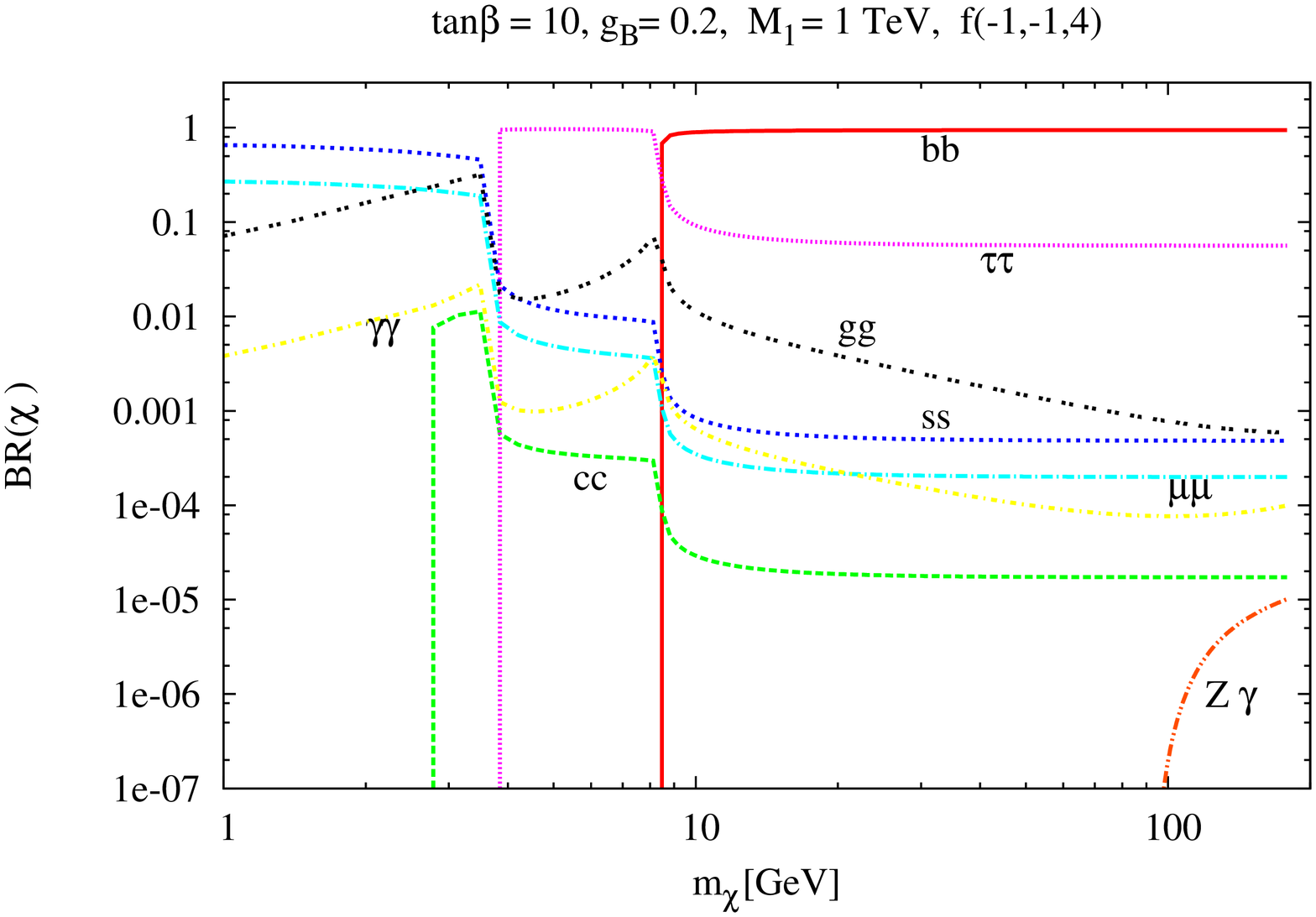}}
{\includegraphics[
width=6cm,
angle=-90]{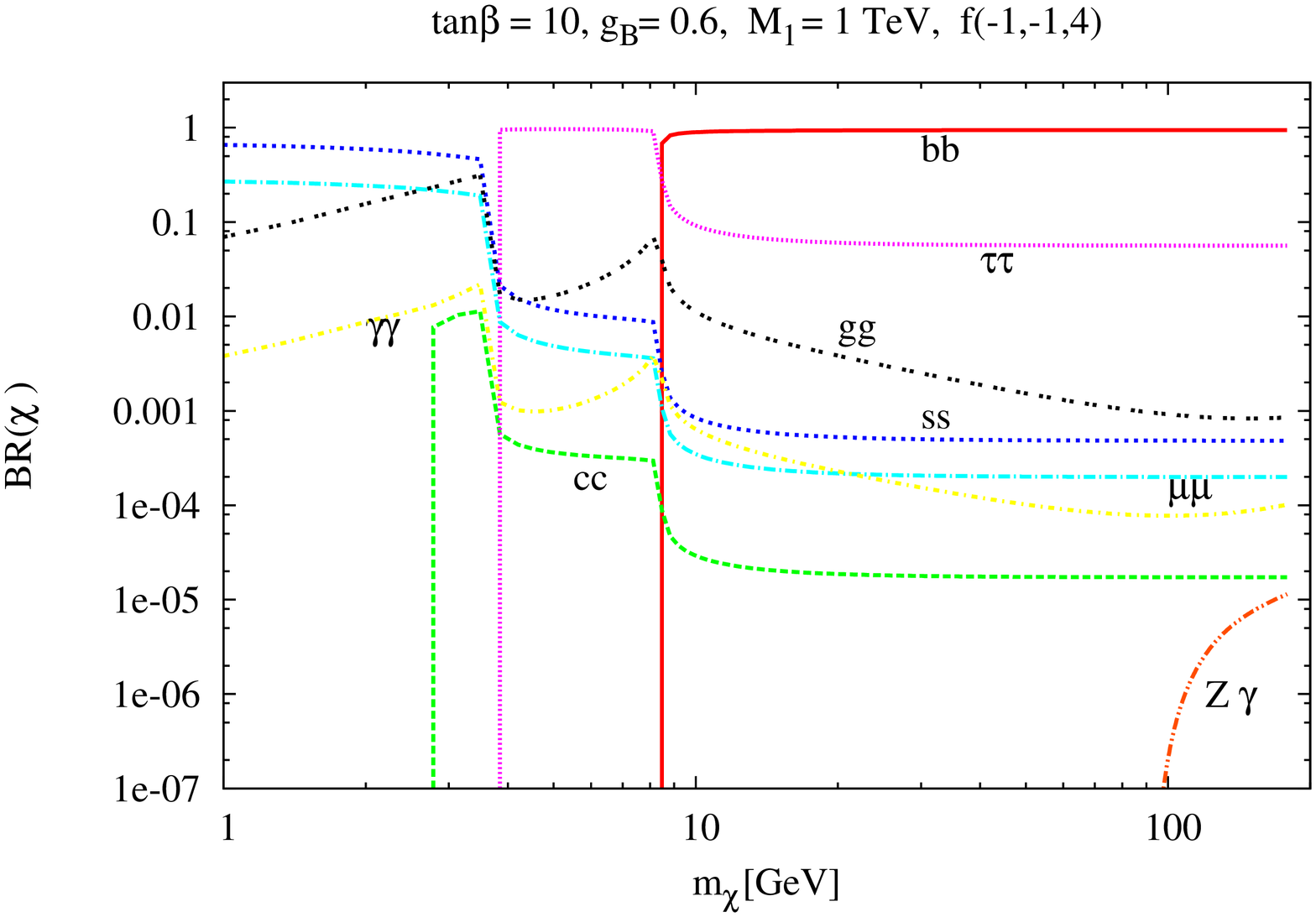}}
\caption{\small Study of the branching ratios of the axi-Higgs. 
We analyze the dependence on the free parameters $g^{}_B, \tan \beta$.} 
\label{Br_ratios}
\end{figure}

\begin{figure}
\begin{center}
{\includegraphics[
width=6cm,
angle=-90]{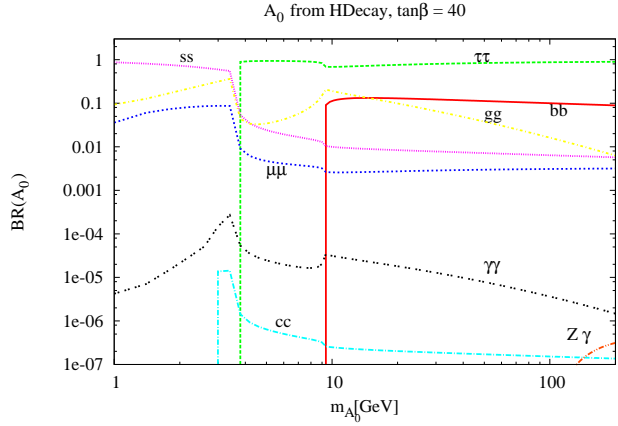}}
\caption{Branching ratios for the CP-odd scalar $A_0$ of the MSSM}
\label{Br_ratiosMSSM}
\end{center}
\end{figure}

\begin{itemize}
\item{\bf The decay $\chi \rightarrow {\g \cal Z}$}
\end{itemize}

The partial decay rate computed from the corresponding WZ counterterm and fermion loop, analogously to Fig.~\ref{chi_decay}, is
\ba
\Gamma(\chi \rightarrow \g {\cal Z}) &=&  \frac{m^3_\chi}{8 \pi}   \left[ 4 (g^\chi_{\g {\cal Z}})^2 
+  | \sum_f N_c(f) i C_0(m^2_\chi, m^2_{\cal Z}, m_f) e^2 Q_f^2 c^{\chi, f}_{\g {\cal Z}} |^2    \right. \nonumber\\
&&\left. \,\,\,\,\,\,\, \,\,\,\,\,\,\, \,\,\,\,+ \,\,4 g^\chi_{\g {\cal Z}} \sum_f N_c(f) i C_0(m^2_\chi, m^2_{\cal Z}, m_f) e^2 Q_f^2 c^{\chi, f}_{\g {\cal Z}}   \right]  \left( 1- \frac{m_{\cal Z}^2}{m^2_\chi}  \right)^3,
\ea
(${\cal Z}= Z, Z^\prime$) which is well defined only for a mass of the $\cal Z$ boson under the threshold production $m_{\cal Z} < m_\chi$.
The couplings are defined as
\ba
c^{\,\chi, f}_{\g \cal Z} = g^{}_{\cal Z} g_V^{{\cal Z}, f} c^{\,\chi, f}.
\ea 
The vector and axial-vector couplings of the $\cal Z$ bosons to the fermions in the physical basis are related to the charges of the chiral fermions by the expressions
\ba
g_V^{{\cal Z}, f} = \frac{1}{2} (Q_{\cal Z}^{R, f}+ Q_{\cal Z}^{L, f}),  \qquad  g_A^{{\cal Z}, f} = \frac{1}{2} (Q_{\cal Z}^{R, f}- Q_{\cal Z}^{L, f}),
\ea
which are obtained, as detailed in \cite{Armillis:2007tb, Coriano:2007xg}, starting from the interaction basis ($W^3, Y, B$) by means of the following rotations  
\ba
g^{}_{\cal Z} Q_{\cal Z}^{R, f}  &=& g^{}_Y Y^{R, f} O^A_{Y \cal Z} +  g^{}_B Y_B^{R, f} O^A_{B \cal Z}   \nonumber\\
g^{}_{\cal Z} Q_{\cal Z}^{L, f} &=& g^{}_2 T^{3L, f} O^A_{W_3 \cal Z} + g^{}_Y Y^{L, f} O^A_{Y \cal Z} +  g^{}_B Y_B^{L, f} O^A_{B \cal Z}. 
\ea
The $Y^{L/R}_B$, $Y^{L/R}$ and $T^{3L}$ are the generators of the gauge group of the model in the chiral basis. \\
The pseudoscalar triangle $C_0(m^2_\chi, m^2_{\cal Z}, m_f)$ involved in the decay $\chi \rightarrow \g \cal Z$ with both external lines on their mass-shell, $k_1^2=0$ and $k_2^2=m_{\cal Z}^2$, is given by (see \cite{Kniehl:1989qu})
\ba
C_0(m^2_\chi, m^2_{\cal Z}, m_f)= \frac{1}{m_\chi^2-m^2_{\cal Z}} \left[ m_\chi^2 C_0(m^2_\chi, m_f) -  m^2_{\cal Z}  C_0(m^2_{\cal Z}, m_f) \right],
\label{pseudo_phZ}
\ea
where the structure of $C_0(m^2_\chi, m_f)$ has already been studied in (\ref{region1}, \ref{region2}).  In complete analogy, the function $C_0(m^2_{\cal Z}, m_f)$ can be obtained from $C_0(m^2_\chi, m_f)$ just by replacing the first argument $m^2_\chi$ with $m^2_{\cal Z}$. Then, the study of the decay rate  is closely related to the behaviour of the three-point function  (\ref{pseudo_phZ}) in the various physical domains of its definition.
In the domain $0 < m_{\cal Z} < m_\chi < 2 m_f$ the expressions for $C_0(m^2_\chi, m_f)$ and $C_0(m^2_{\cal Z}, m_f)$ can be read from eq.~(\ref{region1}), in particular we obtain
\ba
C_0(m^2_{\cal Z},m_f) = - \frac{m_f}{\pi^2 m^2_{\cal Z}} \arctan^2 \frac{1}{  \sqrt{ - \rho^2_{f \cal Z} } },
\label{region1_Z}
\ea 
with
\ba
\rho^{}_{f \cal Z} = \sqrt{1 -  \left( \frac{2 m_f}{m_{\cal Z}} \right)^2 }.
\ea
As $m_\chi$ grows we can have two possible cases. If $0 < m_{\cal Z} < 2 m_f < m_\chi $, while the function $C_0(m^2_\chi, m_f)$ develops real and imaginary part as shown in eq.~(\ref{region2}), the function $C_0(m^2_{\cal Z}, m_f)$ is still well defined. But finally if $0 < 2 m_f< m_{\cal Z}  < m_\chi $ also
$C_0(m^2_{\cal Z}, m_f)$ develops real and imaginary parts, in particular
 \bea
C_0(m^2_{\cal Z}, m_f)= \mbox{Re} C_0(m^2_{\cal Z}, m_f) + i \mbox{Im} C_0(m^2_{\cal Z}, m_f),
\eea
in analogy to eq.~(\ref{region2}). The massless WZ contribution to the decay rate is 
\ba
\Gamma^{}_{WZ}(\chi \rightarrow \g {\cal Z}) = \frac{m^3_\chi}{2 \pi} (g^\chi_{ \g {\cal Z} })^2 
\left( 1 - \frac{ m^2_{\cal Z} }{ m^2_\chi} \right)^3.   \qquad  \mbox{(${\cal Z}=Z, Z^\prime$) }.
\ea
We just remark that in the calculation of $\Gamma(\chi \rightarrow \g \g)$ and $\Gamma(\chi \rightarrow \g {\cal Z})$ we have neglected the contributions coming from the loops generated by the scalar $H_0$ and $h_0$.

\begin{figure}[h]
\begin{center}
{\includegraphics[
width=8cm,
angle=-90]{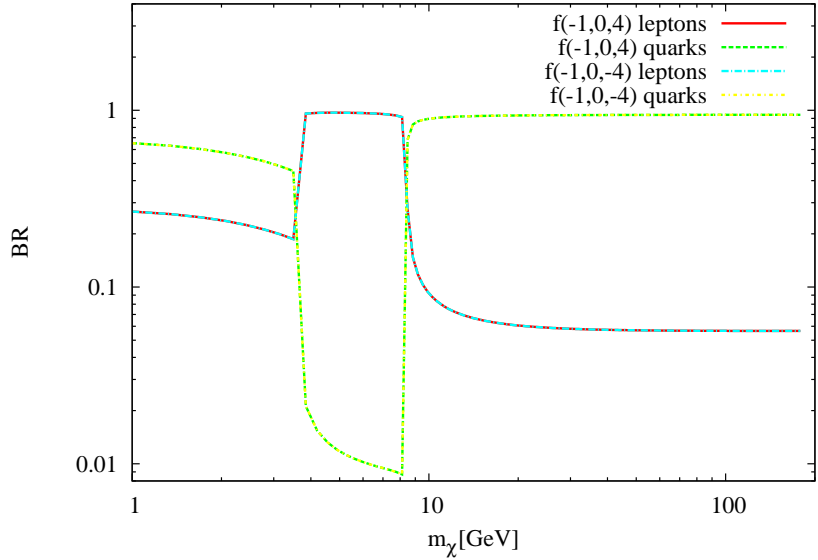}}
\caption{\small Study of the leptonic and the quarks branching ratios of the axi-Higgs. 
We analyze the dependence on the function $f(q^B_{Q_L},q^B_{u_R},\Delta q^B)$.} 
\label{Madrid_model}
\end{center}
\end{figure}

In Fig.~\ref{Br_ratios}, for a given value of the Stueckelberg mass $M_1=1$ TeV, 
we study the dependence on the free parameter $\tan \beta=\{10,40 \}$ 
and on $g_B=\{0.2,0.6\}$. The dependence on $\tan \beta$ strongly affects 
the branching ratio for the decay into a $c\bar{c}$ pair, which appears to be suppressed for a large value of 
$\tan \beta$ ($\tan \beta=40$). The plots clearly show the presence of 3 different main regions in which the decay channels of the axi-Higgs are rather different. In the region  $0 \le m_\chi \le 2.8$ GeV the dominant decay is in the $s\bar{s}$ and $ \mu\bar{\mu}$ channels, with a sizeable gluon channel which becomes very relevant around $m_\chi=3$ GeV. 

For $2.8 \le m_\chi \le 8.5$ GeV the dominant decay channel is the $\tau\bar{\tau}$, followed by a third region 
with $m_\chi >8.5$ GeV in which the $b \bar{b}$ channel opens up. The 4 plots describe different charge assignments. One can notice rather straightforwardly that the leading behaviour in each mass region remains the same in each plot,  while the subleading channels get reshuffled in their separate contributions. We show in Fig. \ref{Br_ratiosMSSM} for a comparison, the branching ratios for the CP-odd scalar of the MSSM as a function of its mass.  In this case the dominant regions are two,  divided approximately into the  two regions by 
$m_\chi = 5$ GeV and the where the dominant decays are into $s\bar{s}$ (in the lower region) 
and into $\tau\bar{\tau}$ (in the higher mass region).

\begin{itemize}
\item{\bf Total rates and dependence on the charge assignments}
\end{itemize}

We show in Fig.~(\ref{Madrid_model}) plots which illustrate the behaviour of the 
(inclusive) branching ratios of the axi-higgs into quarks and leptons as a function of the mass of the physical axion, obtained by varying the charge assignments of the model.  The enhancement of the lepton decay channels for a light axion mass between 4 and 8 GeV, respect to the quark channel, is very stable against these variations.
These changes are described by the function $f(q^B_{Q_L},q^B_{u_R},\Delta q^B)$ and in the various cases are
are almost coincident, and this is due to the fact that the differences in the 
smaller than $10^{-3}$.

\section{CP-even sector: decays and associated production}
We now move to discuss the CP-even sector of the model which involves the two states $H_0$ and $h_0$. 
We include all the relevant channels, such as the $f\bar{f}$, the $WW$, $ZZ$ and $Z\gamma$ and diphoton  channels.  

\begin{itemize}
\item{\bf Decays into $f\bar{f}$}
\end{itemize} 
We start by calculating the tree level decay rate into fermions, which is given by  
\bea
\Gamma_{}(h/H \rightarrow f \bar f)= \frac{ m^{}_{h/H}}{8 \pi} |c^{h/H,f}|^2 N_c(f) \left( 1 - \frac{4 m^2_f}{m^2_{h/H}}\right)^{3/2}
\label{form}
\eea
where the scalar couplings to the fermions $c^{h/H,f}$ have been defined in Eqs.~(\ref{Hff},\ref{hff}).
The decay for the SM scalar Higgs is obtained from Eq.~(\ref{form}) just by substituting the coupling
$c^{h/H,f}$ with the SM one, that is $-m_f/v$, where $v$ is the vacuum expecation value of the 
SM Higgs field ($v\approx 246$ GeV).

\begin{itemize}
\item{\bf Tree level decays of the scalar Higgs bosons into $W^{\pm}$ and $ZZ$}
\end{itemize}
The tree level contributions to the total decay rate of the two Higgs 
due to the decay into a $W^{\pm}$ pair and a $ZZ$ pair are computed similarly. These are found to be 
relevant in the case of $H_0$ for a mass $m_{H_0}$ greater than $100$ GeV. 
In particular we have added the contributions due to $H_0/h_0\rightarrow Z^* Z$ and 
$H_0/h_0\rightarrow W^* W$ that could be significant when the mass of the scalar is close to
the thresholds for $ZZ$ and $WW$ pair production. 

For the case of a $ZZ$ pair we obtain
\begin{displaymath}
\Gamma(H\rightarrow Z Z)=
\left\{\begin{array}{ll}
\Gamma_{Z^* Z}^{H}=\left(C^{H0/h0}_{ZZ} \frac{g_2}{c_w M_z}\right)^2 
\left(7 -\frac{40}{3} s_w^2 +\frac{160}{9}\frac{s_w^4}{c_w^4}\right)\frac{m_{H}F(M_Z/m_H)}{2048\pi^3} & \textrm{if}\;\; M_Z \leq m_H \leq 2 M_Z\\ \\
\left(C^{H0/h0}_{ZZ}\right)^2\frac{\sqrt{1-x_z}}{128 \pi m_H}\left(3+\frac{4}{x_z^2} -\frac{4}{x_z}\right) 
+\Gamma_{Z^* Z}& \textrm{if}\;\; m_H \geq 2 M_Z
\end{array}\right.
\end{displaymath}
where the coupling $C^{H0/h0}_{ZZ}$ has been defined in the previous sections and $x_z=4M_Z^2/m_H^2$. $s_w,c_w$ are short notations for $\sin\theta_W,\cos\theta_W$ respectively. 

For the case of two charged $W$'s we have 
\begin{displaymath}
\Gamma(H\rightarrow W W)=
\left\{\begin{array}{ll}
\Gamma_{W^* W}^{H}=\left(C^{H0/h0}_{WW} \frac{g_2}{M_W}\right)^2 \frac{\bar{n}\; m_{H}}{512\pi^3}F(M_W/m_H) & \textrm{if}\;\; M_W \leq m_H \leq 2 M_W\\ \\
\left(C^{H0/h0}_{WW}\right)^2\frac{\sqrt{1-x_w}}{64 \pi m_H}\left(3+\frac{4}{x_w^2} -\frac{4}{x_w}\right) 
+\Gamma_{W^* W}^{H}& \textrm{if}\;\; m_H \geq 2 M_W
\end{array}\right.
\end{displaymath}
Here the coefficient $\bar{n}$ is equal to 3 if $W^*\rightarrow t b$ is not allowed, while is equal to 4 
if $W^*\rightarrow t b$ is allowed. Again, we have defined the coefficient $x_w=4M_W^2/m_H^2$.
 
In the region $1/2 \leq x \leq 1$ the function $F(x)$ is defined as follows 
\ba
&&F(x)=-|1-x^2|\left( \frac{47}{2}x^2 -\frac{13}{2} +\frac{1}{x^2}\right)
+3(1 - 6 x^2 +4x^4)|\ln(x)| 
\nonumber\\
&&\hspace{1cm}+\frac{3(1 - 8 x^2 + 20 x^4)}{\sqrt{4x^2-1}}\cos^{-1}\left(\frac{3x^2-1}{2x^3} \right).
\nonumber\\
\ea

\begin{itemize}
\item{\bf Two photon decay of the scalar Higgs bosons $h^0, H^0 \rightarrow \g\g$}
\end{itemize}
The computation of the decay rate of a CP-even scalar of the MLSOM 
into a pair of photons is similar to that of the SM. It includes the contribution of the spin 1/2 particles (the fermion loop), 
of the spin 1 (the W loop) and the spin 0 ($H^{\pm}$ loop) and it is given by
\ba
&&\Gamma(H\rightarrow \g\g)=\frac{4\alpha_{em}^2}{1024\pi^3}
m_H^3\left|\sum_{f} N_c(f)Q_f^2\frac{c^{H,f}}{m_f}(-2)\tau_f   
\left[1+(1-\tau_f)f(\tau_f)\right]
\right.\nonumber\\
&&\hspace{2cm}\left. + \left(\frac{C^H_{WW}}{g_2 M_W^2}\right)
[2 + 3\tau_w + 3 \tau_w (2-\tau_w) f(\tau_w)]\right|^2
\ea
where $H$ represents $H_0$ or $h_0$, $\tau_f=4m_f^2/m_H^2$, $\tau_w=4M_W^2/m_H^2$ and the function $f(\tau)$ has been  defined previously. 
The scalar couplings of the lighter Higgs boson $h^0$ to the fermions are shown in ${\cal L}_{\rm Yuk}(h)$ 
and their expressions are
\ba
c^{h_0, u} &=&  - \frac{m_u}{v_u} \sin \a = - \frac{m_u}{v } \frac{\sin \a}{\sin \beta} ,  \qquad 
c^{h_0, d} =  - \frac{m_d}{v_d} \cos \a=- \frac{m_d}{v }  \frac{\cos \a}{\cos \beta},    \nonumber\\
c^{h_0, \nu} &=& - \frac{m_\nu}{v_u}  \sin \a= - \frac{m_\nu}{v }  \frac{\sin \a}{\sin \beta},  \qquad 
c^{h_0, e} =  - \frac{m_e}{v_d}  \cos \a = - \frac{m_\nu}{v }  \frac{\sin \a}{\cos \beta},    
\label{hff}
\ea
while the scalar couplings of the heavier Higgs boson $H^0$ to the fermions are shown in ${\cal L}_{\rm Yuk}(H)$ and are given by
\ba
c^{H_0, u} &=&  \frac{m_u}{v_u} \cos \a =  \frac{m_u}{v} \frac{\cos \a}{\sin \beta},  \qquad 
c^{H_0, d} =  - \frac{m_d}{v_d}  \sin \a= - \frac{m_d}{v}  \frac{\sin \a}{\cos \beta},    \nonumber\\
c^{H_0, \nu} &=&  \frac{m_\nu}{v_u}  \cos \a=  \frac{m_\nu}{v}  \frac{\cos \a}{\sin \beta},  \qquad 
c^{H_0, e} = - \frac{m_e}{v_d}  \sin \a =- \frac{m_e}{v}  \frac{\sin \a}{\cos \beta} .    
\label{Hff}
\ea
Here we have used the relations for the expectation values $v_u=v \sin \beta$ and $v_d=v \cos \beta$ to express these couplings in terms of the couplings of the Higgs boson o the SM.
The calculation of the rate into gluons is similar but we have only the fermion loop
\ba
&&\Gamma(H\rightarrow g g)=\frac{4\alpha_{s}^2}{512\pi^3}
m_H^3\left|\sum_{f}c^{H,f} \frac{\tau_f}{m_f}(-2)  
\left[1+(1-\tau_f)f(\tau_f)\right]\right|^2.
\ea
\begin{itemize}
\item{\bf $Z\g$ decay of the scalar Higgs bosons}
\end{itemize}

The last contribution that we consider in the computation of the total decay rate of $H_0/h_0$
is the decay into $Z\g$. Also in this case we include only the contribution of the 
fermion loop and of the spin-1 loop and we neglect the contribution coming from other loops of scalars
\ba
&&\Gamma(H\rightarrow Z\g)=\frac{m_H^3}{32}\left(1-\frac{M_Z^2}{m_H^2}\right)^3\left|
\sum_f N_c(f) \frac{\alpha_{em}}{2\pi}\frac{c^{H,f}}{m_f}
\frac{(-2)Q_f\left(T^{3L}_f -2 Q_f s_w^2\right)}{s_w c_w}
\left[I_1(\tau_f,\lambda_f)-I_2(\tau_f,\lambda_f)\right]
\right.\nonumber\\
&&\hspace{6cm}\left.
-\frac{\alpha_{em}}{4\pi M_W^2}C^{H}_{WW} \cot\theta_W \left\{4(3-(\tan\theta_W)^2) I_2(\tau_w,\lambda_W)
\right.\right.\nonumber\\
&&\hspace{6cm}\left.\left.
+\left[ \left(1+\frac{2}{\tau_W}\right)(\tan\theta_W)^2 -\left(5+\frac{2}{\tau_W}\right) \right]
I_1(\tau_w,\lambda_W) \right\}
\right|^2
\label{Zgamma}
\ea
where $\lambda_f=4m_f^2/M_Z^2$ and $\lambda_W=4 M_W^2/M_Z^2$, while the functions $I_{1,2}$ are given in
\cite{Bergstrom:1985hp} \cite{Gunion:1989we}. We report them here for completeness
\ba
&&I_1(a,b)=\frac{a b}{2(a-b)} + \frac{a^2 b^2}{2(a-b)^2}[f(a)-f(b)]+\frac{a^2 b}{(a-b)^2}[g(a)-g(b)]
\nonumber\\
&&I_2(a,b)=-\frac{a b}{2(a-b)}[f(a)-f(b)].
\ea
The function $f(\tau)$ has been defined in a previous section while $g(\tau)$ is given by 
\begin{displaymath}
g(\tau)=
\left\{\begin{array}{ll}
\sqrt{\tau-1} \arcsin (1/\sqrt{\tau})& \textrm{if}\;\; \tau \geq 1 \\ \\
\frac{1}{2}\sqrt{1- \tau} \left[\ln\left(\frac{1+\sqrt{1-\tau}}{1-\sqrt{1-\tau}} \right)-i\pi\right]& 
\textrm{if}\;\; \tau < 1
\end{array}\right.
\end{displaymath}
It is important to observe that in the first line of Eq. (\ref{Zgamma}) we have neglected the contribution 
to the the fermion-boson couplings due to the presence of an extra anomalous $U(1)$.
As a matter of fact, in our hypothesis ($M_1=1$ TeV and $g_B=0.1-0.2$), this contribution is found to be very small
and for this kind of study these couplings 
can be considered substantially coincident with those of the SM.

Finally, the total decay rate for $H_0/h_0$ will be given as follows
\ba
\Gamma_{tot}^{H}=\sum_{f}\Gamma_{f\bar{f}} + \Gamma_{\g\g}+ \Gamma_{gg} +
\Gamma_{WW} +\Gamma_{ZZ} +\Gamma_{Z\g}.
\ea
\subsection{Numerical results} 
\begin{figure}[t]
{\includegraphics[
width=6cm,
angle=-90]{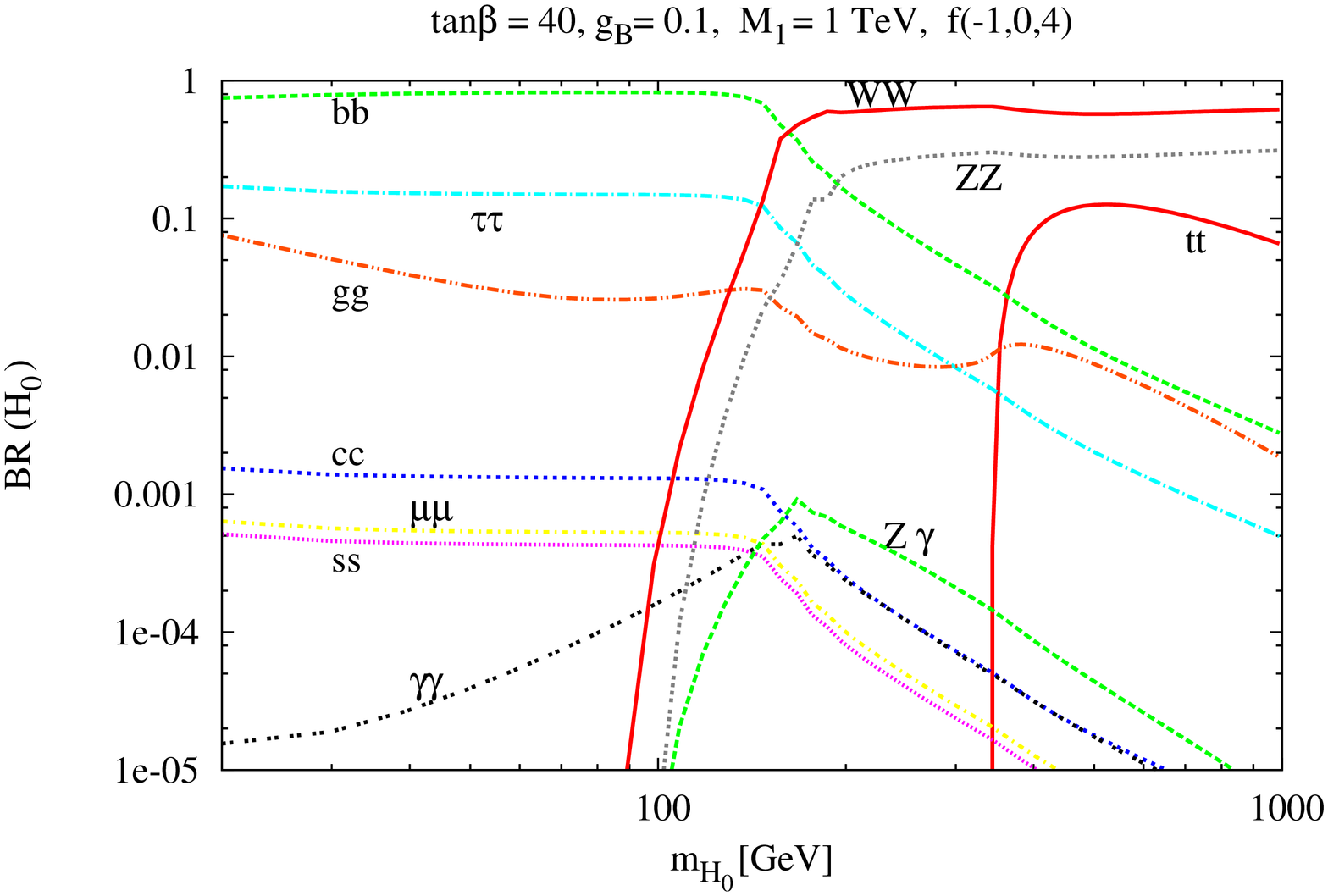}}
{\includegraphics[
width=6cm,
angle=-90]{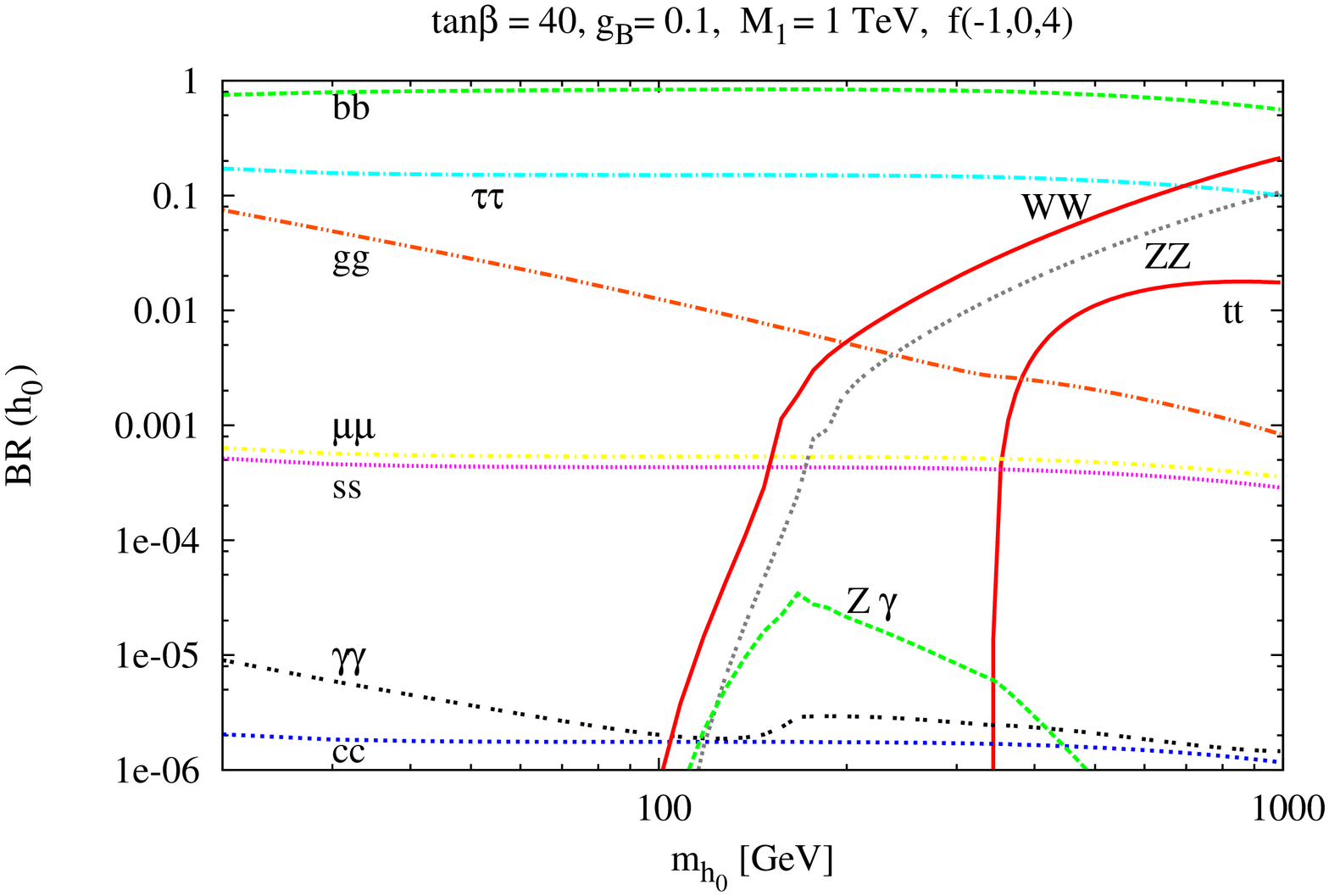}}
\caption{\small Study of the branching ratios of the CP-even sector.} 
\label{Br_ratiosH0}
\end{figure}

\begin{figure}[t]
\begin{center}
{\includegraphics[
width=6cm,
angle=-90]{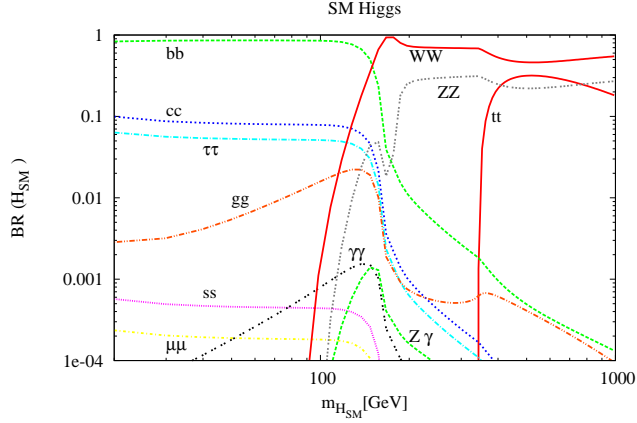}}
 \caption{\small (Study of the branching ratios of the SM-higgs.}
\label{SSMH}
\end{center}
\end{figure}

We shown in Figs.(\ref{Br_ratiosH0}-\ref{SSMH}) a comparative study of the branching ratios of the scalars $H_0$ and $h_0$ in the CP-even sector of the MLSOM and those of the Higgs of the SM. While the $H_0$ and the SM Higgs  appear to be dominated in their decays by the $b\bar{b}$ channel only below the $WW$ region, the preferential decay of the $h_0$ is entirely into this final state for all mass ranges.  Both the $H_0$ and the $h_0$ appear to have a more sizeable decay into $\tau\bar{\tau}$ compared to the SM Higgs. The branching ratio for the decay into $\gamma\gamma$  appears to be rather small for the $h_0$ in all the mass range, while the $H_0$ and the SM Higgs show, for this channel, a similar behaviour. The two-gluons channel also appears to be more significant for both states of the MLSOM compared to the SM Higgs, over the entire mass range, while the $c\bar{c}$ channel appears to be rather suppressed in the case of the $h_0$ compared to the SM Higgs. Smaller $c\bar{c}$ rates are also found for the $H_0$ respect to the ordinary Higgs.

\subsection{\bf Associated Production of the CP-even states with vector bosons}

Another possible way of detecting the Higgs at hadron colliders is through its associated production with a vector boson. Here we calculate the 
LO cross section for the $H_0/h_0$ associated production with a $W$ and $Z$ at the LHC 
and Tevatron and we have made a comparison with the corresponding rates for the ordinary SM Higgs.

\begin{figure}[t]
\includegraphics[width=6cm, angle=-90]{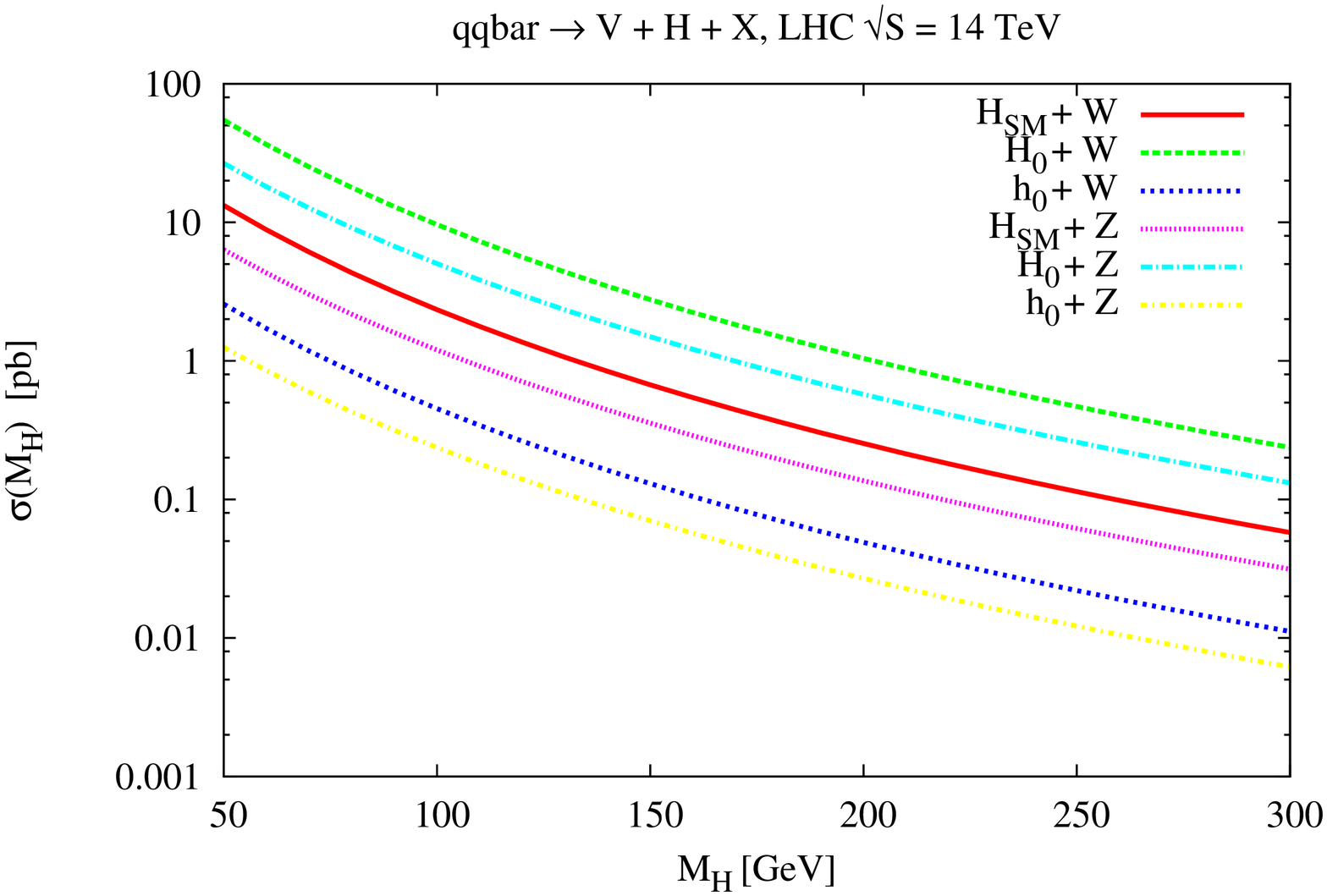}
\includegraphics[width=6cm, angle=-90]{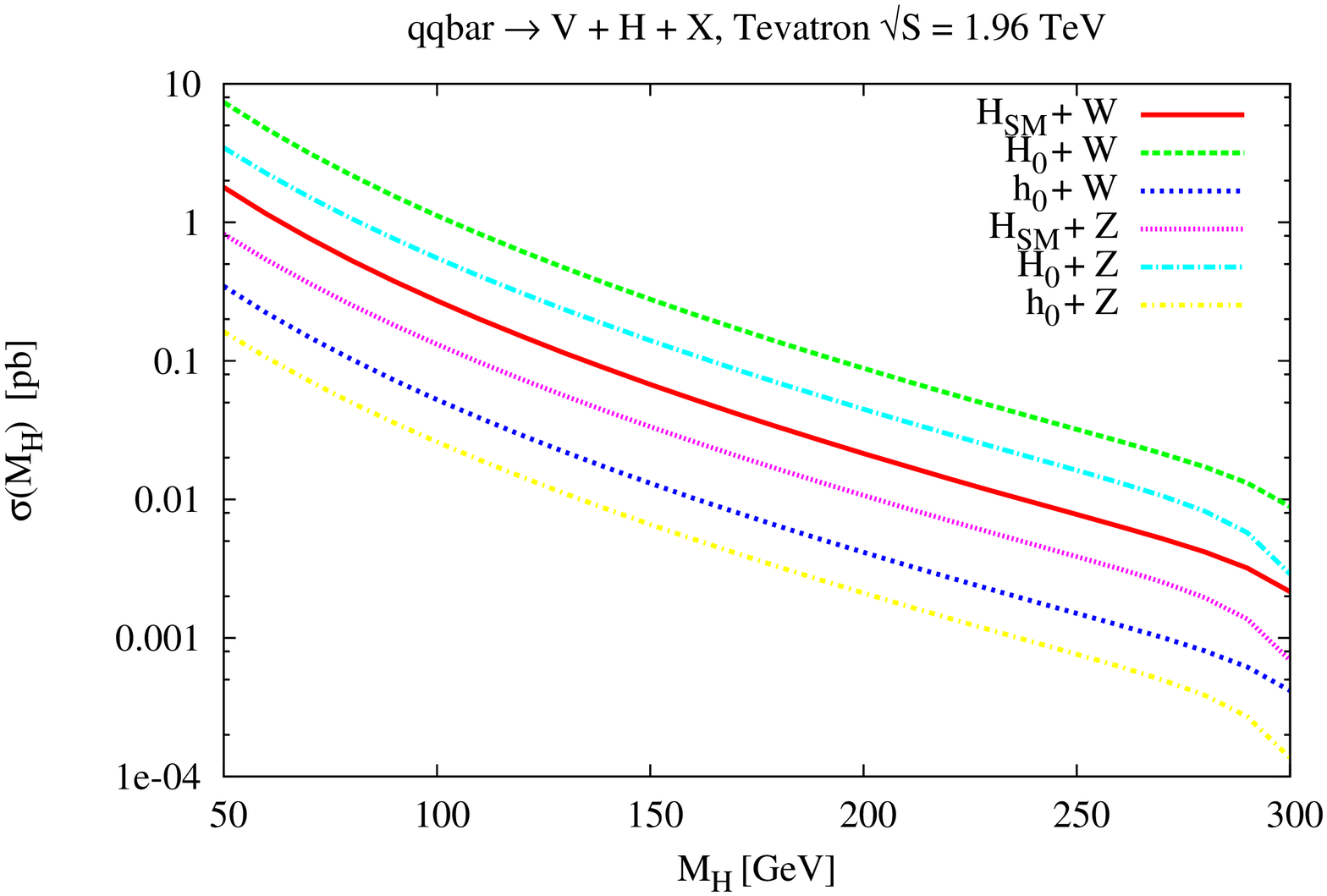}
\caption{\small $q\bar q\rightarrow H + V + X$ at LO at the Tevatron.}
\label{LHC_Teva}
\end{figure}

The partonic cross section can be written as 
\ba
&&\hat{\sigma}(q\bar{q}\rightarrow H_0/h_0 + V)= \frac{g_2^2}{32 M_V^2}(C^{H}_{VV})^2\frac{1}{288\pi \hat{s}}
\left[(g_A^f)^2 +(g_V^f)^2\right] \times
\nonumber\\
&&\hspace{2cm}\lambda^{1/2}(M_V^2,m_H^2,\hat{s})\frac{\lambda(M_V^2,m_H^2,\hat{s}) 
+12M_V^2/\hat{s}}{\left(1-M_V^2/\hat{s}\right)^2}
\ea 
where $V$ represents $W$ or $Z$, the couplings to the fermions are defined as 
$g_A^f=2T^{3L}_f$, $g_V^f=2 T^{3L}_f -4Q_f s_w^2$ for the Z, while $g_A^f=g_V^f=\sqrt{2}$ for the W.  
The phase space coefficient is defined as $\lambda(x,y,z)=(1-x/z-y/z)^2 -4xy/z^2$.
The total cross section as a function of the mass of the Higgs is given by the convolution
of the partonic cross section with the PDFs luminosity of the quark-antiquark pair produced 
in the initial state which is given by
\ba
\Phi_{q\bar{q}}(\tau,\mu_F,\mu_R)=\int_{\tau}^1\sum_{q,\bar{q}} \frac{dx}{x}\left[f^{q}_{H_1}(x,\mu_F,\mu_R)f^{\bar{q}}_{H_2}(\tau/x,\mu_F,\mu_R)+\{H_1 \leftrightarrow H_2\}\right]
\ea
where $\mu_F,\mu_R$ are the factorization and renormalization scales and $f^{q}_{H_1}$ represents the
quark probability relative to the hadron $H_1$, etc.
We have performed the PDF evolution with CANDIA \cite{Cafarella:2008du} 
and we have used the set MRST 2001 as input distributions, evolved up to $\mu_F=\mu_R=Q$.
The total cross section is given by
\ba
\sigma_{LO}(m_H,\mu_F,\mu_R)=\int_{\tau_0}^{1} 
\Phi_{q\bar{q}}(\tau,\mu_F,\mu_R)           
\hat{\sigma}( \tau S) d \tau 
\ea
where $\tau_0=(M_V+m_H)^2/S$ and $S$ is the center of mass energy of the two incoming hadrons.
In Fig.~(\ref{LHC_Teva}) we have shown the plots of the total cross section
for the LHC and the Tevatron. In the $W$-channel the cross section of the SM Higgs is smaller that the similar one of the MLSOM due to the $H_0$, while the same cross section for the $h_0$ is more suppressed. A similar behaviour is found both at the LHC and at the Tevatron. The cross section in the case of the $Z$ follows a similar pattern in all the three cases. 

\section{Axi-Higgs production at hadron colliders}
The study of the production of the axi-Higgs at hadron colliders is particularly interesting, especially for the possibility of having sizeable branching ratios of the two Higgs $H_0$ and $h_0$ into final state axions.  
 Before we come to this study, we pause for some observations regarding the scalar potential of the MLSOM, stressing on  the similarities and on the differences respect to the 2-Higgs doublets Model (2HDM) of type II, which is sufficiently general to describe most of the scalar extensions which can be envisioned for LHC applications, and to the potential of the MSSM (see ref. \cite{Barger:1989fj}). 
 
 Naturally, the most problematic feature of the 2HDM is the presence of a
large number of free parameters that affect the possibility of unique and specific predictions, 
due to the different scenarios that may emerge at future experiments in regard to the scalar sector. The MLSOM potential is also affected by the same problem.
In the case of the MSSM instead, the presence of supersymmetry allows some relations between
the masses and the couplings and between the mass of the gauge bosons
and their interaction parameters, which provide further constrains on the allowed parameter space.
In the scalar sector, in this case, there are only two free
parameters, which can be identified with $\tan\beta$ and with the
mass of one of the two Higgs bosons \cite{Gunion:1989we}.
As a result of this, for instance, in the MSSM, some Higgs-to-Higgs
decays (see ref. \cite{Kanemura:2009mk}) which are possible in the MLSOM, are avoided.
Other features of the $CP$-odd sector of the MLSOM are, for instance, the independence of the mass of the 
axi-Higgs from the parameters of the $CP$-even sector and the existence of a sum rule 
relating $H_0$ and $h_0$ with the vector bosons (V), which is also typical of the 2HDM

\ba
\sum_i g_{h^0_i VV}^2=g_{H_{SM} VV}^2.
\ea
\subsection{Axion-like interactions}
As we have discussed above, in the MLSOM the specific feature of the CP-odd sector is the presence of axion-like interactions which are not found in the 2HDM and which are the true novelty of the entire construction. It is important to remark that while in models containing CP-odd scalars effective interactions such as $A_0 \gamma\gamma$ induced by the fermion loops are indeed present, they turn out to be proportional to the mass of the fermion running in the loop. This mass-dependence, obviously, is completely absent in the MLSOM, since the origin of the Wess Zumino terms, which provide these interactions, is related to the restoration of the 
gauge symmetry of the anomalous effective theory and not to a mechanism of symmetry breaking.

In complete analogy to the case of the SM Higgs, the most relevant sector to look for in 
the production of an axion-like particle is the gluon-gluon fusion channel.
It is important to point out that given the presence of free parameters that are involved in the generation
of its mass appearing in the PQ-breaking potential, the axion can be searched for in different kinematical domains
because the model allows both a very light axion with a mass of the order of 
1 GeV or less, and a heavier one. 
As stated before, the particular features of the scalar 
potential render the predictions of the MLSOM 
different respect to the general 2HDM, due to the presence of the $b$ field,
and this of course imposes some differences in the treatment of the experimental constraints
on the allowed parameter space.

\subsection{The parameters}
The free parameters of the scalar potential can be identified by the coefficients
$(\lambda_{uu},\lambda_{dd},\lambda_{ud},\lambda'_{ud})$ that are contained in the
$PQ$ potential and by $(b_1,\lambda_1,\lambda_2,\lambda_3)$, that are contained in
the $PQ$-breaking potential.
The other free parameters are the ratio of the higgs vevs, identified with $\tan\beta$,
the Stueckelberg mass $M_1$ and the coupling constant $g_B$.

We start our analysis by considering a scenario in which the mass
of the $Z$ boson is exactly reproduced at $M_Z=91.1876$ GeV and the
bounds on the mass of the extra $Z'$ are required to be compatible with the current
Tevatron data. These conditions can be obtained by
fixing the value of the anomalous coupling $g_B\approx 0.1$, the value of $v_u\approx 246$ GeV,
the value of the Stueckelberg mass $M_1$ in the TeV range and $\tan\beta=40$.
These requirements induce also a small mixing parameter between $Z$ and $Z'$ (below $10^{-3}$), which is also in agreement with current data.
Thus, the mass of the particles of the scalar sector are identified
by the eight parameters listed above.
The value of the mass of the axi-Higgs is completely governed by the $PQ$-breaking
sector of the potential and one can always find a combination of its parameters  so that the axion is very light.
The other parameters enter in the structure of the mass of the two neutral Higgs
and the eigenvalues are found to be very sensitive to the selection of these parameters. 
In our case, these have been chosen as follows:\\
$\left\{\lambda_1,\lambda_2,\lambda_3,b_1,\lambda_{uu},\lambda_{dd},\lambda_{ud}\right\}
=\left\{-9~10^{-5},-1~10^{-6},-1~10^{-5},5~10^{-3},6~10^{-2},5,0.9\right\}$,
and we have obtained the following values for the masses of the CP-even and the CP-odd
sectors: \\ $\left\{m_{H_0}\approx 122,m_{H_0}\approx 15,m_{\chi}\approx 5\right\}$ (GeV).

\begin{figure}[t]
\begin{center}
\includegraphics[width=6cm]{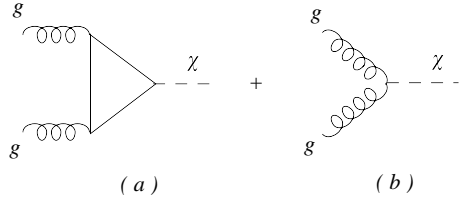}
\caption{\small The two contributions to the $gg \rightarrow \chi$ production channel.
\label{chi_production}}
\end{center}
\end{figure}

\subsection{The invariant mass distribution}
To quantify the cross section of the processes 
that we are considering, we introduce the invariant mass distributions
that must be convoluted with the gluon luminosity in order to obtain predictions at hadron level.  
In general, the total cross section for each process can be 
determined by using the following factorization formula
\ba
\sigma(S,\mu_R^2,\mu_F^2)=\int_{0}^{1}d\xi_1 \int_{0}^{1}d\xi_2 \,
g(\xi_1,\mu_F^2) g(\xi_2,\mu_F^2) \hat{\sigma}(\alpha_s(\mu_R^2),Q^2/\mu_R^2,Q^2/\mu_F^2)
\ea
where $\tau=Q^2/S$, $g(\xi_2,\mu_F^2)$ is the gluon density, function of the Bjorken variable 
$\xi$ and of the factorization scale $\mu_F$.  A similar expression holds for the invariant mass distributions for the production of a pseudoscalar with an invariant mass $Q$, which is given at parton level by 

\ba
\frac{d\hat{\sigma}}{d Q^2} = \sum_{pol,spin}\frac{|{\cal M}|^2}{2 Q^2}d\Phi_n 
\frac{1}{\xi_1 \xi_2 S}\delta\left(1 - \frac{\tau}{\xi_1 \xi_2} \right).
\ea
Here $|{\cal M}|^2$ represents the square of the matrix element for the production of $n$
scalar particles in the final state, the variables $\xi_1,\xi_2$ represent 
the fraction of the momentum carried by the partons in the collision and $d\Phi_n$ is the Lorentz invariant
phase space. The invariant mass $Q^2$ is defined as $\hat{s}+\hat{t}+\hat{u}=Q^2$,  
while the fraction $1/Q^2$ is the partonic flux.
Then we can write  at hadron level
\ba
\frac{d\sigma}{dQ^2}=\frac{\hat{\sigma}(Q^2)}{S}\Phi_{gg}(\tau)
\ea
where 
\ba
\hat{\sigma}(Q^2)=\sum_{pol,spin}\frac{|{\cal M}(\alpha_s(\mu_R^2))|^2}{2 Q^2} d\Phi_n,
\ea
and the gluon luminosity is given by the following convolution product  
\bea
\Phi^{}_{gg}(x,\mu^2_F)= \int_{x}^{1} \frac{dy}{y} g(y, \mu^2_F) g(\frac{x}{y}, \mu^2_F).  
\eea
The computation of this cross section for the production of the axi-Higgs  
$pp \rightarrow gg \rightarrow \chi +X$ via gluon fusion involves two contributions: the fermion loop correction and the direct (contact) decay due to the Wess-Zumino term, as shown in Fig. \ref{chi_production}, with the 
WZ counterterm suppressed as $1/M_1$ and therefore quite subleading respect to the first.

\begin{figure}[t]
\includegraphics[width=6cm, angle=-90]{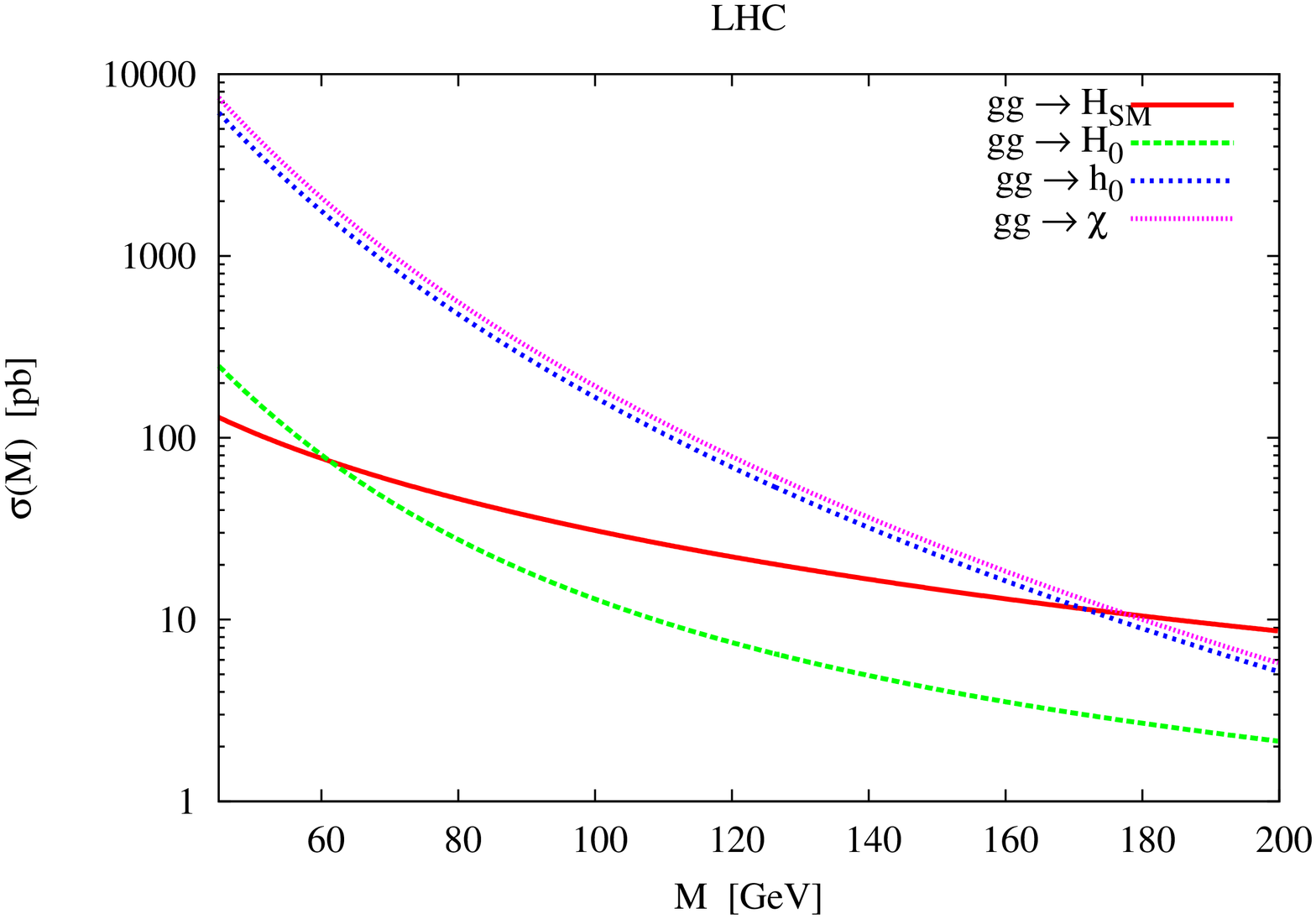}
\includegraphics[width=6cm, angle=-90]{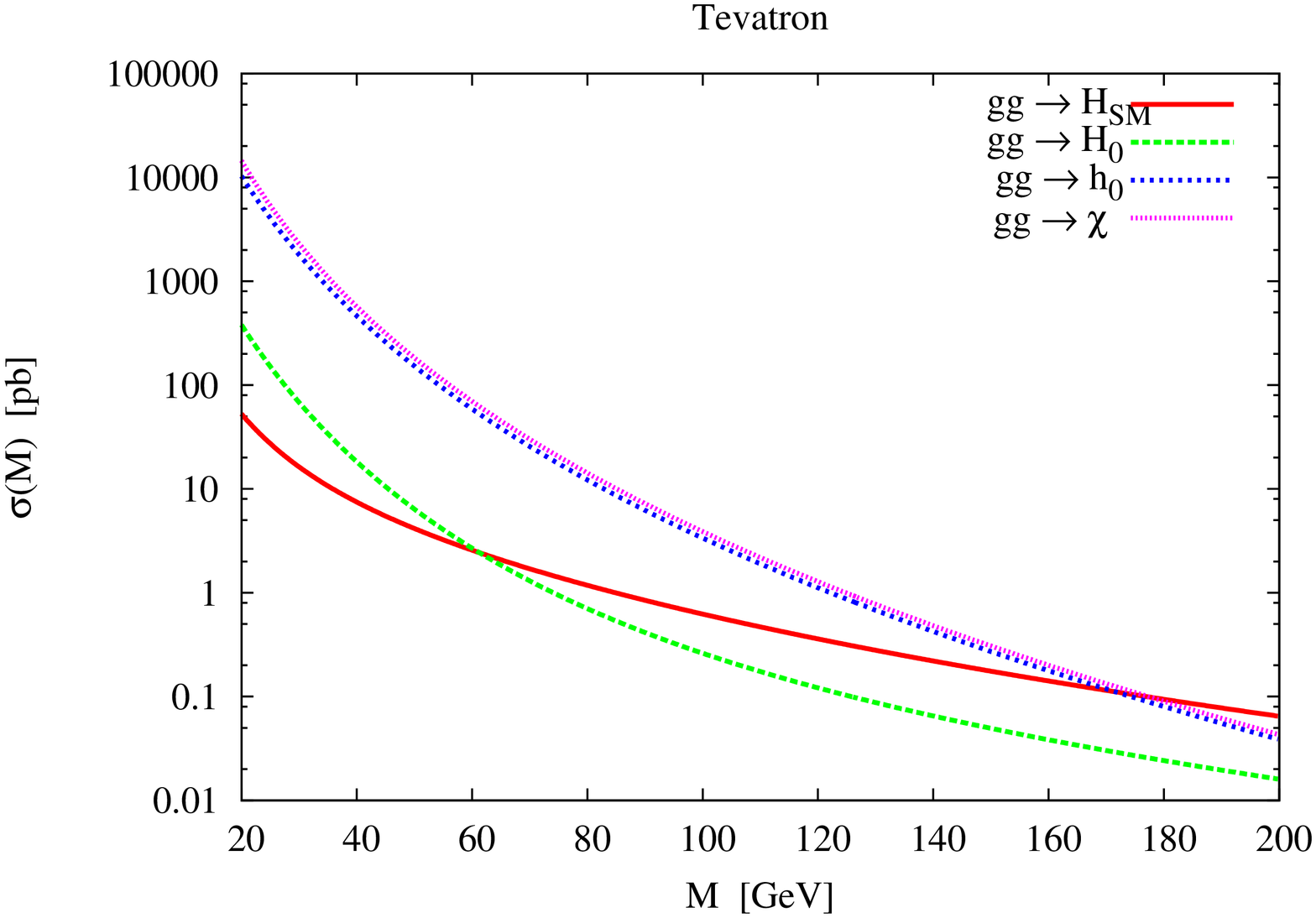}
\caption{\small  Cross section for the production of the two Higgs $h_0$ and $H_0$ and the axi-Higgs via gluon-gluon fusion at LO at the LHC (left panel) and at the Tevatron (right panel).}
\label{compares}
\end{figure}

At parton level the production cross section
for the axi-Higgs via gluon fusion is related to the 
decay rate by the following relation
\bea
\sigma_{g g \rightarrow \chi}(\hat s) = \frac{8 \pi^2}{m_\chi N_c^2} \Gamma(\chi \rightarrow gg) \delta(\hat s - m^2_\chi)
=\sigma^0_{g g \rightarrow \chi} \delta(\hat s - m^2_\chi)
\eea
where $\hat s$ is the squared partonic c.m. energy and $N_c=8$ 
is the color factor for the gluons. 
At hadron level the total cross section for the inclusive 
axi-Higgs production is given by
\bea
\sigma(pp \rightarrow gg \rightarrow \chi + X) = \int_{m_\chi^2/S}^{1} d \tau \,\Phi^{}_{gg}(\tau) \sigma_{g g \rightarrow \chi}(\tau S) \,
= \frac{1}{S}  \sigma^0_{g g \rightarrow \chi}  \Phi^{}_{gg}(\tau) |_{\tau=m^2_\chi/S}  \qquad \tau =\frac{Q^2}{S}\nonumber \\
\eea 
where the variables $S$ and $\sqrt {Q^2}$ stand for the squared c.m. energy of the incoming hadrons and the invariant mass of the gluon pair, respectively.
In Figs. \ref{compares} we show the plots of the total
cross section at LO at the LHC and at the Tevatron respectively, for the production
of the axion and of each of the CP-even $H_0,h_0$ Higgs, and the corresponding plots for the SM Higgs. Notice that the result shows a ratehr sharp rise of the production cross section with a decrease of the axion mass, larger by a factor of 10 compared to the case of other CP-even scalars. A similar rise is found also for the CP-odd sector of the 2HDM, being typical of the CP-odd sector.

\section{Axion plus photon production}

\begin{figure}[t]
\begin{center}
\includegraphics[width=9cm]{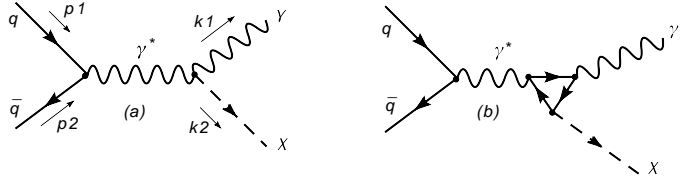}
\caption{\small Production channel for a single axion plus a photon}
\label{X_prod}
\end{center}
\end{figure}

In this section we compute the production of an axion
plus one photon at the LHC in leading order (LO), given by 
the diagrams in Fig.\ref{X_prod}. The computation of the amplitude requires the three-point correlator between two photon and 
one axion, with one off-shell photon and with $m_f\neq 0$. This can be achieved by using the 
parametrization of the trilinear vertex with two off-shell external legs and away from the chiral limit
(see \cite{Armillis:2009sm}).   

Denoting by $T^{\lambda\mu\nu}$ the correlator with outgoing momenta $k_1^{\mu},k_2^{\nu}$ 
and incoming momentum $k^{\lambda}$, the generalized WI gives the following relation
\ba
k_{\lambda} T^{\lambda\mu\nu}= 2 m_f T^{\mu\nu} + a_n\varepsilon\left[k_1,k_2,\mu,\nu\right]
\ea
where the tensor $T^{\mu\nu}$ is defined by
\ba
T^{\lambda\mu\nu}=-\frac{i\bar{C}_0(k^2,k_2^2,m_f^2) m_f^2}{\pi^4}\varepsilon\left[k_1,k_2,\mu,\nu\right].
\ea
Performing the change of the momenta $k_1\rightarrow k_1, k_2\rightarrow -q,k\rightarrow k_2$, 
we obtain the expression for the three-point correlator between two photons (one off-shell) 
and one pseudoscalar, suitable for our calculation.
The function $\bar{C}_0$ has the following expression
\ba
\bar{C}_0(s,m_{\chi}^2,m_f^2)=\frac{i\pi^2}{2(s^2 - m_{\chi}^2)}\left[\log^2\left(\frac{a_2+1}{a_2-1}\right)
-\log^2\left(\frac{a_3+1}{a_3-1}\right) \right],
\ea
where we have defined 
\ba
a_2=\sqrt{1-\frac{4m_f^2}{s}} && a_3=\sqrt{1-\frac{4 m_f^2}{m_{\chi}^2}}. 
\ea
We can identify four kinematic regions in which the function $\bar{C}_0$ can be analytically continued:
\begin{itemize}
\item Region I  $q^2>4m_f^2$, $m_{\chi}^2>4m_f^2$,
where $a_2-1<0$ and $a_3-1<0$ and 
\ba
\bar{C}_0(s,m_{\chi}^2,m_f^2)=\frac{i\pi^2}{2(s^2 - m_{\chi}^2)}
\left\{\left[\log\left(\frac{a_2+1}{1-a_2}\right) +i\pi\right]^2
-\left[\log\left(\frac{a_3+1}{a_3-1}\right) +i\pi\right]^2\right\}
\ea
\item Region II  $q^2<4m_f^2$, $m_{\chi}^2<4m_f^2 $
where $a_2\rightarrow i\sqrt{-a_2^2}$ and $a_3\rightarrow i\sqrt{-a_3^2}$
\ba
\bar{C}_0(s,m_{\chi}^2,m_f^2)=\frac{i\pi^2}{2(s^2 - m_{\chi}^2)}
\left\{\left[-2i\arctan\left(\frac{1}{\sqrt{-a_2^2}}\right)\right]^2 - \left[-2i\arctan\left(\frac{1}{\sqrt{-a_3^2}}\right)\right]^2
\right\}
\ea
\item Region III  $q^2>4m_f^2$, $m_{\chi}^2<4m_f^2 $
where $a_2-1<0$ and $a_3\rightarrow i\sqrt{-a_3^2}$
\ba
\bar{C}_0(s,m_{\chi}^2,m_f^2)=\frac{i\pi^2}{2(s^2 - m_{\chi}^2)}
\left\{\left[\log\left(\frac{a_2+1}{1-a_2}\right) +i\pi\right]^2 - \left[-2i\arctan\left(\frac{1}{\sqrt{-a_3^2}}\right)\right]^2
\right\}
\ea
\item Region IV  $q^2<4m_f^2$, $m_{\chi}^2>4m_f^2 $
where $a_2-1<0$ and $a_3\rightarrow i\sqrt{-a_3^2}$
\ba
\bar{C}_0(s,m_{\chi}^2,m_f^2)=\frac{i\pi^2}{2(s^2 - m_{\chi}^2)}
\left\{\left[-2i\arctan\left(\frac{1}{\sqrt{-a_2^2}}\right)\right]^2 
- \left[\log\left(\frac{a_3+1}{a_3-1}\right) +i\pi\right]^2\right\}.
\ea
\end{itemize}

The squared and averaged partonic contributions are given by
\ba
&&\sum_{spin}\left|{\cal M}_1\right|^2=\frac{1}{36}\left[\frac{t^2 +u^2}{s}\right]
Q_f^2 e^2  \left|\sum_{f'}Q_{f'}^2 e^2 \frac{O^{\chi}_{f'}}{v_{f'}}\bar{C}_{0}(s,m_{\chi^2},m_{f'}^2) \frac{m_{f'}^2}{\pi^4}\right|^2
\nonumber\\
&&\sum_{spin}\left|{\cal M}_{WZ}\right|^2=\frac{1}{36}\left[\frac{t^2 +u^2}{s}\right]
Q_f^2 e^2 (g_{\gamma\gamma}^{\chi})^2
\nonumber\\
&&\sum_{spin} 2 \textrm{Re}\left[{\cal M}_1 {\cal M}_{WZ}^*\right]=\frac{1}{36}\left[\frac{t^2 +u^2}{s}\right]
Q_f e 2 g_{\gamma\gamma}^{\chi} \sum_{f'}Q_{f'}^2 e^2\textrm{Re}\left[\bar{C}_0(s,m_{\chi^2},m_{f'}^2)\right]\frac{O^{\chi}_{f'}}{v_{f'}}\frac{m_{f'}^2}{\pi^4}
\ea
where $O^{\chi}_{f'}/v_{f'}$ is $O^{\chi}_{11}/v_u$ for an up type quark, while $O^{\chi}_{21}/v_d$ for a
down type quark.

Integrating over the two-particle phase space we obtain
\ba
&&\hat{\sigma}_1(s,m_{\chi}^2)=\frac{1}{48 \pi N_c^2}\frac{1}{2 s}\frac{(s-m_{\chi}^2)^2(s+m_{\chi}^2)}{s^2}
Q_f^2 e^2 \left|\sum_{f'}Q_{f'}^2 e^2 \frac{O^{\chi}_{f'}}{v_{f'}}\bar{C}_{0}(s,m_{\chi^2},m_{f'}^2) \frac{m_{f'}^2}{\pi^4}\right|^2,
\nonumber\\
&&\hat{\sigma}_{WZ}(s,m_{\chi}^2)= \frac{1}{48 \pi N_c^2}\frac{1}{2 s}\frac{(s-m_{\chi}^2)^2(s+m_{\chi}^2)}{s^2}
Q_f^2 e^2 (g_{\gamma\gamma}^{\chi})^2,
\nonumber\\
&&\hat{\sigma}_{int}(s,m_{\chi}^2)=\frac{1}{48 \pi N_c^2}\frac{1}{2 s}\frac{(s-m_{\chi}^2)^2(s+m_{\chi}^2)}{s^2}
Q_f e 2 g_{\gamma\gamma}^{\chi} \sum_{f'}Q_{f'}^2 e^2\textrm{Re}\left[\bar{C}_0(s,,m_{\chi^2},m_{f'}^2)\right]\frac{O^{\chi}_{f'}}{v_{f'}}\frac{m_{f'}^2}{\pi^4},
\nonumber\\
\ea
where $\hat{\sigma}_{int}$ denotes the interference term.
Introducing the invariant mass distribution at hadron level, we have
\ba
\frac{d\sigma}{dQ^2}=\frac{\hat{\sigma}(Q^2,m_{\chi}^2)}{S}\Phi_{q\bar{q}}(\tau)
\ea
where the parton luminosity $\Phi_{q\bar{q}}$ has been previously defined and $Q$ represents the invariant mass of the final state.

\begin{figure}[t]
\begin{center}
\includegraphics[width=6cm,angle=-90]{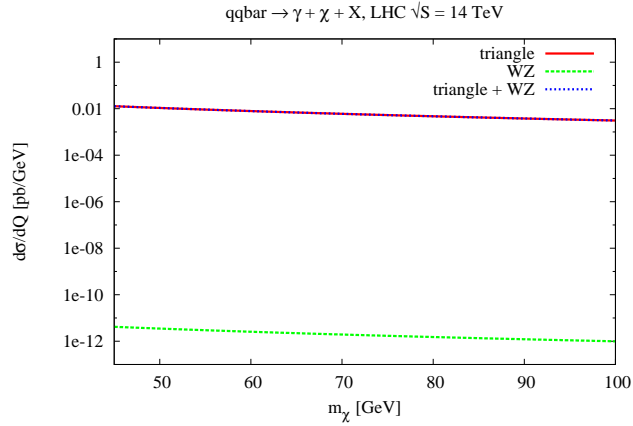}
\caption{\small Invariant mass distribution for the associated production of an
axion plus one photon at the LHC.
\label{crossXph}}
\end{center}
\end{figure}

We show in Fig.~\ref{crossXph} a plot of the cross section for the production of an axion and one photon at the LHC as a function of the mass of the $\chi$. The mass dependence of the result is quite small, except for a 
larger mass of the particle, in a region where it is Higgs-like. For an ultralight axion the value of the cross section is around $10^{-2}$ pb. We have shown the contribution from the triangle and the Wess-Zumino terms combined and separately, in order to show the dominance of one channel respect to the other. The Wess-Zumino term is indeed strongly suppressed (by a factor of $10^{10}$).

\section{Multi axion production}
One of the peculiarities of a light axion-like particle is its possibility to generate cascade decays with multi-lepton final states which are more sizeable especially for a mass of $\chi$ in the GeV range. We have indeed seen that for 
$m_\chi$ around few GeV's, the largest contribution to the branching ratio of its decay is 
predominantly into leptons, and for this reason we are going to investigate systematically this particular interval in parameter sace.  Our analysis will include two types of vertices, the trilinear $\chi\chi H_0,h_0$ vertex and the $\chi^4$ vertex. As we are going to see, multilepton decays will be sizeable even in the presence of a considerable phase space suppression and we will quantify them rather accurately. 

We consider both the production of axions in combination with a scalar of the CP-even sector of the MLSOM, and final states made entirely of several light axions which branch primarily into leptons. We consider the gluon fusion channel, in which the production of the CP-even scalars ($h_0,H_0$) is mediated by the top and bottom loops. 
The sizeable values of the multi-axion cross sections for the invariant mass distributions are related to the large production cross sections which are typical of pseudoscalar channels and to the large values
of the reduced couplings - normalized to the SM ones - of the trilinear interactions
of the scalars. The leading contribution to the production cross section comes from the fermion loop graph 
with a final state axion. In the model, each contribution is accompanied by the corresponding WZ counterterm, which is suppressed by a factor of $10^{5}$ compared to the loop graph (see Fig. \ref{chi_production}). 
 \begin{figure}[t]
\begin{center}
\includegraphics[width=9cm]{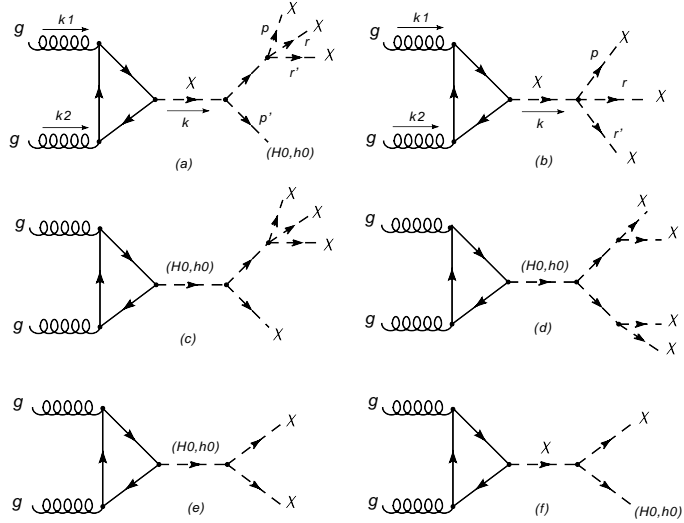}
\caption{\small Channels for multi axion production from gluon gluon fusion. }
\label{sixmu}
\end{center}
\end{figure}

Channels involving several final state axions can be built rather easily.  list of several diagrams contributing to these channels is given in Fig. \ref{sixmu}. For instance, the simplest process involves a 
$gg-h_0$ production channel combined with the $h_0-\chi\chi$ vertex. In this case the WZ counterterm is absent. 
A similar process is the $gg-\chi$ triangle vertex, followed by the $\chi\chi h_0$ vertex, which gives the combination of a $\chi$ and of a CP-even Higgs $(h_0)$ in the final state. In this case the channel is accompanied by a WZ term $gg-\chi$ describing the direct interaction of the two gluons of the initial state with the axion. Cascading channels can be easily obtained by combining trilinear 
($\chi\chi h_0$) and quadrilinear $(\chi^4)$ vertices, which are more involved and that we will study below. 

\subsection{ $H_0, h_0- 3 \chi$ decay}   
The amplitude for the on-shell production of three $\chi$ and one scalar Higgs
- through the process $gg\rightarrow h^0\chi\chi\chi$ -  is given by the
sum of a part containing the fermion triangle plus the counterterm
\ba
{\cal M}={\cal M}_{loop} + {\cal M}_{count}\, .
\ea
Defining $s=q^2=(k_1+k_2)^2$ we can write the square of the matrix element as
\ba
&&\sum_{spin, pol.}\vert{\cal M}_{h,3\chi}\vert^2=\left\{\frac{(4\pi\alpha_s)^2}{(N_c^2-1)^2}\,
\sum_{f}(c^{f\bar{f}}_{\chi})^2 \, C_{\chi^2 higgs}^2 \, C_{\chi^4}^2  \frac{4 m_f^2}{\pi^4 q^4}N_c^2\,|f(\tau_f)|^2 
+\frac{(g^{\chi}_{GG})^2\, C_{\chi^2 higgs}^2 \, C_{\chi^4}^2}{8(N_c^2-1)^2}
\right.\nonumber\\
&&\hspace{3cm}\left.
+\frac{4\pi\alpha_s \,\textrm{Re}[N_c \,g^{\chi}_{GG} f(\tau_f)]}{(N_c^2-1)^2} 
\frac{m_f c^{f\bar{f}}_{\chi} \, C_{\chi^2 higgs} \, C_{\chi^4}}
{\pi^2 q^2} 
\right\}\frac{1}{(q^2-m_{\chi}^2)^2(1 - x_2 +\frac{\rho_1 -\rho_2}{4})^2},
\nonumber\\
\ea
where $x_2= p'\cdot q/q^2$.
The coefficient $\tau_f$ is defined as $4m_f^2/q^2$, while $\rho_1= 4m_h^2/q^2$, and $\rho_2=4 m_{\chi}^2/q^2$. The coefficient $g_{\chi GG}$ of the counterterm is defined in the
previous sections and the couplings of the axion to the up-type and down-type quarks
are given by
\beq
c^{u\bar{u}}_{\chi}=\frac{m_u}{\sqrt{2}v_u}O^{\chi}_{11}=-\frac{m_u v_d}{\sqrt{2}v}
\hspace{1cm}
c^{d\bar{d}}_{\chi}=\frac{m_d}{\sqrt{2}v_d}O^{\chi}_{21}=\frac{m_d v_u}{\sqrt{2}v}.
\eeq
The details of the computation
can be found in an appendix.

\subsection{$4-\chi$ decay}
We move to discuss the possibility of producing four axions in the final state
mediated by a CP-even higgs $(H_0,h_0)$. 
At parton level, the squared amplitude for the process $gg\rightarrow H \rightarrow 4\chi$ 
is given by
\ba
\sum_{spin, pol.}\vert{\cal M}_{4\chi}\vert^2=
\frac{(4\pi\alpha_s)^2}{(N_c^2-1)^2}\,
\sum_{f}\frac{(c^{f\bar{f}}_{H})^2 \, C_{\chi^2 higgs}^2 \, C_{\chi^4}^2}{(q^2-m_{H}^2)^2(1 - x_2)^2}  \frac{4 m_f^2}{\pi^4 q^4} N_c^2 \, |1+(1-\tau_f)f(\tau_f)|^2
\ea
where $H=H_0,h_0$ and the couplings of the higgs to the quarks are given by
\ba
c^{u\bar{u}}_{H}= \frac{m_u}{v} R_{12} \hspace{0.5cm} c^{u\bar{u}}_{H}= \frac{m_u}{v} R_{22}
\nonumber\\
R_{12}=-\cos{\alpha} \hspace{0.5cm} R_{22}=\sin{\alpha}.
\ea
The coefficients $R_{12},R_{22}$ are the matrix elements for passing from the
interaction eigenstate basis to the physical basis, already defined
in the previous sections.

\begin{figure}[t]
\includegraphics[width=6cm, angle=-90]{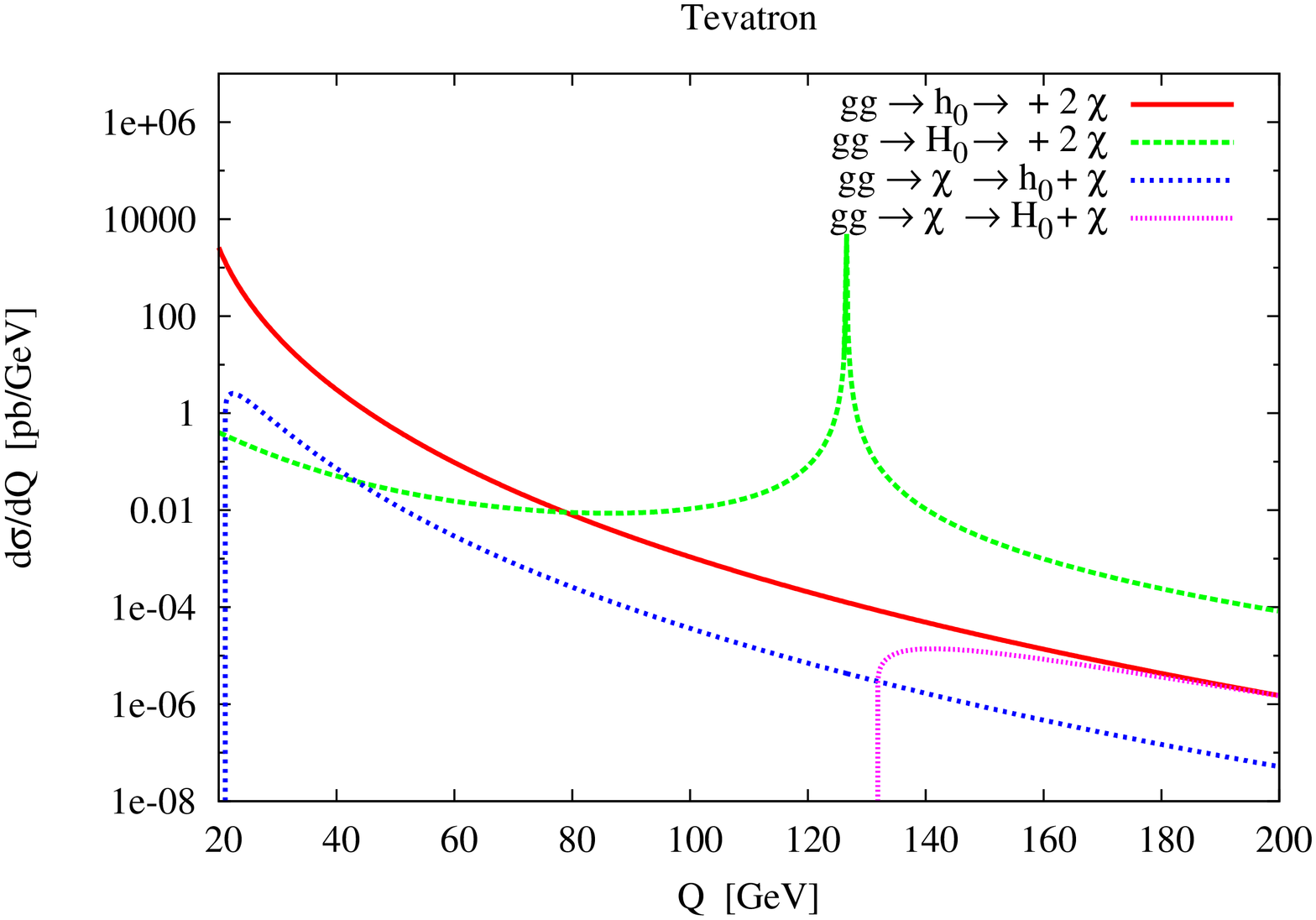}
\includegraphics[width=6cm, angle=-90]{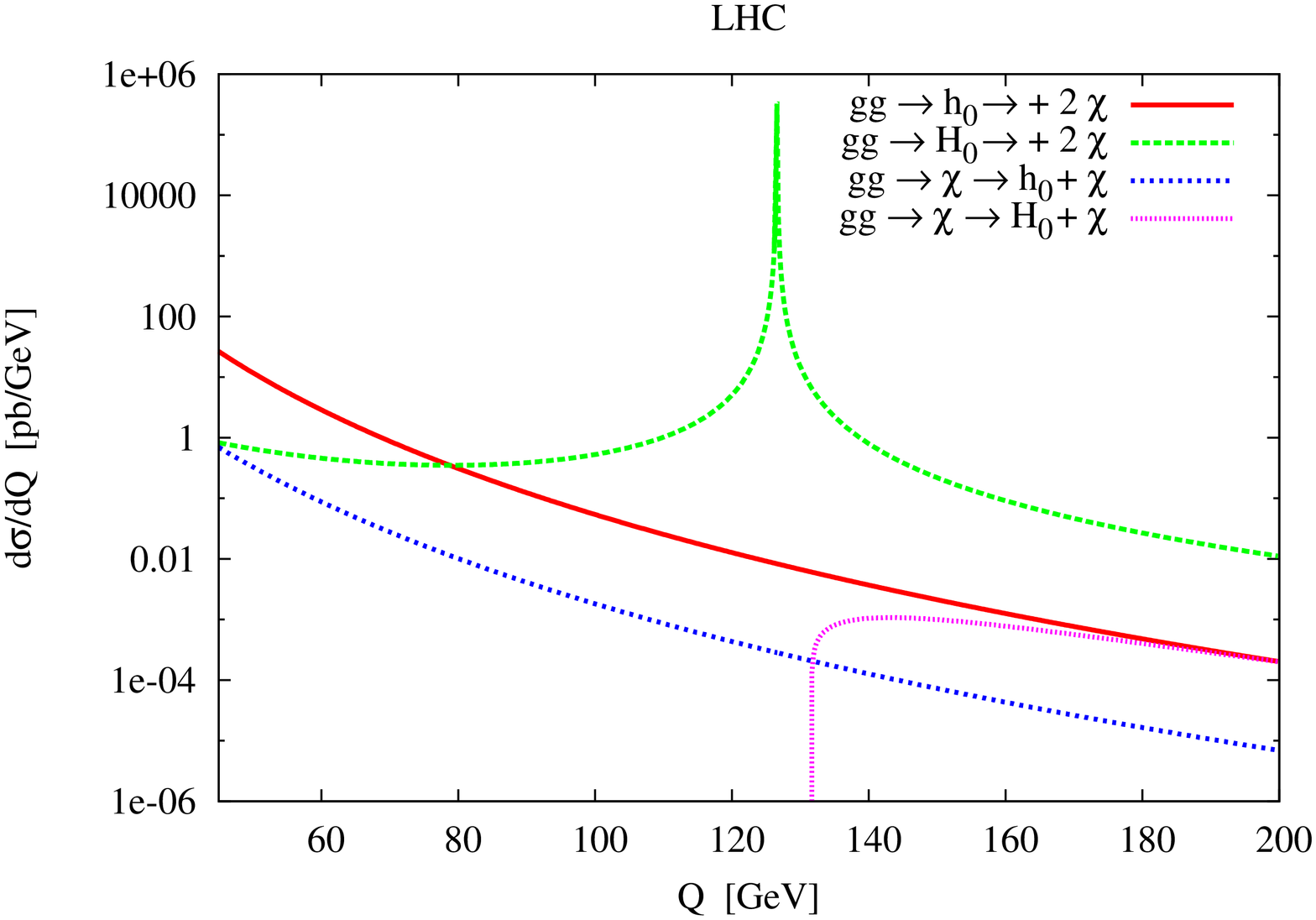}
\caption{\small $gg\rightarrow$ 2-scalar reactions mediated by trilinear vertices.}
\label{2scalars}
\end{figure}

\begin{figure}[t]
\includegraphics[width=6cm, angle=-90]{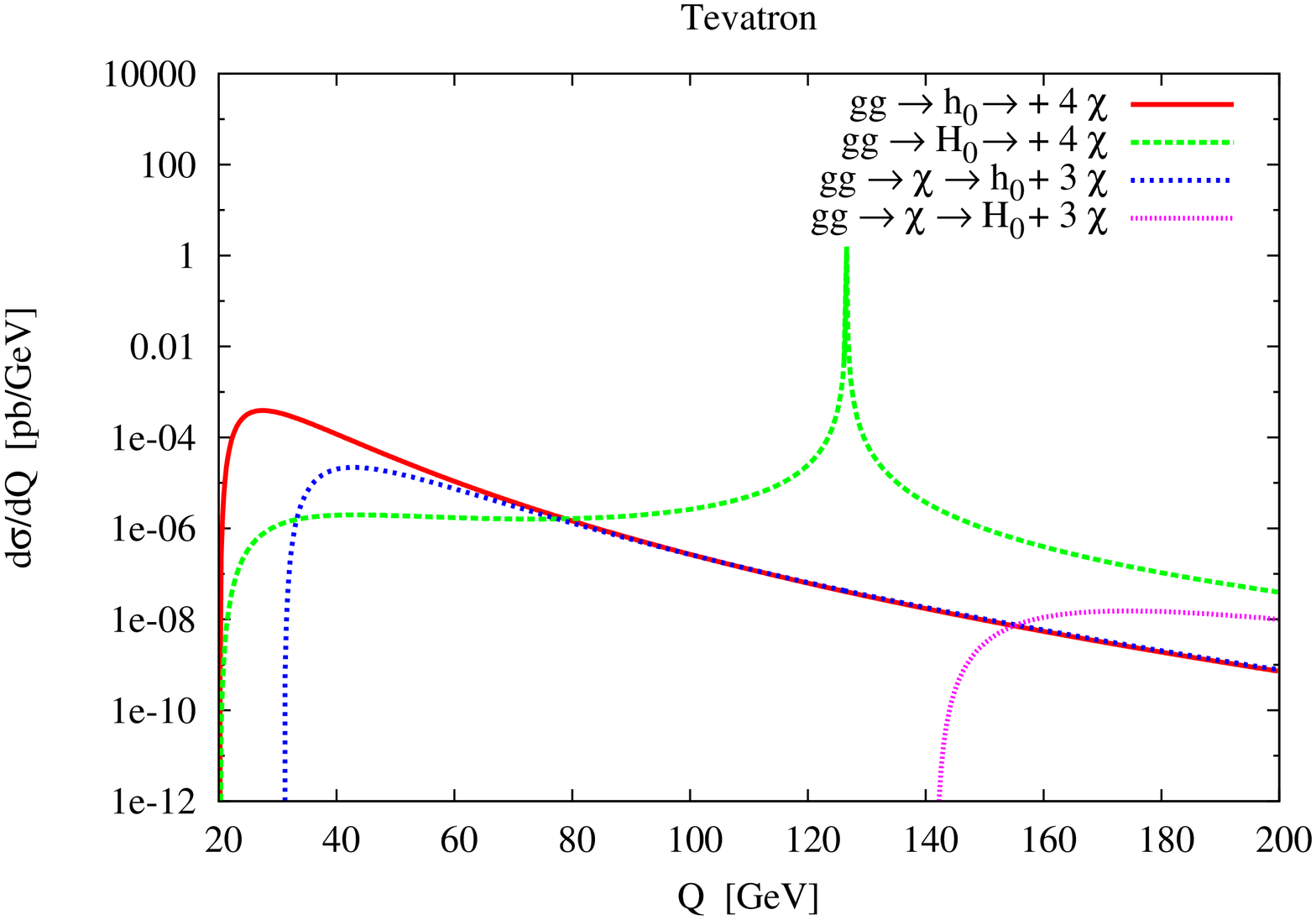}
\includegraphics[width=6cm, angle=-90]{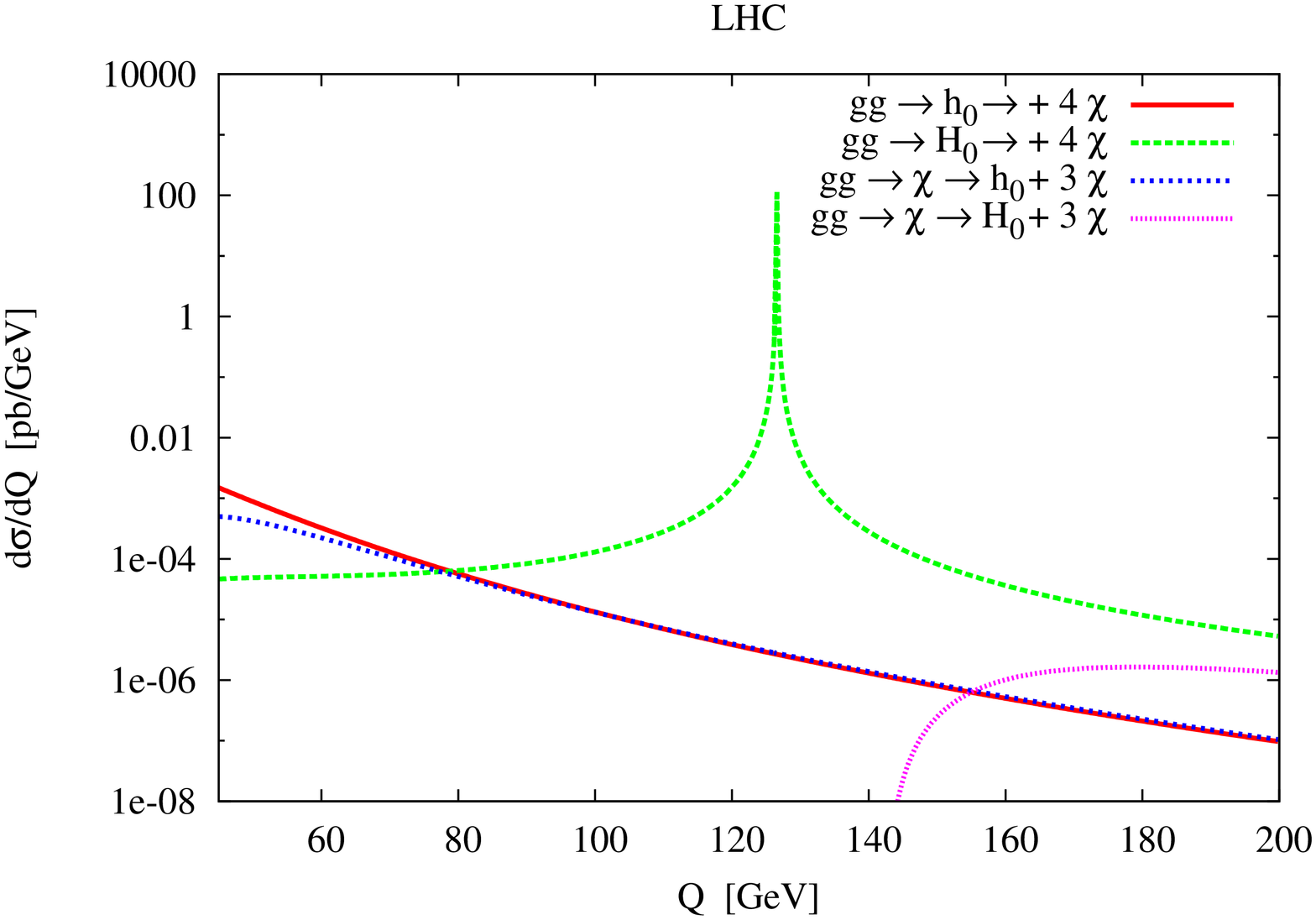}
\caption{\small Production of 3 and 4 scalars from gluon gluon fusion mediated by trilinear and quadrilinear vertices.}
\label{4scalar}
\end{figure}

The plots for the production of four scalar particles via gluon-gluon fusion are shown 
in Fig. \ref{4scalar}. Notice that the production of four axions and 
that of three axions and one $h_0$ show invariant mass distribution which are rather similar in their sizes. 
This is due to the fact that in this study we have chosen $h_0$ to be not too much heavier  
than $\chi$ ($m_{h_0}\approx 15$ GeV). Details on the computation of the 4-particle phase space can be found in an appendix. We have performed a direct computations of the phase space integrals, which have been reduced into a 2-dimensional form and then have been integrated numerically.  
The results of this study are shown in Fig. \ref{2scalars} for the Tevatron and the LHC respectively.
 The plots presented in the two figures show sizeable rates which become large on the Higgs ($H_0$) resonance, chosen to be at 120 GeV. At the LHC the peak value of the cross section for $pp\to \chi\chi$, mediated by the $H_0$ is larger by a factor of about $10-100$ compared to the Tevatron and would be significant. In the same figures, the same production channel, mediated by the $h_0$, is also resonant at 15 GeV, but is not shown in our study since it involves an extrapolation of the parton distributions towards the small-x region, which we have not included in our analysis. 

 Coming to the 4-axions final state, the numerical values of the various distributions are shown in Fig. \ref{4scalar}, where they appear to be down by a factor of approximately $10^4$ compared to the analogous ones with 2 $\chi$'s or with one $\chi$ and one CP-even Higgs in the final state. 
We have summarized in Tab.(\ref{Processes}) the numerical value of the cross sections at a representative 
value of $Q$ at which they appear to be sizeable, within the parametric choices used in our analysis. The largest values shown are those on the resonances of the two neutral Higgs. The multilepton channels, for a GeV axion, appear to be rather small even on the largest production resonance, which is on the peak of the $H_0$, due to a large phase space suppression. Typical resonant rates are $10^{-5}$ pb/GeV for 4 muons and $10^{-16}$ pb/GeV for the production of 8 muons. For final states with 8 muons mediated by the $h_0$ in the non-resonant region and coming from the pairwise decays of 4 axions, the rates are much smaller ($\sim 10^{-20}$ GeV pb/GeV).

\begin{table}[h]
\begin{center}
\begin{tabular}{|c|c|c|c|c|}
\hline
Process & $Q$& $d\sigma/dQ$ (LHC)& $Q$ & $d\sigma/dQ$ (T) \\
\hline
$gg\rightarrow h_0\rightarrow 4\chi$ & $45$ & $\approx 10^{-3}$  & $22$ & $ 4\cdot 10^{-4}$  \\
\hline
$gg\rightarrow H_0\rightarrow 4\chi$ & $M_{H_0}$ & $ 103$  & $M_{H_0}$ & $ 1.56$ \\
\hline
$gg\rightarrow \chi\rightarrow 3\chi + h_0$ & $50$ & $ 5\cdot 10^{-4}$ & $40$ & $ 2\cdot 10^{-5}$ \\
\hline
$gg\rightarrow \chi\rightarrow 3\chi + H_0$ & $150$ & $ 2\cdot 10^{-7}$  & $150$ & $ \approx 10^{-8}$ \\
\hline
$gg\rightarrow h_0 \rightarrow 2\chi$ & $45$ & $ 26$  & $20$ & $ 2.5 \cdot 10^3$ \\
\hline
$gg\rightarrow H_0 \rightarrow 2\chi$ & $M_{H_0}$ & $ 324\cdot 10^{3}$  & $20$ & $ 4.9\cdot 10^3$ \\
\hline
$gg\rightarrow \chi \rightarrow h_0 + \chi$ & $45$ & $0.69$  & $20$ & $ 2.5\cdot 10^3$ \\
\hline
$gg\rightarrow \chi \rightarrow H_0 + \chi$ & $150$ & $\approx 10^{-3}$  & $150$ & $\approx 10^{-5}$ \\
\hline
$gg\rightarrow H_0 \rightarrow h_0 +h_0 \rightarrow 4 \chi$ & $M_{H_0}$ & $5\cdot 10^3$  & $150$ & $82$ \\
\hline
\end{tabular}
\end{center}
\caption{\small A list of processes analyzed at hadron colliders at the LHC and at the Tevatron (T). $Q$ is in GeV 
and $d\sigma/dQ$ in pb/GeV. 
\label{Processes}}
\end{table}

\begin{figure}[t]
\begin{center}
\includegraphics[width=7cm, angle=-90]{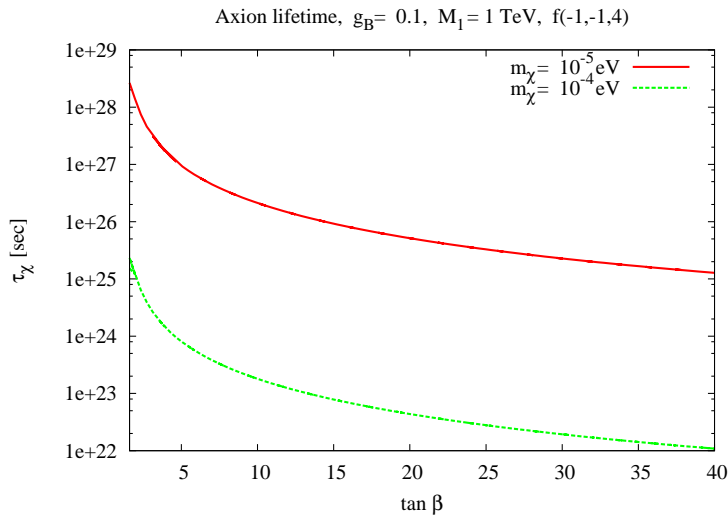}
\caption{\small Lifetime for an ultralight axion as a function of $\tan\beta$}
\label{multimuon}
\end{center}
\end{figure}

\section{The light mass region of the axion and its lifetime} 
One obvious question to ask is whether the axi-Higgs, which takes the role of a valid example of a gauged axion, has any chance of being a dark matter candidate, with properties which remain quite distinct from those of the axion of the PQ model.

As we have already remarked in the introduction, axion-like particles originate from the gauging of anomalous symmetries, and take the role of phases in the scalar potential, which is characterized by a small curvature in these variables. We have seen that for a particle mass in the GeV range the branching ratios for its decay into leptons appear to be too large for the particle to be long lived. We can pause for a moment and try to understand the origin of this result.  

The axion interaction with the fermions is generated by the ordinary Yukawa couplings, being the particle part of the scalar sector of the model. In particular, the CP-odd contributions are re-expressed in the physical basis by the elements of the rotation matrix $O_\chi$ together with other parameters, the most significant of them being $\beta$, as shown by Eq. (\ref{chif}). Notice that the matrix elements of this matrix are $O(1)$, which means that we can't expect a large suppression of its coupling to the fermions just from  its mixing with the other CP-odd components of the Higgs sector.

If we look more closely into the two contributions which appear in the decay of an axion, the triangle diagram and the WZ term, one finds that the contribution from the triangle is $O(m_f/ v)$, where $v $ is the vev which represents the symmetry breaking scale of the symmetry to which the axion is associated as a phase of a complex scalar.  Consider, for instance, the mechanism of chiral decoupling, that we have described in the previous sections. In this case, the only interaction of the axion with the gauge fields takes place through the WZ terms, since there are no Yukawa couplings between the light fermions and the pseudoscalar. 
Then, if we assume that the decoupling scale $M_D=g_B M_S$ is around $10^{10}$ GeV, which is the decoupling scale of a right handed neutrino in a typical leptogenesis scenario, the decay rate is simply given by the relation

\ba
\Gamma_{\chi}=\frac{m_{\chi}^3}{4\pi}\left[(g^{\chi}_{\g\g})^2 + 2 (g^{\chi}_{gg})^2\right]
\ea  
where $g^{\chi}_{\g\g}$ and $g^{\chi}_{gg}$ are proportional to $M_D^{-1}$
\ba
g^{\chi}_{\g\g}\propto \frac{g_B}{M_D^{-1}}
\ea
and is dominated by the 2-photons and 2-gluons channels. For a very weakly coupled axion, with a small value of 
the coupling constant ($g_B\approx 10^{-5}$), we have indeed a 
long lived particle of around $1$ GeV with a rather long lifetime
\ba
\tau_{\chi}=\frac{1}{\Gamma_{\chi}}\approx 10^{26}\,\, \textrm{s}.
\ea

In the MLSOM instead, the suppression comes from the St\"uckelberg mass $M_1$ while 
the Yukawa couplings remain unsuppressed. Therefore, in this model, the structure of the axion-fermion-fermion interaction
is proportional to $m_f/v\, \times O^{\chi}$, where $v$ is of the order of the electroweak scale and $O^{\chi}$
is of order $1$ if $M_1$ is in the TeV region.
In these conditions, the MLSOM allows a long lived axion only if this is very light,
with a mass $m_{\chi}\approx 10^{-5}$ eV, which is again, specific of this construction.

 We show in Fig. \ref{multimuon} plots of the lifetime of a very light axion ($10^{-4}-10^{-5}$ eV) of the MLSOM as a function of $\tan\beta$, which shows that in  both cases the particle is very long lived, with features which resemble quite closely those of the traditional Peccei-Quinn axion.

\section{Conclusions}

In this work we have an analyzed the phenomenology of the physical axion 
that emerges in several extensions of the Standard Model and which include 
an anomalous $U(1)$ gauge symmetry. We have focused our study on a mass 
window characterized by an axion of a light-to-intermediate mass, which is probably 
easier to detect at colliders, although windows for a particle of even lower mass can be analyzed in a similar fashion.  One of the most appealing features of the class of models that we have presented consists in the possibility to justify in a natural way a particle in the CP-odd sector of such a small mass, which would be more difficult to motivate at theoretical level in other constructions. We have shown that the origins of the class of effective actions that are characterized by the presence of such a state could be quite different. For instance, in the case of brane models, the small mass of the axions is parameterized by extra terms in the potential which are identified by the symmetry of the low energy model and in which the axion appears as a complex phase. These terms may induce a small tilting on the scalar potential, giving a small mass to the physical axion, extracted after electroweak symmetry breaking. A similar tilting is induced by the instanton vacuum in the case of the Peccei-Quinn axion, and as such, it is possible, given the strong analogy between our case and the PQ case, to borrow most of the results - well known in the case of the invisible axion model - and extend them to this more general model. A very light axion would be, with no doubt, a good candidate for dark matter. 

We have also shown, although in a simplified model, that effective actions  which resemble quite closely the MLSOM, can be obtained by a completely different approach, using the decoupling of a chiral fermion - due to a large vev of a Higgs to which this fermion is coupled - from the effective theory. The charge assignments of generalizations of the MLSOM can be obtained by this approach. In this second case  our analysis has to be considered rather preliminar and needs further extensions, although we expect that most of the features of the special form of chiral 
decoupling  that we have proposed can be worked out more closely in the context of a Grand Unified Theory. 
The generalization of this analysis to the supersymmetric case appears to be rather interesting as are the cosmological implications of the presence of a gauged axion (with or without supersymmetry) in the 
low energy spectra of these theories which deserve further studies.

\centerline{\bf Acknowledgements}
We thank  Simone Morelli for contributing to this analysis and to Roberta Armillis, Antonio Mariano and Nikos Irges for discussions. This work was supported (in part) by the EU grants INTERREG IIIA (Greece-Cyprus) and by the European Union through the Marie Curie Research and Training Network ``Universenet'' (MRTN-CT-2006-035863). 
\section{Appendix A: The Lagrangean}

The classical lagrangean of the model is explicitly given by

\begin{eqnarray}
{\cal L}_0\; =&-&\frac{1}{2} Tr\;[ F^{G}_{\mu\nu}F^{G\mu\nu} ] - \frac{1}{2} Tr[ \; F^{W}_{\mu\nu} F^{W\mu\nu}]
-\frac{1}{4} F^{B}_{\mu\nu} F^{B\mu\nu} -\frac{1}{4} F^{Y}_{\mu\nu} F^{Y\mu\nu}   \nonumber \\
&+&| ( \partial_{\mu} + i g^{}_{2} \frac{ \tau^j }{ 2 } W_{\mu}^j
+i g^{}_{Y} q^{Y}_{u} A_{\mu}^{Y} +i g^{}_{B} \frac{q^{B}_{u}}{2} B_{\mu} ) H_u|^2    \nonumber\\
&+& | ( \partial_{\mu} + i g^{}_{2} \frac{  \tau^j }{ 2 }  W_{\mu}^j
+i g^{}_{Y} q^{Y}_{d} A_{\mu}^{Y} +i g^{}_{B} \frac{q^{B}_{d}}{2} B_{\mu} ) H_d|^2
\nonumber \\
&+&\overline{Q}_{Li} \, i \gamma^{\mu} \left( \partial^{}_{\mu} 
+i g^{}_{3} \frac{\lambda^{a}}{2} G^{a}_{\mu}+ i g^{}_{2} \frac{\tau^{j}}{2} W^{j}_{\mu} 
+ i g^{}_{Y} q^{(Q_L)}_{Y} A^{Y}_{\mu} + i g^{}_{B} q^{(Q_L)}_{B} B_{\mu} \right) Q_{Li} \nonumber\\
&+& \overline{u}_{Ri}  \, i \gamma^{\mu}  \left( \partial_{\mu} + i g^{}_{Y} q^{(u_R)}_{Y}A^{Y}_{\mu} 
+ i g^{}_{B} q^{(u_R)}_{B}  B_{\mu}   \right) {u}_{Ri} 
+ \overline{d}_{Ri}  \, i \gamma^{\mu}  \left( \partial_{\mu} + i g^{}_{Y} q^{(d_R)}_{Y}A^{Y}_{\mu} 
+ i g^{}_{B} q^{(d_R)}_{B}   B_{\mu}  \right)    {d}_{Ri} \nonumber \\
&+& \overline{L}_{i} \, i \gamma^{\mu} \left( \partial^{}_{\mu} + i g^{}_{2} \frac{\tau^{j}}{2} W^{j}_{\mu} 
+ i g^{}_{Y} q^{(L)}_{Y} A^{Y}_{\mu} + i g^{}_{B} q^{(L)}_{B} B_{\mu} \right) L_{i} \nonumber\\
&+& \overline{e}^{}_{Ri} \, i \gamma^{\mu}  \left( \partial_{\mu} + i g^{}_{Y} q^{(e_R)}_{Y}A^{Y}_{\mu} 
+ i g^{}_{B} q^{(e_R)}_{B} B_\mu\right)  {e}_{Ri} +
\overline{\nu}_{Ri} \, i \gamma^{\mu} \left( \partial_{\mu} + i g^{}_{Y} q^{(\nu_R)}_{Y}A^{Y}_{\mu} 
+ i g^{}_{B} q^{(\nu_R)}_{B}  B_\mu   \right) {\nu}_{Ri}\nonumber \\
&+& \frac{1}{2}(\partial_{\mu}b + M_{St} B_{\mu})^2 \nonumber\\ 
&+& V(H_u,H_d,b),
\eeqa
which generates $\mathcal{S}_0$. We have summed over  $SU(3)$ index $a=1,2,...,8$, over the $SU(2)$ index $j=1,2,3$ and over the fermion index $i=1,2,3$ denoting a given generation. 
We have denoted with $F^{G}_{\mu\nu}$ the field-strength for the
gluons and with $F^{W}_{\mu\nu}$ the field strength of the weak gauge bosons $W_{\mu}$. 
$F^{Y}_{\mu\nu}$ and $F^{B}_{\mu\nu}$ are the field-strengths related to the abelian hypercharge and the extra abelian gauge boson, B, which has anomalous interactions with a typical generation of 
the Standard Model.
The fermions are either left-handed or right-handed Dirac spinors $f_L$, 
$f_R$ and they fall in the usual $SU(3)_C$ and 
$SU(2)_W$ representations of the Standard Model.

\section{Appendix B. Matrices of the potential}
The mass matrix in the CP-even sector is given by
\beqa
{\cal N}_2(1,1)&=&
-2 (-4 {v^2\lambda_{uu}} \sin ^2\beta +v^2{\lambda_3} \cos ^2
\beta \cot \beta -\frac{3}{2} v^2{\lambda_2} \sin 2 \beta +b \cot
   \beta ) \nonumber \\
{\cal N}_2(1,2)&=&
2 \left(3 v^2{\lambda_3} \cos ^2\beta +3 v^2{\lambda_2} \sin ^2\beta\ +
2 v^2{\lambda_1} \sin 2 \beta -2 {v^2\lambda_{ud}} \sin 2 \beta +b \right)\nonumber \\
{\cal N}_2(2,2)&=&
-2 \sec \beta \left(-4 {\lambda_{dd}} v^2 \cos ^3\beta-3 {\lambda_3}
v^2 \sin \beta \cos ^2\beta+{\lambda_2} v^2 \sin ^3\beta+b \sin \beta\right). \nonumber \\
\eeqa

In the CP-odd sector we have 
\bea{\cal N}_3 =  - \frac{1}{2} v_u v_d \, c_{\chi^\prime}
\pmatrix{\cot{\b} & -1 & v_d\frac{q_u^{I} - q_d^{I}}{M_I} &  \cr
-1 & \tan{\b} & -v_u\frac{q_u^I - q_d^I}{M_1} & \cr
v_d \frac{ q_u^I - q_d^I }{ M_I } & -v_u \frac{ q_u^I - q_d^I }{ M_I } & v_u v_d \frac{ ( q_u^I - q_d^I )^2 }{ M_I^{2} }\cr}.
\label{enne_tre}
\eea

In the charged sector, the mass matrix elements are
\beqa
{\cal N}_1(1,1)&=&
-2 \cot \beta \left({\lambda_3} \cos ^2\beta+ ( {\lambda_1} -{\lambda'}_{ud})
\sin 2 \beta +{\lambda_2} \sin ^2\beta\right) v^2  -2 b  \cot \beta
\nonumber \\
{\cal N}_1(1,2)&=&
2 \left({\lambda_3} \cos ^2\beta+ ( {\lambda_1} -{\lambda'}_{ud}) \sin 2 \beta +{\lambda_2} \sin
   ^2\beta\right) v^2+2 b \nonumber \\
{\cal N}_1(2,2)&=&
-2 \left({\lambda_3} \cos ^2\beta+ ( {\lambda_1} -{\lambda'}_{ud}) \sin 2 \beta +{\lambda_2} \sin
   ^2\beta\right) v^2 \tan \beta  -2 b \tan \beta.
\eeqa

\section{Appendix C. Matrix $O^\chi$ and quadrilinear interactions }

We report for completeness the matrix $O_\chi$, which is given by
\bea
\left( O^{\chi}  \right)_{11} &=&  - \frac{1}{  \frac{ -  ( q_u^B - q_d^B ) }{M_1} v_u
   \sqrt{ \frac{M_1^{\,2} }{  ( q_u^B - q_d^B )^2 } \frac{ v^2 }{ v_u^2 v_d^2 } + 1 } }   \nonumber\\
   &=&     - \frac{1}{ v_u \, \frac{ v  }{v_u v_d}   } \, N      = -N\cos{\b}    \\
\left( O^{\chi}   \right)_{21}&=&    \frac{1}{  \frac{ - ( q_u^B - q_d^B ) }{M_1} v^{}_d
   \sqrt{ \frac{M_1^{\,2} }{  ( q_u^B - q_d^B )^2 } \frac{ v^2 }{ v_u^2 v_d^2 }+1 } }    \nonumber\\
 &=&     \frac{1}{ v_d \, \frac{ v }{v_u v_d}   } N            =  N\sin{\b}    \\
\left( O^{\chi}  \right)_{31}  &=&  \frac{1}{ \sqrt{ \frac{ M_1^{\,2} }{  ( q_u^B - q_d^B)^2} 
\frac{ v^2 }{ v_u^2 v_d^2 } + 1 } }          \nonumber\\  
  &=&     \frac{1}{  \frac{M_1}{-  (q_u^B - q_d^B) \, v^{}_u} \,\, v^{}_u  
\sqrt{ \frac{  (q_u^B - q_d^B)^2}{M_1^{\,2} } + 
          \frac{ v^2 }{v_u^2 v_d^2} }   }     = NQ_1 \cos{\b}  \\
\nonumber\\
\left( O^{\chi} \right)_{12}&=&  \frac{v^{}_u}{\sqrt {v^{\,2}_u + v^{\,2}_d} }   = \sin{\b}     \\
\left( O^{\chi} \right)_{22}&=&  \frac{v^{}_d}{\sqrt {v^{\,2}_u + v^{\,2}_d} }    =   \cos{\b}                                                     \\
\left( O^{\chi} \right)_{32}&=&    0         \label{coeffic1}    \\
\nonumber\\
\left( O^{\chi} \right)_{13}  &=&   \frac{1}{ \sqrt{ 1 + \frac{  ( q_u^B - q_d^B )^2  }{ M_1^{\,2} } 
    \frac{  v_u^{\,2} v_d^{\,2} }{  v^2  } } }   \left( 
 - \frac{  ( q_u^B - q_d^B ) }{ M_1 } \right)  \frac{v_u v_d^2}{  v^2 }  \nonumber\\
& =&   N   \left[  - \frac{ ( q_u^B - q_d^B ) }{ M_1 } v_u \cos\beta \right] \cos\beta  
 =  N {\overline Q}_1 \cos{\b}  \label{coeff_higgs_up}  \\
\left( O^{\chi}  \right)_{23} &=&  -   \frac{1}{ \sqrt{ 1 + \frac{  ( q_u^B - q_d^B )^2  }{ M_1^{\,2} } 
    \frac{  v_u^{\,2} v_d^{\,2} }{ v^2  } } }   \left( -
  \frac{  ( q_u^B - q_d^B ) }{ M_1 } \right) \frac{  v_u^2 v_d }{ v^2}       \nonumber\\
 &=&   -   N   \left[ 
  \frac{ -  ( q_u^B - q_d^B ) }{ M_1 } v_u \cos\beta  \right] \sin\beta 
 = -  N {\overline Q}_1  \sin\beta    \label{coeff_higgs_down}     \\
\left( O^{\chi}  \right)_{33}  &=&   \frac{1}{ \sqrt{ 1 + \frac{  ( q_u^B - q_d^B )^2  }{ M_1^{\,2} } 
    \frac{  v_u^{\,2} v_d^{\,2} }{  v^2 } }}  =   N.        \label{coeffic2}.         
\eea
The coefficients appearing in the quadrilinear vertices are given by
\ba
&&R_1^{\chi^2 H^0 h^0}=\sin{\alpha}\cos{\alpha} \left[\l_{dd} (O^{\chi}_{21})^2 - \l_{uu} (O^{\chi}_{11})^2 \right]
\nonumber\\
&&R_1^{\chi^2 H^0 H^0}=\frac{1}{2}\cos^2{\alpha} \left[\l_{dd} (O^{\chi}_{21})^2 + \l_{uu} (O^{\chi}_{11})^2 \right]
\nonumber\\
&&R_1^{\chi^2 h^0 h^0}=\frac{1}{2}\cos^2{\alpha} \left[\l_{dd} (O^{\chi}_{21})^2 + \l_{uu} (O^{\chi}_{11})^2 \right]
\nonumber\\
&&R_2^{\chi^2 H^0 h^0}=\sin{\alpha}\cos{\alpha} \l_{ud} \left[(O^{\chi}_{21})^2 - (O^{\chi}_{11})^2 \right]
\nonumber\\
&&R_2^{\chi^2 H^0 H^0}=-\frac{1}{2}\cos^2{\alpha}\l_{ud} \left[(O^{\chi}_{21})^2 + (O^{\chi}_{11})^2 \right]
\nonumber\\
&&R_2^{\chi^2 h^0 h^0}=-\frac{1}{2}\sin^2{\alpha} \l_{ud} \left[(O^{\chi}_{21})^2 + (O^{\chi}_{11})^2 \right]
\nonumber\\
&&R_3^{\chi^2 H^0 h^0}=\sin{\alpha}\cos{\alpha} \l_1 \left[(O^{\chi}_{21})^2 - (O^{\chi}_{11})^2 \right]
-4 \sin{\alpha}\cos{\alpha}\l_1 \frac{\Delta q^B}{M_1}O^{\chi}_{31}
\left(O^{\chi}_{21} + O^{\chi}_{11} \right) +O(1/M_1^2)
\nonumber\\
&&R_3^{\chi^2 H^0 H^0}=
-\frac{1}{2}\cos^2\alpha\l_{1}\left[(O^{\chi}_{21})^2 + (O^{\chi}_{11})^2 +4 O^{\chi}_{11}O^{\chi}_{21} \right]
\nonumber\\
&&\hspace{2cm}+ \cos^2{\alpha} \l_1 \frac{\Delta q^B}{M_1}\left[ v_d(4 O^{\chi}_{11}O^{\chi}_{31} 
+ 2 O^{\chi}_{21}O^{\chi}_{31}) - v_u(4 O^{\chi}_{21}O^{\chi}_{31}+2 O^{\chi}_{11}O^{\chi}_{31})\right]+O(1/M_1^2)
\nonumber\\
&&R_3^{\chi^2 h^0 h^0}=-\frac{1}{2}\sin^2\alpha\l_{1}\left[(O^{\chi}_{21})^2 + (O^{\chi}_{11})^2 -4 O^{\chi}_{11}O^{\chi}_{21} \right]
\nonumber\\
&&\hspace{2cm}+\sin^2{\alpha} \l_1 \frac{\Delta q^B}{M_1}\left[v_d(-4 O^{\chi}_{11}O^{\chi}_{31} 
+ 2 O^{\chi}_{21}O^{\chi}_{31}) - v_u(-4 O^{\chi}_{21}O^{\chi}_{31} + 2 O^{\chi}_{11}O^{\chi}_{31})\right]+O(1/M_1^2)
\nonumber\\
&&R_4^{\chi^2 H^0 h^0}=\sin{\alpha}\cos{\alpha}O^{\chi}_{21}O^{\chi}_{11}(\l_3-\l_2)
+\sin{\alpha}\cos{\alpha} \frac{\Delta q^B}{M_1}O^{\chi}_{31}
\left[v_d O^{\chi}_{11}(\l_2 -3\l_3)+v_u O^{\chi}_{21}(\l_3 -3\l_2)\right]
\nonumber\\
&&R_4^{\chi^2 H^0 H^0}=\frac{1}{2}\cos^2\alpha \left\{\l_2\left[O^{\chi}_{21}O^{\chi}_{11}-(O^{\chi}_{11})^2\right]
+\l_3\left[O^{\chi}_{21}O^{\chi}_{11}-(O^{\chi}_{21})^2\right]\right\}
\nonumber\\
&&\hspace{2cm}+\cos^2{\alpha} \frac{\Delta q^B}{2 M_1}O^{\chi}_{31}
\left\{v_u\left[O^{\chi}_{21}(\l_3+3\l_2)+2 \l_2 O^{\chi}_{11}\right]
-v_d\left[O^{\chi}_{11}(\l_2+3\l_3)+2 \l_3 O^{\chi}_{21}\right]\right\}
\nonumber\\
&&R_4^{\chi^2 h^0 h^0}=\frac{1}{2}\sin^2\alpha \left\{\l_2\left[O^{\chi}_{21}O^{\chi}_{11}+(O^{\chi}_{11})^2\right]
+\l_3\left[O^{\chi}_{21}O^{\chi}_{11}+(O^{\chi}_{21})^2\right]\right\}
\nonumber\\
&&\hspace{2cm}+\sin^2{\alpha} \frac{\Delta q^B}{2 M_1}O^{\chi}_{31}
\left\{v_u\left[O^{\chi}_{21}(\l_3+3\l_2)-2 \l_2 O^{\chi}_{11}\right]
-v_d\left[O^{\chi}_{11}(\l_2+3\l_3)-2 \l_3 O^{\chi}_{21}\right]\right\}
\nonumber\\
\ea

\section{Appendix C: Axi-Higgs Trilinear Interactions}

\ba
&&R_1^{\chi^2 H^0}=\cos{\alpha}\left[(O^{\chi}_{21})^2 v_d \l_{dd} - (O^{\chi}_{11})^2 v_u \l_{uu}\right],
\nonumber\\
&&R_1^{\chi^2 h^0}=\sin{\alpha}\left[(O^{\chi}_{21})^2 v_d \l_{dd} + (O^{\chi}_{11})^2 v_u \l_{uu}\right],
\nonumber\\
&&R_2^{\chi^2 H^0}=\cos{\alpha}\l_{ud}\left[(O^{\chi}_{21})^2 v_u  - (O^{\chi}_{11})^2 v_d \right],
\nonumber\\
&&R_2^{\chi^2 h^0}=-\sin{\alpha}\l_{ud}\left[(O^{\chi}_{21})^2 v_u  + (O^{\chi}_{11})^2 v_d \right],
\nonumber\\
&&R_3^{\chi^2 H^0}=-b_1 \cos{\alpha}O^{\chi}_{31}\frac{\Delta q^B}{M_1}\left(O^{\chi}_{11}+O^{\chi}_{21} \right),
\nonumber\\
&&R_3^{\chi^2 h^0}=b_1 \sin{\alpha}O^{\chi}_{31}\frac{\Delta q^B}{M_1}\left(O^{\chi}_{21}-O^{\chi}_{11} \right),
\nonumber\\
&&R_4^{\chi^2 H^0}=\cos{\alpha} \l_1\left[O^\chi_{21} (2 O^\chi_{11} + O^{\chi}_{21}) v_u - 
O^{\chi}_{11} (O^{\chi}_{11} + 2 O^{\chi}_{21}) v_d\right]  
\nonumber\\
&&\hspace{1.5cm}+2\cos{\alpha}\l_1\frac{\Delta q^B}{M_1}O^{\chi}_{31}\left[ O^{\chi}_{11} v_d (v_d - 2 v_u) 
+ O^{\chi}_{21} v_u (v_u - 2 v_d)\right] +O(1/M_1^2),
\nonumber\\
&&R_4^{\chi^2 h^0}=-\sin{\alpha} \l_1\left[O^{\chi}_{11}(O^{\chi}_{11}-2 O^{\chi}_{21})v_d -O^{\chi}_{21}(2O^{\chi}_{11}+O^{\chi}_{21})v_u \right]
\nonumber\\
&&\hspace{1.5cm}+ 2 \sin{\alpha}\l_1\frac{\Delta q^B}{M_1}O^{\chi}_{31}\left[ O^{\chi}_{21} v_u (2v_d + v_u)
-O^{\chi}_{11} v_d (v_d+2v_u)\right]+O(1/M_1^2),
\nonumber\\
&&R_5^{\chi^2 H^0}=\frac{1}{2}\cos{\alpha}\left[O^{\chi}_{21}\l_3\left(2O^{\chi}_{11}v_d +O^{\chi}_{21}(v_u-v_d)\right)
-O^{\chi}_{11}\l_2\left(O^{\chi}_{11} (v_d-v_u) +2O^{\chi}_{21}v_u\right)\right]
\nonumber\\
&&\hspace{1.5cm}+
\cos{\alpha}\frac{\Delta q^B}{2 M_1}O^{\chi}_{31}
\left[-v_u \l_2 \left( 3 O^{\chi}_{21}v_u+O^{\chi}_{11}(v_u-2v_d)\right)
-v_d\l_3\left( 3 O^{\chi}_{11}v_d+O^{\chi}_{21}(v_d-2v_u)\right)\right],
\nonumber\\
&&R_5^{\chi^2 h^0}=\frac{1}{2}\sin{\alpha}\left[O^{\chi}_{11}\l_2\left(2O^{\chi}_{21}v_u +O^{\chi}_{11}(v_u+v_d)\right)
+O^{\chi}_{21}\l_3\left(O^{\chi}_{21} (v_d+v_u) +2O^{\chi}_{11}v_d\right)\right]
\nonumber\\
&&\hspace{1.5cm}+\sin{\alpha}\frac{\Delta q^B}{2 M_1}O^{\chi}_{31}
\left[v_d \l_3 \left(O^{\chi}_{21}(v_d+2v_u)-3 O^{\chi}_{11}v_d\right)
+v_u\l_2\left(3 O^{\chi}_{21}v_u-O^{\chi}_{11}(2v_d + v_u)\right)\right],
\nonumber\\
\ea

\section{Appendix D:  Quadrilinear self interactions in the CP-even sector}

For $H_0^4$ we have
\ba
&&R_1^{H_0^4}=(\cos{\alpha})^4 \frac{1}{4}\left(\lambda_{uu}+\lambda_{dd}\right),
\nonumber\\
&&R_2^{H_0^4}=-(\cos{\alpha})^4 \frac{1}{2}\lambda_{ud},
\nonumber\\
&&R_3^{H_0^4}=(\cos{\alpha})^4 \frac{1}{2}\lambda_{1},
\nonumber\\
&&R_4^{H_0^4}=(\cos{\alpha})^4 \frac{1}{2}\left(\lambda_{2}+\lambda_{3}\right),
\ea
while for $h_0^4$ we have
\ba
&&R_1^{h_0^4}=\frac{1}{4} (\sin{\alpha})^4 \left(\lambda_{uu}+\lambda_{dd}\right),
\nonumber\\
&&R_2^{h_0^4}=-\frac{1}{2} (\sin{\alpha})^4 \lambda_{ud},
\nonumber\\
&&R_3^{h_0^4}=\frac{1}{2} (\sin{\alpha})^4 \lambda_{1},
\nonumber\\
&&R_4^{h_0^4}=\frac{1}{2} (\sin{\alpha})^4 \left(\lambda_{2}+\lambda_{3}\right).
\ea

For the interactions of the type ${h_0}^2 {H_0}^2$ we obtain
\ba
&&R_1^{H_0^2 h_0^2}=\frac{3}{2}(\sin{\alpha})^2(\cos{\alpha})^2 \left(\lambda_{uu}+\lambda_{dd}\right),
\nonumber\\
&&R_2^{H_0^2 h_0^2}=(\sin{\alpha})^2(\cos{\alpha})^2 \lambda_{ud},
\nonumber\\
&&R_3^{H_0^2 h_0^2}=-(\sin{\alpha})^2(\cos{\alpha})^2 \lambda_{1}.
\ea

For the interactions of the type ${h_0}^3 {H_0}$ we obtain
\ba
&&R_1^{H_0 h_0^3}=(\sin{\alpha})^3 \cos{\alpha} \left(\lambda_{dd}-\lambda_{uu}\right),
\nonumber\\
&&R_2^{H_0 h_0^3}=(\sin{\alpha})^3 \cos{\alpha} \left(\lambda_{3}-\lambda_{2}\right),
\ea
while for ${h_0} {H_0}^3$ we obtain
\ba
&&R_1^{h_0 H_0^3}=(\cos{\alpha})^3 \sin{\alpha} \left(\lambda_{dd}-\lambda_{uu}\right),
\nonumber\\
&&R_2^{h_0 H_0^3}=-(\cos{\alpha})^3 \sin{\alpha} \left(\lambda_{2}-\lambda_{3}\right),
\ea

\section{Appendix E: Trilinear self interactions in the CP-even sector }
For $H_0^3$ we have
\ba
&&R_1^{H_0^3}=(\cos{\alpha})^3 (v_d\lambda_{dd} - v_u\lambda_{uu}), 
\nonumber\\
&&R_2^{H_0^3}=(\cos{\alpha})^3 (v_d- v_u)\lambda_{ud}, 
\nonumber\\
&&R_3^{H_0^3}=(\cos{\alpha})^3 (v_d- v_u)\lambda_{1}, 
\nonumber\\
&&R_4^{H_0^3}=\frac{1}{2}(\cos{\alpha})^3 \left[v_u (3\lambda_2+\lambda_3)- v_d (\lambda_2+3\lambda_3)\right],
\ea
while for $h_0^3$ we have
\ba
&&R_1^{h_0^3}=(\sin{\alpha})^3 (v_d\lambda_{dd} + v_u\lambda_{uu}), 
\nonumber\\
&&R_2^{h_0^3}=-(\sin{\alpha})^3 (v_d+ v_u)\lambda_{ud}, 
\nonumber\\
&&R_3^{h_0^3}=(\sin{\alpha})^3 (v_d + v_u)\lambda_{1}, 
\nonumber\\
&&R_4^{h_0^3}=\frac{1}{2}(\sin{\alpha})^3 \left[v_u (3\lambda_2+\lambda_3)+ v_d (\lambda_2+3\lambda_3)\right].
\ea
For the case $h_0^2 H_0$ we have
\ba
&&R_1^{h_0^2 H_0}=3\cos{\alpha}(\sin{\alpha})^2 (v_d\lambda_{dd} - v_u\lambda_{uu}), 
\nonumber\\
&&R_2^{h_0^2 H_0}=\cos{\alpha}(\sin{\alpha})^2 \lambda_{ud}(v_d - v_u),
\nonumber\\
&&R_3^{h_0^2 H_0}=\cos{\alpha}(\sin{\alpha})^2 \lambda_1 (v_u - v_d),
\nonumber\\
&&R_4^{h_0^2 H_0}=-\frac{3}{2}\cos{\alpha}(\sin{\alpha})^2(v_u + v_d)(\lambda_2-\lambda_3),
\ea
while for the case $h_0 H_0^2$ we have
\ba
&&R_1^{h_0 H_0^2}=3\sin{\alpha}(\cos{\alpha})^2 (v_d\lambda_{dd} + v_u\lambda_{uu}), 
\nonumber\\
&&R_2^{h_0 H_0^2}=\sin{\alpha}(\cos{\alpha})^2 \lambda_{ud}(v_d + v_u),
\nonumber\\
&&R_3^{h_0 H_0^2}=-\sin{\alpha}(\cos{\alpha})^2 \lambda_1 (v_u + v_d),
\nonumber\\
&&R_4^{h_0 H_0^2}=\frac{3}{2}\sin{\alpha}(\cos{\alpha})^2(v_d - v_u)(\lambda_2-\lambda_3).
\ea

\section{Appendix F. The axion Lagrangian in the physical basis}

We have seen that after symmetry breaking, in the scalar sector we isolate a physical axion, 
$\chi$, also called the axi-Higgs. Here we present the axion Lagrangian rotated on the basis of the 
mass eigenstates. In particular, the $W_3, A^Y$ and $B$ gauge bosons become 
linear combinations of the physical states $A_\g, Z, Z^\prime$. Indeed, 
the mass-matrix in the neutral gauge sector is given by
\begin{displaymath}
{\cal L}_{mass}=\left(W_3,~Y,~B\right){\bf M}^2\left(\begin{array}{c}
W_3\\
Y\\
B\end{array}\right)
\end{displaymath}
where $B$ is the St\"uckelberg field and the mass matrix is defined as
\bea
{\bf M}^2 = {1\over 4} \pmatrix{
{g^{}_2}^{2} v^2 & - {g^{}_2} \, {g^{}_Y} v^2 &  - {g^{}_2} \,  x^{}_B \cr
 - {g^{}_2} \,{g^{}_Y} v^2 &  {g^{}_Y}^{2} v^2 & {g^{}_Y}  x^{}_B \cr
 -{g^{}_2} \, x^{}_B &{g^{}_Y}  x^{}_B  & 2 M_1^2 + N^{}_{BB}}
\label{massmatrix}
\eea
with
\bea
N^{}_{BB}=  \left( q_u^{B\,2} \,{v^{\,2}_u} + q_d^{B\,2} \,{v^{\,2}_d} \right)\, g_B^{\,2},
&& x^{}_B=  \left(q_u^B {v^{\,2}_u} + q_d^B {v^{\,2}_d}  \right)\, g^{}_B.
\eea
Here $v_u$ and $v_d$ denote the vevs of the two Higgs fields $H_u,H_d$ while
$q_u^{B}$ and $q_d^{B}$ are the Higgs charges under the extra anomalous $U(1)_B$.
We have also defined $v=\sqrt{v_u^2+v_d^2}$ and $g=\sqrt{g_2^2+g_Y^2}$.
The mass-squared eigenstates of the mass matrix corresponding
to one zero mass eigenvalue for the photon $A^{}_{\gamma}$ and two non-zero mass
eigenvalues for the $Z$ and for the ${Z}^{\prime}$ vector bosons, are respectively given by
\bea
M_{Z}^2 &=&  \frac{1}{4} \left( 2 M_1^2 + g^2 v^2 + N^{}_{BB}
- \sqrt{\left(2  M_{1}^2 - g^2 v^2 + N^{}_{BB} \right)^2 + 4
   g^2 x_{B}^2} \right)    \\
&\simeq&     \frac{g^2 v^2}{2} - \frac{1}{M_{1}^2} \frac{g^2 x_{B}^{2}}{4}
 + \frac{1}{M_{1}^4}\frac{g^2 x_{B}^2}{8 } (N^{}_{BB} - g^2 v^2) , \nn
\label{massZ}
\eea
\bea
M_{{Z}^\prime}^2 &=&   \frac{1}{4} \left( 2 M_1^2 + g^2 v^2 + N^{}_{BB}
+ \sqrt{\left(2  M_{1}^2 - g^2 v^2 + N^{}_{BB} \right)^2 + 4   g^2 x_{B}^2} \right)   \\
&\simeq&    M^{2}_{1} +  \frac{N^{}_{BB}}{2} .  \nn
\label{massZp}
\eea
The mass of the $Z$ gauge boson gets corrections of the order $v^{2}/M_1$ 
converging to the SM value as $M_1\to \infty$, while the mass of the $Z'$ 
gauge boson can grow large with $M_1$.
The physical gauge fields can be obtained from the rotation matrix $O^A$
\ba
\pmatrix{A_\g \cr Z \cr {{Z^\prime}}} =
O^A\, \pmatrix{W_3 \cr A^Y \cr B} 
\label{OA}
\ea
which can be approximated at the first order as
\bea
O^A  \simeq  \pmatrix{
\frac{g^{}_Y}{g}           &     \frac{g^{}_2}{g}         &      0   \cr
\frac{g^{}_2}{g} + O(\epsilon_1^2)          &     -\frac{g^{}_Y}{g} + O(\epsilon_1^2) &      \frac{g}{2} \epsilon_1 \cr
-\frac{g^{}_2}{2}\epsilon_1     &     \frac{g^{}_Y}{2}\epsilon_1  &   1 + O(\epsilon_1^2) } .
\label{matrixO}
\eea
Moreover, after symmetry breaking, as we have already shown in eq.~(\ref{CPodd}), the St\"uckelberg field $b$ is 
rotated by means of the matrix $O^\chi$ as follows
\bea
b &=&  O_{31}^{\chi} \chi +  O_{32}^{\chi} G_1^0 + O_{33}^{\chi} G_2^0,       
\label{rot12}
\eea
where the elements of the rotation matrix have the following expressions 
\bea
O^{\chi}_{31}  = \frac{1}{ \sqrt{ \frac{ M_1^{\,2} }{  ( q_u^B - q_d^B)^2} 
\frac{ v^2 }{ v_u^2 v_d^2 } + 1 } }, \qquad
O^{\chi}_{33}  =  \frac{1}{ \sqrt{ 1 + \frac{ ( q_u^B - q_d^B )^2  }{ M_1^{2} } 
\frac{  v_u^2 v_d^2}{v^2} }},  \qquad O^{\chi}_{32}= 0.
\eea
Then, starting from eq.~(\ref{rot12}), the Goldstone modes $G^Z$ and $G^{Z^\prime}$ in the $\g$-basis are obtained by the combination
\bea
G^0_2= C^\prime_Z G^Z + C^\prime_{Z^\prime} G^{Z^\prime}. 
\eea
More details can be found in \cite{Coriano:2008pg}. Starting from the WZ Lagrangian in the Y-basis  
\beqn
{\cal L}^{axion}_{Y-basis} &=& D\, b \, Tr [ F^G \wedge F^G ] + F \,b \, Tr [ F^W \wedge F^W ]   \nonumber\\
&& + C_{YY} \, b \, F^{Y} \wedge F^{Y}  + C_{BB} \, b \, F^{B} \wedge F^{B} 
 + C_{YB} \, b \, F^{Y} \wedge F^{B}
 \eeqn
and rotating into the physical mass eigestates using eqs. (\ref{OA}) and (\ref{rot12})
we obtain the axion-like terms of the WZ Lagrangian 
\ba
{\cal L}^{axion}( \chi )&=&  g^{\chi}_{gg} \, \chi\, Tr\,  [ F^G\wedge {F^G} ] + g^{\chi}_{+-} \, \chi\, Tr\, [ F^{W+}\wedge F^{W^-} ] 
+ g^{\chi}_{\g \g} \, \chi\, F^{\g} \wedge F^{\g}  \nonumber\\
&+&g^{\chi}_{ZZ} \, \chi\, F^{Z} \wedge F^{Z}+  g^{\chi}_{Z'Z'} \, \chi\, F^{Z'} \wedge F^{Z'} 
+ g^{\chi}_{\g Z} \, \chi\, F^{\g} \wedge F^{Z}   \nonumber\\
&+&  g^{\chi}_{\g Z'} \, \chi\, F^{\g} \wedge F^{Z'} +  g^{\chi}_{ZZ'} \, \chi\, F^{Z} \wedge F^{Z'}, 
\ea
\ba
 {\cal L}^{axion}(G^{\cal Z})&=& c^{\cal Z}_{gg} G^{\cal Z} Tr\, [ F^G\wedge F^{G} ] +  c^{\cal Z}_{+-} G^{\cal Z}  Tr\, [ F^{W+}\wedge F^{W-} ]  
+ c^{\cal Z}_{\g \g} \, G^{\cal Z}\, F^{\g} \wedge F^{\g}   \nonumber\\
&+&   c^{\cal Z}_{ZZ} \, G^{\cal Z}\, F^{Z} \wedge F^{Z}  +  c^{\cal Z}_{Z'Z'} \, G^{\cal Z}\, F^{Z'} \wedge F^{Z'}  
+     c^{\cal Z}_{\g Z} \, G^{\cal Z}\, F^{\g} \wedge F^{Z}   \nonumber\\
&+&  c^{\cal Z}_{\g Z'} \, G^{\cal Z}\, F^{\g} \wedge F^{Z'}  
+  c^{\cal Z}_{ZZ'} \, G^{\cal Z}\, F^{Z} \wedge F^{Z'}, 
\ea
where ${\cal Z}$ stays for $Z,Z^{\prime}$.
Finally the WZ Lagrangian in the physical basis is given by the sum of three contributions
\ba
{\cal L}^{axion}_{\g-basis}= {\cal L}^{axion}( \chi )+ {\cal L}^{axion}(G^{Z})+ {\cal L}^{axion}(G^{Z'}),
\ea
where we have identified the physical couplings of the axi-Higgs $\chi$ to the gauge bosons as
\beqn
\label{phys_couplings}
&& g^{\chi}_{gg}\,= \,  D \, O^{\chi}_{31}      \nonumber\\
&& g^{\chi}_{+-}\, = \,  F \,   O^{\chi}_{31}      \nonumber\\
&& g^{\chi}_{\g \g}\, = \, \left(  F O^A_{ W_3 \gamma} O^A_{ W_3 \gamma} 
+ C_{YY}  O^{A}_{Y \gamma  } O^A_{Y \gamma  } \right)\,  O^{\chi}_{31} \nonumber\\
&&  g^{\chi}_{ZZ}\, = \, \left(  F O^A_{W_3 Z } O^A_{ W_3 Z} +
C_{YY}  O^{A}_{Y Z } O^A_{Y Z } + C_{BB}  O^{A}_{B Z } O^A_{B Z } 
+ C_{YB} O^{A}_{YZ} O^{A}_{BZ} \right)\,  O^{\chi}_{31} \nonumber\\
&& g^{\chi}_{Z'Z'}\, = \, \left(  F O^A_{  W_3 Z'} O^A_{ W_3 Z'} +
C_{YY}  O^{A}_{Y Z'  } O^A_{Y Z'  } + C_{BB}  O^{A}_{B Z'  } O^A_{B Z' } 
+  C_{YB} O^{A}_{Y Z'} O^{A}_{B Z'} \right)\,  O^{\chi}_{31} \nonumber\\
&& g^{\chi}_{\gamma Z}\, = \, \left( 2 F O^A_{ W_3 \gamma} O^A_{ W_3 Z} +
2 C_{YY}  O^{A}_{Y \gamma  } O^A_{Y Z  } +  C_{YB} O^{A}_{Y \gamma } O^{A}_{BZ} \right)\,  
O^{\chi}_{31} \nonumber\\
&& g^{\chi}_{\gamma Z'}\, = \, \left( 2 F O^A_{  W_3 \gamma} O^A_{ W_3 Z'} +
2 C_{YY}  O^{A}_{ Y \gamma } O^A_{ Y Z' } +  C_{YB} O^{A}_{Y \gamma } 
O^{A}_{BZ'} \right)\,  O^{\chi}_{31} \nonumber\\
&& g^{\chi}_{Z Z'}\, = \, \left( 2 F O^A_{W_3 Z} O^A_{W_3 Z' } +
2 C_{YY}  O^{A}_{Y Z  } O^A_{Y Z'  } + 2 C_{BB}  O^{A}_{ B Z } O^{A}_{ B Z'}  \right. \nonumber\\
&& \,\,\,\,\,\,\,\,\,\,\,\,\,\,\,\,\,\,\,\,\,\,\,\,\,\,\,\,\,\, \left. + \, C_{YB} O^{A}_{ Y Z } O^{A}_{B Z'} 
+ C_{YB} O^{A}_{ Y Z' } O^{A}_{B Z} \right) \, O^{\chi}_{31}
\eeqn
and the interactions of the NG bosons $G^{\cal Z}$ (${\cal Z}=Z, Z^\prime$) with the gauge bosons
\beqn
&&c^{\cal Z}_{gg} =  D   O^{\chi}_{33} C'_{\cal Z}   \nonumber\\
&&c^{\cal Z}_{+-} =    F \,   O^{\chi}_{33} C'_{\cal Z}   \nonumber\\
&&c^{\cal Z}_{\gamma \gamma} =  \, \left(  F O^A_{  W_3 \gamma} O^A_{ W_3 \gamma} 
+ C_{YY}  O^{A}_{Y \gamma  } O^A_{Y \gamma } \right)\,  O^{\chi}_{33}  C'_{\cal Z}  \nonumber\\
&& c^{\cal Z}_{ZZ} =  \, \left(  F O^A_{W_3 Z} O^A_{W_3 Z } +
C_{YY}  O^{A}_{Y Z  } O^A_{Y Z  } + C_{BB}  O^{A}_{B Z } O^A_{B Z }
 +  C_{YB} O^{A}_{YZ} O^{A}_{BZ} \right)\,  O^{\chi}_{33}  C'_{\cal Z}   \nonumber\\
&&c^{\cal Z}_{Z'Z'} =  \, \left(  F O^A_{W_3 Z' } O^A_{W_3 Z'} +
C_{YY}  O^{A}_{Y Z' } O^A_{Y Z' } + C_{BB}  O^{A}_{B Z' } O^A_{B Z' } 
+  C_{YB} O^{A}_{Y Z'} O^{A}_{B Z'} \right)\,  O^{\chi}_{33}  C'_{\cal Z}  \nonumber\\
&& c^{\cal Z}_{\gamma Z} =  \, \left( 2 F O^A_{W_3 \gamma } O^A_{W_3 Z } +
2 C_{YY}  O^{A}_{Y \gamma  } O^A_{Y Z  } 
+  C_{YB} O^{A}_{Y \gamma } O^{A}_{B Z} \right)\,  O^{\chi}_{33}  C'_{\cal Z}    \nonumber\\
&& c^{\cal Z}_{\gamma Z'} =  \, \left( 2 F O^A_{W_3 \gamma} O^A_{W_3 Z'} +
2 C_{YY}  O^{A}_{Y \gamma  } O^A_{Y Z' } 
+  C_{YB} O^{A}_{Y \gamma } O^{A}_{B Z'} \right)\,  O^{\chi}_{33} C'_{\cal Z}    \nonumber\\
&&c^{\cal Z}_{ZZ'} =   \, \left( 2 F O^A_{W_3 Z } O^A_{W_3 Z' } +
2 C_{YY}  O^{A}_{Y Z  } O^A_{Y Z'} + 2 C_{BB}  O^{A}_{BZ} O^{A}_{BZ'}  \right. \nonumber\\
&& \left.   \,\,\,\,\,\,\,\,\,\,\,\,\,\,\,\,\,\,\,\,\,\,\,\,\,\,\,\,\,\, + \, C_{YB} O^{A}_{YZ} O^{A}_{BZ'}   
 + C_{YB} O^{A}_{ Y Z' } O^{A}_{B Z}  \right) \,  O^{\chi}_{33} C'_{\cal Z}.   \nonumber\\
\eeqn 
where ${\cal Z}$ stays for $Z,Z^{\prime}$.
We also summarize for convenience the coefficients of the WZ counterterms 
\beqn
&& F=  \frac{g^{}_{B}}{M_1}i  g^{2}_{2} \frac{a^{}_{n}}{2}   D^{(L)}_{B}, \nonumber \\
&&D =  \frac{g^{}_{B}}{M_1} i  g^{2}_{3} \frac{a^{}_{n}}{2}  D^{(Q)}_{B}, \nonumber \\
&& C_{BB} = \frac{g^{\,3}_{B} }{M_{1}}  \frac{i }{3!}  a^{}_{n}  D^{}_{BBB},\nonumber \\
&&C_{YY} =  \frac{g^{}_{B}}{ M^{}_1 }  i  g^{\,2}_{Y}  \frac{a^{}_{n}}{2}  D^{}_{BYY},\nonumber\\
&& C_{YB} =  \frac{g^{\,2}_{B}}{ M^{}_1 } i g^{}_{Y}  \frac{a^{}_{n}}{2}  D^{}_{YBB}, 
\label{WZcoeff}
\eeqn
with $a_n=- \frac{i}{2 \pi^2}$ and the chiral asymmetries have been defined, for brevity, in the following way
\beqn
D^{(L)}_{B} &=&   \frac{1}{8}  \sum_{f} \theta_{fL}^B= - \frac{1}{8} \sum_{f} q_{fL}^{B}, \nonumber\\
D^{(Q)}_{B} &=&  \frac{1}{8}  \sum_{Q} \theta_Q^B   = \frac{1}{8} \sum_{Q} \left[ q_{Q_R}^{B}
 - q_{QL}^B  \right] , \nonumber\\
D^{}_{BBB}&=&    \frac{1}{8}  \sum_{f} \theta_f^{BBB}  = \frac{1}{8} \sum_{f} \left[ (q_{fR}^{B})^{3}
 - (q_{fL}^B)^{3}  \right], \nonumber   \\
D^{}_{BYY}&=&   \frac{1}{8}  \sum_{f} \theta_f^{BYY}  =
                \frac{1}{8} \sum_{f} \left[ q_{fR}^{B} (q_{fR}^{Y})^{2}  - q_{fL}^{B} (q_{f_L}^{Y})^{2}  \right],   \nonumber    \\
D^{}_{YBB}&=&  \frac{1}{8}  \sum_{f} \theta_f^{YBB}  =
                \frac{1}{8} \sum_{f} \left[ q_{fR}^{Y} (q_{fR}^{B})^{2}  - q_{fL}^{Y} (q_{f_L}^{B})^{2}   \right].     
\label{asymmetries}
\eeqn

 \section{Appendix G. Three- and Four-particle phase space}

The three and four body phase space in the case of massive particles can be computed directly
in four dimensions since there are no soft and collinear divergences.
The reactions that we are considering are 
$g(k_1) + g(k_2)\rightarrow \chi(q^2) \rightarrow \chi(p)\, \chi(r)\,\chi(p')\, H(p')$ and 
$g(k_1) + g(k_2)\rightarrow H(q^2) \rightarrow \chi(p)\, \chi(r)\,\chi(p')\, \chi(p')$,
where the on-shell conditions are given by $r^2=r'^2=p^2=m_{\chi^2}$ and $p'^2=m_{H}^2$
for the first reaction, while for the second we have $r^2=r'^2=p^2={p'}^2=m_{\chi^2}$.
The computation follows closely \cite{Oleari:1997az}, with some modifications due to our specific case, given the 
three axions and one higgs boson in the final state.

\subsection{Phase space for the three axions and one scalar higgs final state}

In four dimensions we can write the most general formula as follows
\ba
d\Phi_4=\frac{1}{2!}\int \frac{d^3 p}{2p_0(2\pi)^3}\frac{d^3 p'}{2p'_0(2\pi)^3}
\frac{d^3 r}{2r_0(2\pi)^3}\frac{d^3 r'}{2r'_0(2\pi)^3}(2\pi)^4\delta^4(q-p-p'-r-r'),
\label{4body}
\ea
where $1/2!$ is a statistical factor that takes into account the fact that a pair
of identical particles are produced in the final state.
The reference frame in the CM of $r,r'$ can be chosen as
\ba
&&r=(r_0,|\vec{r}|\sin{\theta}\sin{\phi},|\vec{r}|\sin{\theta}\cos{\phi},|\vec{r}|\cos{\theta})
\nonumber\\
&&r=(r_0,-|\vec{r}|\sin{\theta}\sin{\phi},-|\vec{r}|\sin{\theta}\cos{\phi},-|\vec{r}|\cos{\theta})
\nonumber\\
&&p=p_0(1,0,0,\sqrt{1-\frac{m_{\chi}^2}{p_0^2}})
\nonumber\\
&&p'=p'_0(1,0,\sqrt{1-\frac{m_{H}^2}{{p'}_0^2}}\sin{\alpha},
\sqrt{1-\frac{m_{H}^2}{{p'}_0^2}}\cos{\alpha}).
\ea
We introduce the following variables
\ba
&&x_1=2\frac{q\cdot p}{q^2}, \hspace{0.5cm} x_2=2\frac{q\cdot p'}{q^2},
\hspace{0.5cm} y=2\frac{(r+r')^2}{q^2}, \hspace{0.5cm} \theta,\hspace{0.5cm} \phi,
\nonumber\\
&&r_0=\sqrt{q^2}\frac{\sqrt{y}}{2}, \hspace{0.5cm} |\vec{r}|=\sqrt{r_0^2-m_{\chi}^2}, 
\nonumber\\
&&\rho_1 =4\frac{m_{H}^2}{q^2}, \hspace{0.5cm} \rho_2 =4\frac{m_{\chi}^2}{q^2}.
\ea
From the momentum conservation equations $(q-p)^2=(p'+r+r')$ and $(q-p')^2=(p'+r+r')$
we derive the expression of $p_0$ and $p'_0$ as a function of the variables $(x_1,x_2,y,\sqrt{q^2},m_\chi,m_H)$
as follows
\ba
&&p_0=\frac{(1-x_2-y)\sqrt{q^2}}{2\sqrt{y}} + \frac{m_H^2-m_{\chi}^2}{2\sqrt{q^2}\sqrt{y}},
\nonumber\\
&&{p'}_0=\frac{(1-x_1-y)\sqrt{q^2}}{2\sqrt{y}} + \frac{m_H^2-m_{\chi}^2}{2\sqrt{q^2}\sqrt{y}}.
\ea
Using the equation $q^2=(p +p' +r+r')^2$, we obtain the expression of $\cos{\alpha}$ 
in terms of the kinematic variables defined above
\ba
\cos{\alpha}=\frac{p_0 + {p'}_0 + \sqrt{q^2} \sqrt{y} -(p_0^2-m_{\chi}^2) -({p'}_0^2-m_{H}^2)-q^2}
{2\sqrt{p_0^2-m_{\chi}^2}\sqrt{{p'}_0^2-m_{H}^2} }.
\ea
In order to integrate the expression given in Eq. (\ref{4body}) it is useful to introduce 
the following identities 
\ba
\int \frac{d^4 t}{(2\pi)^4} (2\pi)^4 \delta^4(t-r-r')=1, &&  
q^2 \int \frac{d y}{(2\pi)} (2\pi) \delta(t^2-q^2 y)=1
\ea
which allow us to incorporate $r$ and $r'$ in the $t$ state. Thus, we obtain
\ba
&&d\Phi_4=\frac{1}{2!}\int \frac{d^3 t}{2 t_0 (2\pi)^3}
\int \frac{d^3 p}{2p_0(2\pi)^3}\int\frac{d^3 p'}{2p'_0(2\pi)^3}(2\pi)^4\delta(q-t-p-p')
\nonumber\\
&&\hspace{2cm}
q^2\int \frac{d y}{(2\pi)}\Theta(y)  
\int\frac{d^3 r}{2r_0(2\pi)^3}\frac{d^3 r'}{2r'_0(2\pi)^3}(2\pi)^4\delta^4(t-r-r'),
\ea  
where $\Theta(y)$ is the Heaviside step function. In this way we have factorized the expression
of $d\Phi_4$ phase space as a product of $d\Phi_3 \times d\Phi_2$
\ba
&&d\Phi_2=\int\frac{d^3 r}{2r_0(2\pi)^3}\frac{d^3 r'}{2r'_0(2\pi)^3}(2\pi)^4\delta^4(t-r-r')
\nonumber\\
&&d\Phi_3=\int \frac{d^3 t}{2 t_0 (2\pi)^3}
\int \frac{d^3 p}{2p_0(2\pi)^3}\int\frac{d^3 p'}{2p'_0(2\pi)^3}(2\pi)^4\delta(q-t-p-p').
\ea
Integrating over $d\Phi_2$ we obtain 
\ba
d\Phi_2=\frac{1}{4}\frac{1}{(2\pi)^2}\sqrt{1-\frac{\rho_2}{y}}\int_0^1 d v \int_0^{2\pi} d \phi 
\ea
where we have defined $v=1/2(1-\cos\theta)$, while the integration over $d\Phi_3$ brings us to 
\ba
d\Phi_3 = \int \frac{(2\pi)}{2 t_0} \frac{|p|^2 d|p|\Omega_3}{2p_0(2\pi)^3}
\frac{|p'|^2 d|p'|\sin{\beta}d\beta\Omega_2}{2{p'}_0(2\pi)^3}\delta(q_0-t_0-p_0-{p'}_0),
\ea
where $t_0$ and $\beta$ have been computed below 
\ba
&&t_0 =\sqrt{|\vec{t}|^2 +q^2 y}=\sqrt{|\vec{p}|^2 +|\vec{p'}|^2 +2|\vec{p}| |\vec{p'}|\cos{\beta} +q^2 y},
\nonumber\\
&&\cos{\beta}=\frac{[(2-x_1-x_2)^2 -4y] -(x_1^2-\rho_2)-(x_2^2-\rho_1)}{2\sqrt{x_1^2-\rho_2}\sqrt{x_2^2-\rho_1}}.
\ea
Finally we obtain
\ba
d\Phi_3 = \frac{q^2}{2 (4\pi)^3}\int dx_1 dx_2,
\ea
and the final result for the $d \Phi_4$ phase space is given by
\ba
d\Phi_4= \frac{q^4}{2! (4\pi)^6}\int_{\rho_2}^{\bar{y}_{+}} \sqrt{1-\frac{\rho_2}{y}} dy \int_{\sqrt{\rho_2}}^{\bar{x}_{1+}} dx_1 \int_{\bar{x}_{2-}}^{\bar{x}_{2+}} dx_2 
\int_0^1 dv \int_0^{2\pi}d\phi,
\ea
where the integration limits are discussed in the next section.

\subsection{Integration limits}

At this stage we need to define the integration limits of the integrals appearing in the four body 
phase space. 
From the definitions of $x_1,x_2$ it is clear that $0\leq x_1,x_2\leq 1$, but imposing the reality
condition of the square root we obtain 
\ba
x_1\geq \sqrt{\rho_2} && x_2\geq \sqrt{\rho_1}.
\ea  
Solving the condition $-1\leq \cos{\beta}\leq 1$ with respect to $x_2$ 
we obtain a bound on this variable which is given by
\ba
&&\bar{x}_{2\pm}=\frac{1}{8x_1 -2(\rho_2 + 4)}\left\{[(x_1 -2)(\rho_1 -4(x_1 + y -1)) +(x_1-2) \rho_2]
\right.\nonumber\\
&&\hspace{1cm}\left.\pm \sqrt{(x_1^2 -\rho_2)[16 x_1^2 +8 x_1(4 y  + \rho_1-\rho_2 -4) + 16 y^2 + (\rho_2-\rho_1+4)^2 
-8 y(\rho_1 +\rho_2 +4)]}\right\}.
\nonumber\\
\label{x2bound}
\ea
Again, we have to impose the condition  $\bar{x}_{2\p} \geq \sqrt{\rho_1}$ which gives us a condition 
on the variable $x_1$ 
\ba
x_1(y)\leq \frac{-4 y +\rho_1 +\rho_2 -4 \sqrt{\rho_1} +4}{4 -2\sqrt{\rho_1}},
\ea
but $x_1$ must be such that the square root in Eq. (\ref{x2bound}) is real 
\ba
&&x_1(y)\leq \frac{1}{4}(-4y -4\sqrt{\rho_1}\sqrt{y} -\rho_1 +\rho_2 +4)
\nonumber\\
&&x_1\geq \sqrt{\rho_2}.
\ea
From these three conditions we can extrapolate some conditions on the $y$ variable
\beq
\rho_2\leq y \leq \frac{1}{4}(\sqrt{\rho_1} + \sqrt{\rho_2} -2)^2
\eeq

\subsection{Phase space for a four axions final state}

In the case of a four axions final state we have a simplification in the computation since $\rho_2=\rho_1=\rho$.
Thus, the four body phase space is computed exactly as in \cite{Oleari:1997az} and the final result
is given by
\ba
d\Phi_4= \frac{q^4}{4! (4\pi)^6}\int_{\rho}^{\bar {y}_{+}} \sqrt{1-\frac{\rho}{y}} dy 
\int_{\sqrt{\rho}}^{\bar{x}_{1+}} dx_1 \int_{\bar{x}_{2-}}^{\bar{x}_{2+}} dx_2 \int_0^1 dv \int_0^{2\pi}d\phi
\ea 
where the factor $1/4!$ is a statistical factor that takes into account the four identical particles
in the final state and the integration bounds are defined as
\ba
&& \bar{x}_{2\pm}=\frac{1}{4(1-x_1) +\rho}\left[(2-x_1)(2+\rho -2y -2x_1) 
\pm 2\sqrt{(x_1^2 -\rho)[(x_1 -1+y)^2 -\rho y]} \right]
\nonumber\\
&&x_1\leq 1-y -\sqrt{\rho y}
\nonumber\\
&&\bar{y}_{1+}= (1-\sqrt{\rho})^2. 
\ea
These integrals have been computed numerically.

\end{document}